\documentclass[12pt]{article}%
\usepackage{amssymb}
\usepackage{amsfonts}
\usepackage{amsmath}
\usepackage{amsthm}
\usepackage[nohead]{geometry}
\usepackage[singlespacing]{setspace}
\usepackage[bottom]{footmisc}
\usepackage{indentfirst}
\usepackage{graphicx}%
\usepackage{rotating}
\usepackage{enumerate}
\usepackage{booktabs}
\usepackage[flushleft]{threeparttable}
\usepackage{float}
\usepackage{verbatim}
\usepackage{epstopdf}
\usepackage{pdflscape}

\newlength{\dhatheight}

\newlength{\dtildeheight}

\newtheorem{tm}{Theorem}[section]

\newtheorem{example}{Example}[section]
\newtheorem{pn}{Proposition}[section]
\newtheorem{cy}{Corollary}[section]
\newtheorem{assumption}{Assumption}[section]

\newtheorem{remark}{Remark}[section]
\newtheorem{appxlem}{Lemma}[section]

\usepackage[usenames,dvipsnames]{color}
\usepackage[round]{natbib}
\usepackage[hyperfootnotes=false]{hyperref}
\hypersetup{
  colorlinks,
  citecolor=Blue,
  linkcolor=Blue,
  urlcolor=Blue}

\numberwithin{equation}{section}

\ifdefined\pdfminorversion\pdfminorversion=4\fi
\begin{document}

\title{Time-Varying Tails: Estimation, Inference, and Risk Measurement}
\author{ Shuo Li\thanks{School of Statistics, Tianjin University of Finance and Economics, Tianjin 300222, China. E-mail: shuoli@tjufe.edu.cn.} ~and
Liuhua Peng\thanks{Corresponding author. School of Mathematics and Statistics,
The University of Melbourne, Victoria 3010, Australia.
E-mail: liuhua.peng@unimelb.edu.au. }}
\date{\today}

\maketitle

\begin{center}
\textbf{Abstract}
\end{center}
Motivated by evidence that the tails of economic and financial time series need not be stable over time, we develop a framework for studying time-varying tails. We model the extreme value index as a smooth function of time in the locally stationary framework of \citet{ZhouWu2009AS}. We estimate it by a ratio-form kernel Hill estimator in which a single intermediate quantile thresholds the whole local window. We establish consistency and asymptotic normality, and we construct uniform confidence bands. We also propose a consistent test of tail stability under extremal clustering. Combined with the Weissman extrapolation method, the estimator of the extreme value index yields time-varying extreme Value-at-Risk and Expected Shortfall. In an empirical application, we find the tail of daily loss of S\&P 500 is compatible with a constant shape, whereas that of volatility-filtered loss varies significantly.\\

\vspace{-0.5em}

\vspace{-0.5em}

\thispagestyle{empty}

\newpage \pagenumbering{arabic} \setcounter{page}{1} \setcounter{footnote}{0}

\section{Introduction}
The tails of economic and financial time series determine the likelihood of extreme events, such as market crashes, large insurance losses, and systemic disruptions. They therefore lie at the heart of modern risk management. Tail estimates feed directly into risk measurement, which underlies margin setting and capital requirements \citep{McNeilFrey2000JEF}. The standard summary of tail behavior is the extreme value index $\gamma$, which governs the decay rate of the survival function; see \citet{Embrechts1997} and \citet{deHaanFerreira2006} for comprehensive treatments. A large literature estimates the index from the largest order statistics of a stationary sample, beginning with \citet{Hill1975AS} and developed by \citet{Hsing1991AS} and \citet{Drees2000AS}, among others. Nearly all of this literature maintains that the index is constant over time.

A growing body of evidence challenges the assumption of stable tail shape. \citet{Quintos2001RES} document instability of the tail index of Asian equity and currency markets around the 1997--1998 crisis. \citet{deHaanZhou2021JASA} estimate a smoothly trending extreme value index and reject constancy for S\&P 500 losses over 1963--2012, although the trend is not detectable over the shorter 1988--2012 subsample. \citet{KellyJiang2014RFS} construct a time-varying measure of aggregate tail risk from the cross-section of U.S. stock returns and show that exposure to it is priced in expected returns. \citet{Massacci2017MS} documents that tail risk in U.S. equity returns is countercyclical and highly time-varying, and that it is linked to macroeconomic fundamentals and uncertainty. \citet{DInnocenzoLSZ2024JBES} find that both parameters of the generalized Pareto tail vary significantly over time for the S\&P 500 and for IBM. Taken together, these studies point to the same question. Does the tail shape itself move over time, and if so, where and by how much?

Answering the question requires four ingredients. The first is an estimator of the whole path $\gamma(u)$, whose sampling variability is understood under temporal dependence and extremal clustering. The second is simultaneous inference, so that the question of which movements are real can be answered without cherry-picking dates. The third is scale invariance of the estimator, so that shape variation is isolated from volatility variation. The fourth is a formal test of tail stability. This paper provides these four ingredients, together with the resulting time-varying risk measures.

In our framework, the extreme value index is modeled as a smooth deterministic function of time. We embed the data-generating process in the locally stationary framework of \citet{ZhouWu2009AS}. Specifically, we observe $\{X_{t,n}\}_{t=1}^{n}$ and the data are generated by $X_{t,n}=H(t/n,\mathcal{F}_{t})$ with $H$ measurable and smoothly varying in its first argument. Then $\gamma(u)$ is a function of rescaled time $u=t/n$. Temporal dependence is quantified by the geometric moment contraction condition of \citet{Wu2005PNAS}. At each time $u$, the marginal tail is regularly varying with index $\gamma(u)$ and satisfies a uniform second-order condition. With a bandwidth of a few years of daily data, the framework nests gradual drift and approximates sudden changes in the integrated sense relevant to estimation and testing. More importantly, it supports infill asymptotics under which estimation, uniform inference, and testing can all be analyzed rigorously.

This paper makes four contributions. First, we develop the inference theory for the common-threshold local Hill construction. At each evaluation point $u$, a single intermediate quantile thresholds all observations in the local window. The estimation error of the threshold then cancels between the numerator and the denominator of the estimator. As a result, the estimator attains the classical Hill variance, scaled by a kernel constant, the local exceedance count, and a long-run variance factor that captures extremal clustering. The same cancellation makes the smoothing bias free of any first-order scale component. Shape and scale are separated by construction.

Second, we construct a uniform confidence band for the path $\gamma(\cdot)$. The studentized estimation error process is approximated by a stationary Gaussian process whose correlation function is the kernel convolution, with a vanishing Kolmogorov error. The approximation is $h$-dependent, and any fixed Gaussian limit would be inconsistent. Dependence enters the band only through a local heteroscedasticity-and-autocorrelation-consistent estimator. In practice, we simulate the critical values from the pivotal finite-sample experiment at the working configuration. These pivotal critical values restore the nominal coverage in the independent designs we study.

Third, we propose tests of tail stability based on quadratic functionals of the deviation of the estimated path from a projection centering, namely the precision-weighted average of the path over the tested interval. Projection centering is essential for validity under scale-varying nulls. The integral statistic has nontrivial local power against alternatives of size $m^{-1/2}h^{-1/4}$. Ours isolates the extreme value index.

Fourth, we construct time-varying estimators of extreme Value-at-Risk and Expected Shortfall by the extrapolation method of \citet{Weissman1978JASA}, and we derive the joint limit of the threshold and the index estimators. At the extrapolation horizons used in practice, the threshold error is not negligible relative to the index error. Accounting for it inflates the standard error of the log-VaR by up to $62\%$ at the horizons of our application.

A series of Monte Carlo experiments confirm our theoretical results. The estimator is nearly unbiased under exact Pareto tails and exhibits the classical second-order bias under Student-$t$ tails. With pivotal critical values, the uniform band attains its nominal coverage under exact Pareto tails, and the stability test has approximately correct size under independent nulls and several dependent nulls, although it is liberal under strong GARCH-type clustering. Power is substantial against breaks of the size documented in the application. Control experiments with a constant index and a moving scale show that per-time-threshold alternatives manufacture spurious shape variation, while the ratio-form estimator does not.

In an empirical application, we estimate the extreme value index of daily S\&P 500 losses over 1990--2026. The estimated path varies between $0.22$ and $0.42$ around the level $0.30$. The omnibus test does not reject stability for the raw series ($p=0.22$), and a direct crisis-versus-calm contrast is insignificant. Extremal clustering masks all but the strongest signals. However, filtering the losses through a GARCH model removes the clustering and reveals the shape variation. The index of the standardized residuals varies between $0.14$ and $0.31$, and stability is rejected ($p\leq0.02$). For risk measurement, the decomposition is clear. Movements of the extreme risk measures are dominated by the intermediate-quantile component. Updating the threshold and the index over time cuts the violation rate of a static $99\%$ Value-at-Risk in the 2008--2009 crisis window from $10.7\%$ to $4.8\%$.

From the methodological perspective, this paper is related to three strands of literature. First, \citet{deHaanZhou2021JASA} estimate a smoothly trending extreme value index by a local Hill estimator with a common order-statistic threshold, the construction closest to ours. Their focus is pointwise estimation of the trend under independence, whereas we develop inference for the whole path under serial dependence and extremal clustering. our estimator is also invariant to arbitrary smooth scale variation, which isolates the tail shape from the scale. Second, the regression literature links the conditional index to observed covariates \citep{WangTsai2009JASA,GardesStupfler2014E,Youngman2019JASA}, and observation-driven models let the tail parameters follow an updating equation \citep{Massacci2017MS,DInnocenzoLSZ2024JBES}. These approaches deliver a conditional index given covariates or a parametric updating equation, and they require correct specification of the tail dynamics. Our nonparametric estimator is free of such commitments. It can therefore serve as a benchmark, and the stability tests can serve as primary specification checks for parametric tail dynamics. Third, \citet{Einmahl2016JRSSB} model heteroscedastic extremes, in which the scale varies while the index is constant, and \citet{BucherJennessen2024B} develop a bootstrap test of that null under serial dependence. Their question, stability of the scale, is the complement of ours. \citet{ZhangShao2026JBES} study time-varying high-quantile estimation for nonstationary tail-dependent series. They target the quantile path itself, whereas we separate its shape and scale components.

On the theory side, our estimator localizes the classical Hill estimator \citep{Hill1975AS,Hsing1991AS,Hill2010ET}, the second-order bias analysis follows \citet{Hall1982AS} and \citet{Drees1998SPA,Drees2000AS}, and extremal dependence enters through a local long-run variance factor in the spirit of the extremal index literature \citep{Leadbetter1983AS,Basrak2002SPA}. The locally stationary framework and the coupling-based tools are those of \citet{ZhouWu2009AS,ZhouWu2010JRSSB} and \citet{Wu2005PNAS}, and volatility filtering of extremes in the spirit of \citet{McNeilFrey2000JEF} appears in our application as a device to separate scale from shape.

The remainder of the paper is organized as follows. Section \ref{sec:data} presents a motivating example based on daily S\&P 500 losses and turns it into a set of methodological requirements. Section \ref{sec:model} sets up the framework, namely local stationarity, the time-varying tail model, and examples. Section \ref{sec:est} develops the estimator and its asymptotic properties. Sections \ref{sec:band}, \ref{sec:test}, and \ref{sec:risk} then develop the uniform band, the stability tests and the risk measures. Section \ref{sec:mc} reports simulations, Section \ref{sec:emp} the empirical study, and Section \ref{sec:concl} concludes. All proofs are relegated to the Appendix.

\section{A motivating example}\label{sec:data}
This section uses a rolling estimate of the extreme value index to introduce the methodological requirements for the rest of the paper.
Let $P_{t}$ denote the daily closing level of the S\&P 500 index and define percentage log-losses $X_{t}=-100\Delta\log P_{t}$, so that large positive values correspond to large market declines. Our sample runs from January 2, 1990 to June 29, 2026, $n=9179$ observations (details in Section \ref{sec:emp}). The series is heavy-tailed by any measure. Its kurtosis is $13.8$, its minimum is $-11.0\%$, and its maximum is $12.8\%$.

\begin{figure}[!htbp]
\centering
\includegraphics[width=0.98\textwidth]{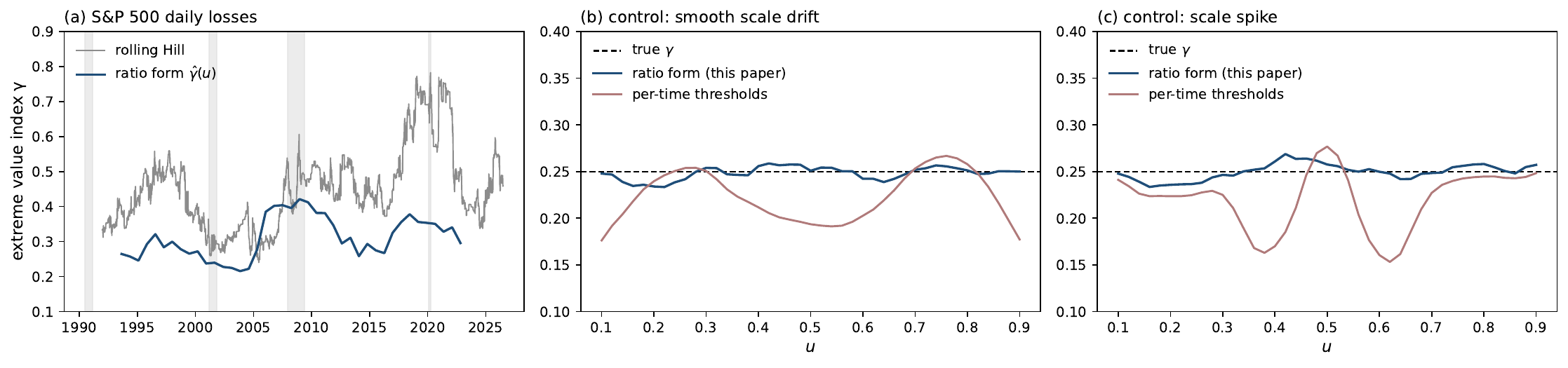}
\caption{Panel (a): naive rolling Hill estimator of the extreme value index of S\&P 500 daily losses (500-day windows, 50 order statistics, grey), and the ratio-form smoothed estimator $\hat{\gamma}(u)$ of Section \ref{sec:est} (blue, $m=200$, $h=0.10$), 1990--2026. Panels (b) and (c): control experiment on $9000$ simulated independent Pareto observations with constant extreme value index $\gamma=0.25$ and a deterministic scale process. Mean estimated paths over $60$ replications for the ratio-form estimator (blue) and the per-time-threshold estimator (red). Panel (b): smooth scale drift $1+0.8\sin^{2}(2\pi u)$. Panel (c): localized scale spike of factor $2$ on $u\in[0.45,0.55]$.}
\label{fig:motivation}
\end{figure}

A practitioner who suspects time-varying tails would typically begin with a rolling Hill plot. The grey path in Figure \ref{fig:motivation}(a) does exactly that. For each window of $500$ consecutive trading days, we compute the Hill estimator from the $50$ largest losses in the window. The path ranges from $0.24$ (mid-2000s) to $0.78$ (2020), with a correlation-like momentum of multi-year excursions rather than day-to-day jitter.

There are four explanations for the rolling plot, and they call for very different responses. We consider them in turn.

\emph{(i) Dependence.} Financial losses, and especially extreme losses, are naturally dependent. Crash days arrive in bunches, so large losses cluster in time. Although this produces more extreme observations in a given window, those observations are dependent on each other and therefore carry less independent information than the same number of independent extremes. A valid analysis must therefore model and adjust for this dependence in the framework.

\emph{(ii) Sampling noise.} With $50$ exceedances, the pointwise standard error of a Hill estimate is about $\gamma/\sqrt{50}\approx0.06$ at $\gamma=0.4$, so the pointwise interval has half-width $\approx0.12$. The swing from trough to peak is about $0.5$, which is four times the half-width. If the estimates at different dates were independent, sampling noise alone could hardly account for a swing of this size. However, they are not independent. Overlapping windows share most of their data, so the rolling path is highly autocorrelated by construction, and long excursions arise even under a constant extreme value index. Properly accounting for this correlation is exactly what a simultaneous band does. Moreover, such a band can only be built on a well-defined estimation path, not on an ad hoc sequence of window estimates.

\emph{(iii) The scale confound, and a control experiment.} The third possibility is that the estimator itself translates scale variation into apparent shape variation. ``The tail of the loss distribution'' bundles together the tail of the standardized innovations and the scale that multiplies them. Panels (b) and (c) demonstrate that this channel is quantitatively important for a natural class of smoothed estimators. We generate $9000$ independent Pareto observations with a constant extreme value index $\gamma=0.25$ and multiply them by a deterministic scale process. The scale process is a smooth drift in panel (b) and a spike of factor $2$ in panel (c), mimicking volatility variation. On these constant-shape samples we run two smoothed estimators, namely the ratio-form estimator of Section \ref{sec:est} and a per-time-threshold variant with the expected exceedance count as denominator. The figure reports the mean estimated paths over $60$ replications.

The per-time-threshold estimator dips to $0.15$ around the spike, a spurious deviation of $0.10$, and wanders by $\pm0.07$ under the drift. The ratio-form estimator deviates by at most $0.03$ in either design. The reason is that per-time thresholds create a timing mismatch. A scale surge raises the exceedance count at dates where the threshold has not yet adjusted, and the surplus of moderate exceedances masquerades as a lighter tail. Under a common evaluation-point threshold, each log-excess above the threshold remains exponentially distributed with mean $\gamma$, whatever the local scale, so the shape estimate is untouched. This experiment guides the construction of Section \ref{sec:est}, and it warns that any rolling diagnostic can overstate the amplitude of shape variation when the scale moves.

\emph{(iv) Genuine variation.} If $\gamma(u)$ truly varies, the rolling plot is an inefficient and noisy, though roughly unbiased-in-level, guide to it. The blue path in panel (a) previews the ratio-form estimator developed below. Its range, $0.22$ to $0.42$, is much narrower than the rolling plot suggests. The range is narrower for two reasons. First, the common-threshold construction removes the threshold-induced artifact documented in panels (b) and (c). Second, kernel smoothing pools information across neighboring windows and therefore reduces estimation variance. Whether the remaining variation is statistically significant, and whether it survives volatility filtering, is exactly what the uniform band of Section \ref{sec:band} and the stability tests of Section \ref{sec:test} are designed to answer.

Overall, the example shows that a valid analysis requires consideration of extremal dependence, an estimator of the whole index path, simultaneous inference, and scale invariance built into the estimator. It also requires a formal test for tail stability. We next set up the framework that delivers these ingredients and will return to this application in Section \ref{sec:emp}.\section{A framework for time-varying tails}\label{sec:model}
This section sets up the two ingredients of the model. The first is a dynamic structure that permits rigorous local-in-time asymptotics (Section \ref{sec:ls}). The second is a tail model with a time-varying extreme value index (Section \ref{sec:tail}).

\subsection{Dynamic structure}\label{sec:ls}
Suppose we observe $\{X_{t,n}\}_{t=1}^{n}$. Following \citet{ZhouWu2009AS}, let $\{e_{t}\}_{t=-\infty}^{\infty}$ be i.i.d.\ random variables and $\mathcal{F}_{t}=(\cdots,e_{t-1},e_{t})$ the associated filtrations. We assume
\begin{equation}\label{dgp}
X_{t,n}=H(u_t, \mathcal{F}_{t}),\qquad u_{t}=t/n,
\end{equation}
where $H$ is measurable and $H(u,\mathcal{F}_{t})$ is well defined for each $u\in[0,1]$. For fixed $u$, the process $H(u,\mathcal{F}_{t})$ is stationary. As $u$ varies, the data-generating mechanism changes smoothly, so $\{X_{t,n}\}$ is locally stationary. Infill asymptotics ($n\to\infty$ on the fixed interval $[0,1]$) formalize the idea that more data reveal the local structure at each time.

To quantify temporal dependence, we adopt the physical dependence measure of \citet{Wu2005PNAS}. Let $e_{0}'$ be an independent copy of $e_{0}$ and $\mathcal{F}_{t}^{*}=(\cdots,e_{-1},e_{0}',e_{1},\cdots,e_{t})$ the coupled filtration. For $q>0$ and $j\geq0$, define
\begin{equation}\label{pdm}
\delta_{q}(j)=\sup_{u\in[0,1]}\left\|H(u, \mathcal{F}_{j})-H(u, \mathcal{F}_{j}^{*})\right\|_{q},
\end{equation}
whenever the norm is finite, where $\|Z\|_{q}=(E|Z|^{q})^{1/q}$. The quantity $\delta_{q}(j)$ measures the contribution of the innovation $e_{0}$ to the system $j$ steps ahead, uniformly over rescaled time.

\begin{assumption}[Local stationarity and short-range dependence]\label{as:ls}
(i) The marginal distributions of $H(u,\mathcal{F}_{t})$ satisfy the tail conditions of Section \ref{sec:tail} with $\gamma(u)\in[\underline{\gamma},\bar{\gamma}]\subset(0,1)$ for all $u$. (ii) For some $q>2$, the log-scale process $W_{t,n}=\log\max(X_{t,n},1)$ satisfies $\sup_{u}\|H_{W}(u,\mathcal{F}_{0})\|_{q}<\infty$, where $W_{t,n}$ inherits the measurable representation $W_{t,n}=H_{W}(t/n,\mathcal{F}_{0,t},\ldots)$ from $X_{t,n}$. In addition, there exist $C>0$ and $\varrho\in(0,1)$ such that its physical dependence measure satisfies $\delta_{q}(j)\leq C\varrho^{j}$ for all $j\geq0$ (geometric moment contraction, GMC). (iii) Stochastic Lipschitz continuity holds, namely $\sup_{|u-v|\leq s}\|H_{W}(u,\mathcal{F}_{0})-H_{W}(v,\mathcal{F}_{0})\|_{q}\leq Cs$ for all $s\in[0,1]$.
\end{assumption}

\begin{remark}\label{rm:gmc}
Assumption \ref{as:ls} is standard in the locally stationary literature (see \citet{ZhouWu2009AS} and \citet{ZhouWu2010JRSSB}). Four comments are in order. First, parts (ii)--(iii) are formulated on the log-scale process $W_{t,n}=\log\max(X_{t,n},1)$ rather than on $X_{t,n}$ itself. Under the tail conditions of Section \ref{sec:tail}, $\Pr(W_{t,n}>w)\asymp e^{-w/\gamma(u)}$ up to a slowly varying factor, so $W_{t,n}$ possesses finite moments of \emph{every} order. Consequently, the moment requirement $q>2$ used by the central limit theorem and the exponential inequality (Lemmas \ref{lem:clt} and \ref{lem:expineq} in the Appendix) is compatible with any $\bar{\gamma}>0$. Formulating GMC directly on $X_{t,n}$ would instead force $q\bar{\gamma}<1$, which combined with $q>2$ imposes the extraneous side condition $\bar{\gamma}<1/2$. That restriction is not merely technical, since $\hat{\gamma}$ reaches $0.73$ in the 2016--2026 subsample of our application. Because $x\mapsto\log\max(x,1)$ is $1$-Lipschitz, we have $\delta_{q}^{W}(j)\leq\delta_{q}^{X}(j)$ whenever the latter is finite. The log-scale formulation is therefore strictly weaker than the raw-scale one. It also suffices for all of our results, since every statistic below is a (smoothed) function of log-excesses over high thresholds, which possess finite moments of all relevant orders (Lemma \ref{lem:mom} in the Appendix). 

Second, GMC is a mild short-range dependence condition. It is implied by geometric ergodicity of the underlying Markov structure, and it holds for the AR and GARCH-type recursions in Examples \ref{ex:ar} and \ref{ex:garch} below. Third, the physical dependence framework is particularly convenient for kernel-weighted statistics, since covariance bounds follow directly from coupling arguments (see Lemma \ref{lem:cov}). Fourth, the stochastic Lipschitz condition (iii) is likewise imposed on the log scale, which is the scale on which all bias expansions below are derived.
\end{remark}

Two examples show that time-varying tails arise naturally in standard dynamic models.

\begin{example}[Time-varying AR with heavy-tailed innovations]\label{ex:ar}
Let $X_{t,n}=a(u_{t})X_{t-1,n}+\sigma(u_{t})\varepsilon_{t}$, where $|a(u)|\leq\bar{a}<1$, $\sigma(\cdot)>0$, and $a$, $\sigma$ are Lipschitz, and $\varepsilon_{t}$ are i.i.d.\ regularly varying with index $1/\gamma(u)$, which is allowed to vary with $u$ through a smooth deformation $\varepsilon_{t}=G(u_{t},\eta_{t})$ with $\eta_{t}$ i.i.d. For instance, $\varepsilon_{t}$ can be Student-$t$ with $\nu(u_{t})$ degrees of freedom. The process admits the representation (\ref{dgp}) with $H(u,\mathcal{F}_{t})=\sum_{j\geq0}a(u)^{j}\sigma(u)\varepsilon_{t-j}$. By Karamata's theorem, the marginal extreme value index at time $u$ is $\gamma(u)$. 

The GMC condition follows by replacing the innovation $e_{0}$ (equivalently, $\eta_{0}$) by an independent copy $\eta_{0}'$. The coupled process then satisfies $H(u,\mathcal{F}_{j})-H(u,\mathcal{F}_{j}^{*})=a(u)^{j}\sigma(u)\bigl(G(u,\eta_{0})-G(u,\eta_{0}')\bigr)$, so $\delta_{q}(j)\leq C\bar{a}^{j}$. Note that the coupling acts on the \emph{innovation}, not on the observation. Replacing $X_{0,n}$ directly would violate the recursion at all subsequent dates, whereas replacing $\eta_{0}$ propagates through the recursion and keeps the coupled process marginally stationary at every frozen $u$. The same argument bounds $\sup_{|u-v|\leq s}\|H(u,\mathcal{F}_{0})-H(v,\mathcal{F}_{0})\|_{q}$, because the frozen processes $H(u,\mathcal{F}_{t})$ and $H(v,\mathcal{F}_{t})$ share the same innovation path and differ only through the Lipschitz coefficients. Extremes of this process cluster. The extremal index is $\theta(u)=1-|a(u)|^{1/\gamma(u)}$, so that the long-run variance factor (\ref{lrv}) satisfies $S(u)>1$. The time-varying recursion is coupled to the frozen-coefficient process at each $u$ by driving both with the same innovation sequence. Over a window of radius $h$, the log-scale difference between the time-varying and the frozen process is $O(h)$ uniformly, so the local tail probabilities computed on the frozen process differ from the true ones by a factor $1+o(1)$ under Assumption \ref{as:tailloc}. On the log scale, the same coupling yields $\delta_{q}^{W}(j)\leq C\bar{a}^{j}$ for every $q>0$, since $x\mapsto\log\max(x,1)$ is $1$-Lipschitz and $W$ has finite moments of every order. On the raw scale, the bound holds only for $q\gamma<1$, which is precisely why the log-scale formulation is adopted. Proposition \ref{pn:ar} below verifies Assumption \ref{as:extdep} for this example and computes $s_{j}(u)$ and $S(u)$ explicitly.
\end{example}

\begin{example}[Time-varying GARCH-type dynamics]\label{ex:garch}
Let $X_{t,n}=\sigma_{t,n}\varepsilon_{t}$ with $\sigma_{t,n}^{2}=\omega(u_{t})+\alpha(u_{t})X_{t-1,n}^{2}+\beta(u_{t})\sigma_{t-1,n}^{2}$, where $\omega,\alpha,\beta$ are nonnegative Lipschitz functions with $\sup_{u}(\alpha(u)\kappa+\beta(u))<1$ for the appropriate moment $\kappa$, and $\varepsilon_{t}$ i.i.d.\ with a light (e.g., standard normal) tail. For fixed $u$, the stationary GARCH recursion has Pareto tails. Both $X$ and $\sigma$ are regularly varying with index $\kappa(u)$ solving $E[(\alpha(u)\varepsilon^{2}+\beta(u))^{\kappa(u)/2}]=1$, an application of Kesten's theorem to the squared volatility recursion (see \citet{Basrak2002SPA} and the references therein). In our notation, $\gamma(u)=1/\kappa(u)$. Since the fixed-point equation varies smoothly with $u$, so does $\gamma(u)$. The representation (\ref{dgp}) and the GMC condition follow from the contraction properties of the stochastic recurrence equation. In fact, \citet[Theorem 2.8]{Basrak2002SPA} establishes that the associated Markov chain is geometrically ergodic, hence strongly mixing at a geometric rate, under a positive innovation density near the origin. This example shows that time-varying tails arise naturally from time-varying volatility persistence, even with light-tailed innovations. Shape and scale are linked, but distinct, features of the dynamics.
\end{example}

Time variation in the extreme value index is conceptually distinct from the well-documented time variation in volatility. GARCH-type models \citep{Bollerslev1986JOE} capture clustering and smooth evolution of the conditional scale, and under suitable conditions a stationary GARCH process has a constant extreme value index determined by the volatility recursion \citep{Basrak2002SPA}. Volatility filtering therefore rescales the tails but does not, by itself, change their shape. Movements in the extreme value index instead reflect deeper changes, such as changes in the composition of market participants, in the leverage of financial intermediaries, and in the frequency of truly extreme news. Distinguishing scale variation from shape variation requires an estimator that is localized in time yet invariant to the local scale. The estimator of Section \ref{sec:est} has exactly this property. Moreover, the empirical study of Section \ref{sec:emp} confronts the distinction directly by filtering the losses through a GARCH model in the spirit of \citet{McNeilFrey2000JEF}.

\subsection{The time-varying extreme value index}\label{sec:tail}
For each $u\in[0,1]$, denote by $F(\cdot,u)$ the marginal CDF of $H(u,\mathcal{F}_{t})$ and by $F^{\leftarrow}(\cdot,u)$ its left-continuous inverse. Then the $(1-1/x)$-quantile is
\[
Q(x, u)=F^{\leftarrow}(1-1/x, u).
\]
We assume that $Q(\cdot,u)$ is regularly varying with extreme value index $\gamma(u)>0$. That is, for all $y>0$,
\begin{equation}\label{right evi}
\lim_{x\to\infty}\frac{Q(xy, u)}{Q(x, u)}=y^{\gamma(u)}.
\end{equation}
The extreme value index $\gamma(u)$ is time-varying. For asymptotic theory we need (\ref{right evi}) to hold uniformly in $u$, together with a rate condition.

\begin{assumption}[Uniform regular variation]\label{as:rv}
There exist a positive function $c(\cdot)$ and a bounded continuous function $\gamma(\cdot)$ taking values in $[\underline{\gamma},\bar{\gamma}]\subset(0,1)$ such that the survival function satisfies $\bar{F}(x,u)\equiv 1-F(x,u)=c(u)x^{-1/\gamma(u)}\ell(x,u)$, where $\ell(\cdot,u)$ is slowly varying at infinity uniformly in $u$. That is, for every sequence $x_{n}\to\infty$ and every sequence $y_{n}$ bounded and bounded away from zero, $\sup_{u}|\ell(x_{n}y_{n},u)/\ell(x_{n},u)-1|\to0$.
\end{assumption}

\begin{assumption}[Uniform second-order condition]\label{as:so}
There exist functions $A(\tau,u)$, with $\sup_{u}|A(\tau,u)|\to0$ as $\tau\to\infty$ and $\tau\mapsto|A(\tau,u)|$ regularly varying with index $\rho(u)$, where $\rho(\cdot)$ is continuous and takes values in $[\underline{\rho},0)$ for some $\underline{\rho}<0$, such that, uniformly in $u\in[0,1]$ and locally uniformly in $y>0$,
\begin{equation}\label{so-quantile}
\frac{\bar{F}(b(\tau,u)e^{y},u)}{\bar{F}(b(\tau,u),u)}=e^{-y/\gamma(u)}\left(1+A(\tau,u)\psi(y,u)+o\left(|A(\tau,u)|\right)\right),
\end{equation}
where $b(\tau,u)=F^{\leftarrow}(1-\tau^{-1},u)$ and
\begin{equation}\label{psi-normal}
\psi(y,u)=\frac{e^{y\rho(u)/\gamma(u)}-1}{\rho(u)}.
\end{equation}
The normalization (\ref{psi-normal}) is essential. Without it, the pair $(A,\psi)$ is determined only up to a multiplicative constant, and the bias coefficient in Theorem \ref{tm:an} would not be well defined. The condition is the uniform version of the classical second-order regular variation condition \citep{deHaanFerreira2006,Drees2000AS}. Theorem 2.3.9 of \citet{deHaanFerreira2006} transports it to the quantile function used in the proofs. We also assume the associated Potter-type bound. For every $\iota>0$ there exists $C_{\iota}$ such that $\bar F(b(\tau,u)e^{y},u)/\bar F(b(\tau,u),u)\leq C_{\iota}e^{-y/(\gamma(u)+\iota)}$ for all $y>0$ and all sufficiently large $\tau$, uniformly in $u$. This provides the integrable domination needed when the expansion is integrated up to $y\asymp\log(n/m)$.
\end{assumption}

\begin{remark}\label{rm:so}
Condition (\ref{so-quantile}) is the uniform version of the classical second-order regular variation condition underlying the asymptotic normality and bias analysis of extreme value index estimators \citep{deHaanFerreira2006,Drees2000AS}. Three representative cases are worth noting. First, exact Pareto tails, $\bar{F}(x,u)=c(u)x^{-1/\gamma(u)}$, give $A\equiv0$. Second, Student-$t$ tails with $\nu(u)=1/\gamma(u)$ degrees of freedom give $\rho(u)=-2\gamma(u)$ and $|A(\tau,u)|$ of order $\tau^{-2\gamma(u)}$. This slow decay generates the well-known finite-sample bias of the Hill estimator. Third, the Hall class, $\bar{F}(x,u)=c(u)x^{-1/\gamma(u)}(1+d(u)x^{\rho(u)/\gamma(u)}+o(x^{\rho(u)/\gamma(u)}))$, gives a parametric family of second-order behaviors. The normalization (\ref{psi-normal}) determines $\psi$ uniquely up to the sign convention of $A$. It is integrable against $e^{-y/\underline{\gamma}}$ whenever $\rho(u)<0$, which is why the assumption excludes the borderline case $\rho=0$. That case requires a separate, model-specific treatment.
\end{remark}

Finally, we impose smoothness on the extreme value index and on the intermediate quantile as functions of time.

\begin{assumption}[Smoothness in time]\label{as:smooth}
$\gamma(\cdot)$ is twice continuously differentiable on $[0,1]$. For each $\tau>1$, $b(\tau,\cdot)$ is twice continuously differentiable on $[0,1]$ with $\sup_{u}|\partial^{2}\log b(\tau,u)/\partial u^{2}|=O(\log\tau)$ uniformly in $\tau$, and the density $f(x,u)=\partial F(x,u)/\partial x$ exists, is positive and continuous in a neighborhood of $(b(\tau,u),u)$, and satisfies $b(\tau,u)f(b(\tau,u),u)=\gamma(u)^{-1}\tau^{-1}(1+o(1))$ uniformly in $u$ as $\tau\to\infty$. Moreover, the conditional distribution of $\log X_{t,n}$ given $\{e_{s}:s\neq t\}$ admits a density bounded by $C\bar{F}(b(n/m,t/n),t/n)$ on a neighborhood of $\log b(n/m,t/n)$ of fixed logarithmic width, uniformly in $t$ and in the conditioning values. This is a high-level smoothness requirement on the conditional tail, used only to control exceedance strips of width $O((mh)^{-1/2}\sqrt{\log n})$. For Example \ref{ex:ar} it follows from the corresponding tail-density property of the innovations when $|a(u)|$ is bounded away from one, since the conditional law of $X_{t,n}$ given $\{e_{s}:s\neq t\}$ is a scaled and shifted innovation law. For GARCH-type dynamics with $\sigma_{t}\approx b$, the conditional log-density at the threshold is $f_{\varepsilon}(b/\sigma_{t})/\sigma_{t}=O(1)$ rather than $O(m/n)$, and we therefore impose the weaker bound $C(m/n)^{1-\iota}$ for some $\iota>0$. The weaker bound changes only logarithmic factors in the strip estimates and does not affect the stated rates.
\end{assumption}

\begin{assumption}[Tail locality]\label{as:tailloc}
The marginal tail varies sufficiently smoothly relative to the bandwidth that
\begin{equation}\label{tailloc}
\sup_{u\in(0,1)}\sup_{|v-u|\leq h}\left|\frac{\bar{F}(b(n/m,u),\,v)}{\bar{F}(b(n/m,u),\,u)}-1\right|\to0,
\qquad\text{and}\qquad h\log(n/m)\to0,
\end{equation}
as $n\to\infty$. In addition, the same display holds with $b(n/m,u)$ replaced by $b(n/m,u)e^{y}$, uniformly over $y$ in compact subsets of $\mathbb{R}$.
\end{assumption}

\begin{remark}\label{rm:tailloc}
Assumption \ref{as:tailloc} is a convenient sufficient condition for the local analysis, not a necessary one for consistency of the ratio estimator. Its role is to keep the local exceedance probabilities and the bias expansions tractable. With a common threshold $q$ at the evaluation point, the population ratio $\Gamma_h(u;q)=\int K(v)\gamma(u+hv)\bar F(q,u+hv)dv/\int K(v)\bar F(q,u+hv)dv$ is a nonnegative weighted average of the tail indices near $u$. Hence $|\Gamma_h(u;q)-\gamma(u)|\leq\sup_{|v|\leq1}|\gamma(u+hv)-\gamma(u)|=O(h)$, even if $h\log(n/m)\to c_{0}>0$. The tilt of the exceedance probabilities enters the numerator and the denominator with the same factor and cancels, so no $O(1)$ bias arises. With $m=n^{a}$, the condition $h\log n\to0$ holds for any bandwidth sequence $h\asymp n^{-b}$, $b>0$. The first part of (\ref{tailloc}) follows from Assumptions \ref{as:rv} and \ref{as:smooth} together with $h\log(n/m)\to0$, since $\bar F(b(n/m,u),v)/\bar F(b(n/m,u),u)=(b(n/m,u)/b(n/m,v))^{1/\gamma(v)}(1+o(1))$ and the log-ratio of the quantiles is $O(h\log(n/m))$ by the Lipschitz smoothness of $\log b(\tau,\cdot)$ in $\tau$ and $u$. We state it as an assumption because it is the exact high-level condition used in the proofs, and because it makes the restriction on the tail dynamics transparent.
\end{remark}

\section{Estimation of the extreme value index}\label{sec:est}
\subsection{A smoothed Hill estimator}\label{sec:estdef}
To motivate the estimator, recall the idea behind the estimator of \citet{Hill1975AS}. Suppose that a random variable $X$ with distribution $F$ has extreme value index $\gamma$. Write $x_{+}=\max(x,0)$ and, for any $\tau>1$, $b(\tau)=F^{\leftarrow}(1-\tau^{-1})$. Let $m/n\to0$ as $n\to\infty$. Some manipulations deliver
\begin{eqnarray*}
&&E[\log X/b(n/m)]_{+}^{k}\\
&=&\bar{F}(b(n/m))\int_{0}^{\infty}\frac{\bar{F}(b(n/m)e^{u^{1/k}})}{\bar{F}(b(n/m))}\,du\\
&\to&\frac{m}{n}k!\gamma^{k},
\end{eqnarray*}
which implies that
\[\gamma^{k}\approx\frac{n}{mk!}E[\log X/b(n/m)]_{+}^{k}.\]
The Hill (1975) estimator is simply the corresponding moment estimator with $k=1$. Under dependence, \citet{Hsing1991AS} and \citet{Hill2010ET} establish its asymptotic theory for stationary sequences. In addition, \citet{Hall1982AS} analyzes its mean squared error under a second-order condition, and \citet{Drees1998SPA} develops data-driven choices of the sample fraction.

Under the time-varying model of Section \ref{sec:model}, the same moment conditions hold locally. That is, for each $u\in[0,1]$,
\begin{equation}\label{moment condition}
\frac{n}{mk!}E[\log H(u, \mathcal{F}_{t})-\log b(n/m, u)]_{+}^{k}-\gamma^{k}(u)\approx0,
\end{equation}
where $b(\tau, u)=F^{\leftarrow}(1-\tau^{-1}, u)$. We call $b(n/m, u)$ the intermediate quantile at time $u$. Two features of (\ref{moment condition}) drive the construction. First, the condition is \emph{local}, since it identifies $\gamma(u)$ from observations with $t/n\approx u$. This suggests kernel weighting in time. Second, it is \emph{scale-free}. Multiplying $H(u,\mathcal{F}_{t})$ by a positive scale $\sigma(u)$ shifts $\log b(n/m,u)$ by $\log\sigma(u)$ and leaves the centered log-excesses unchanged. Hence, the moment conditions, and any estimator built on them with a locally estimated threshold, isolate the shape of the tail from its scale (cf.\ Section \ref{sec:data}).

Suppose we have an estimator $\hat{b}(n/m,u)$ of the intermediate quantile $b(n/m,u)$ at the evaluation point $u$ (Section \ref{sec:iq}). Then (\ref{moment condition}) with $k=1$ motivates the ratio-form smoothed Hill estimator
\begin{equation}\label{local index estimator}
\hat{\gamma}(u)=\frac{\sum_{t=1}^{n}\left(\log X_{t,n}-\log \hat{b}(n/m,u)\right)_{+}K_{h}(u-t/n)}{\sum_{t=1}^{n}\mathbf{1}\{X_{t,n}>\hat{b}(n/m,u)\}K_{h}(u-t/n)},
\end{equation}
where $K_{h}(v)=K(v/h)/h$, $K$ is a kernel function, and $h=h_{n}\to0$ is a bandwidth. Two features of (\ref{local index estimator}) are essential. First, the threshold is \emph{common} to all observations in the window. It is evaluated at the evaluation point $u$, not at each observation time $t/n$. Second, the estimator is a \emph{ratio} of two kernel averages, namely the smoothed log-excess moment and the smoothed exceedance count. The classical Hill estimator has exactly this form, with the sample quantile as the common threshold. The estimator (\ref{local index estimator}) is its direct local analogue. It is a local constant (Nadaraya--Watson type) smoother of the exceedance moments. 

Local polynomial extensions, which reduce the boundary bias, are straightforward but are not pursued here. Throughout, estimation and inference are developed for the interior region $u\in[h,1-h]$, and the evaluation ranges of the uniform bands and tests below are chosen accordingly.

The common-threshold ratio form is what delivers both the exact variance constant and the scale invariance, and it is worth contrasting with the natural alternative. If one re-estimates the threshold at every observation time, $\hat{b}(n/m,t/n)$, the stochastic fluctuations of the $n$ individual thresholds survive the kernel averaging. The first-order variance of the resulting estimator is then a convolution of the kernel with itself, $\gamma^{2}(u)S(u)(mh)^{-1}\int(K-K*K)^{2}+\int K^{2}$. This exceeds the classical Hill variance by about $7\%$ for the Epanechnikov kernel. The threshold fluctuations also leak scale variation into the shape estimate (Figure \ref{fig:motivation}). With a common threshold at the evaluation point, there is only \emph{one} threshold fluctuation per evaluation point, and it cancels between the numerator and the denominator of (\ref{local index estimator}), exactly as in the classical Hill estimator. Theorem \ref{tm:an} makes this precise.

\begin{remark}\label{rm:scale-free}
The estimator is asymptotically scale-free. Multiplying $X_{t,n}$ by any positive time-varying scale $\sigma(t/n)$ shifts every log-excess by $\log\sigma(t/n)-\log\sigma(u)+o_{p}(1)=O(h)$ within the window, and it shifts the estimated threshold by the same amount up to $o_{p}((mh)^{-1/2})$. The first-order stochastic expansion of $\hat{\gamma}(u)$ is therefore completely unchanged by smooth scale variation (Theorem \ref{tm:an}). Moreover, the smoothing bias contains no first-order scale component either. The $O(h^{2})$ bias of the numerator and the denominator involve the curvature of the log-threshold path $u\mapsto\log b(n/m,u)$, which includes the curvature of $\log\sigma(\cdot)$, and these curvature terms cancel in the ratio up to a residual of order $h^{2}\log(n/m)$ that depends only on the \emph{shape} derivatives $\gamma'$, $\gamma''$ (see the bias expression (\ref{bias}) and Remark \ref{rm:bias} below). Pure scale variation, deterministic or stochastic, contributes nothing to the estimand and nothing to the first-order limit law. This invariance is what allows the estimator to separate movements in the tail shape from movements in volatility, and it is exactly what the control experiment of Figure \ref{fig:motivation} verifies in finite samples. In finite samples the invariance is approximate, since the threshold is estimated from about $nh$ observations, and we quantify the residual sensitivity in Section \ref{sec:mc}.
\end{remark}

\begin{assumption}[Kernel]\label{as:kernel}
$K$ is a symmetric, nonnegative, bounded density with support $[-1,1]$, and $K$ is Lipschitz continuous. Write $\nu_{2}(K)=\int_{-1}^{1}K^{2}(v)dv$ and $\mu_{2}(K)=\int_{-1}^{1}v^{2}K(v)dv$.
\end{assumption}

\begin{assumption}[Intermediate order and bandwidth]\label{as:rates}
As $n\to\infty$, $m=m_{n}\to\infty$, $m/n\to0$, $h=h_{n}\to0$, $mh\to\infty$, $mh\geq n^{\theta}$ for some $\theta\in(0,1)$, and $(\log n)^{2}/(mh)\to0$.
\end{assumption}

The effective number of exceedances entering $\hat{\gamma}(u)$ is of order $mh$. Indeed, the local window contains about $nh$ observations, and each of them exceeds the local intermediate threshold with probability about $m/n$. Hence $mh\to\infty$ is the minimal consistency condition, and the stochastic order of $\hat{\gamma}(u)-\gamma(u)$ is $(mh)^{-1/2}$, up to bias.

To state the limit distribution we need to quantify the dependence between large exceedances at different time points. Define the exceedance variables at the evaluation point $u$,
\begin{equation}\label{excess}
\xi_{t,n}(u)=\left(\log X_{t,n}-\log b(n/m,u)\right)_{+},\qquad \mathcal{I}_{t,n}(u)=\mathbf{1}\{\xi_{t,n}(u)>0\},
\end{equation}
and the standardized summands $Z_{t,n}(u)=\xi_{t,n}(u)-\gamma(u)\mathcal{I}_{t,n}(u)$. Under Assumptions \ref{as:rv} and \ref{as:tailloc}, $E[Z_{t,n}(u)^{2}]=(m/n)\gamma^{2}(u)(1+o(1))$ uniformly in $t$ with $|t/n-u|\leq h$ (Lemma \ref{lem:mom}). Note that the log-excess is defined only on the exceedance event, so that $\log X_{t,n}$ enters exclusively through positive parts over a positive threshold. Values of $X_{t,n}\leq0$ therefore require no convention.

\begin{assumption}[Extremal dependence]\label{as:extdep}
There exist constants $C>0$ and $\varrho\in(0,1)$, and continuous functions $s_{j}:[0,1]\to[0,\infty)$ with $\sum_{j=1}^{\infty}s_{j}(u)\leq C$, such that, uniformly in $u$ and in $t$ with $|t/n-u|\leq2h$:
(i) for all $j\geq1$ and $x,y\geq1$, with $b=b(n/m,u)$ and $b_{j}=b(n/m,(t+j)/n)$,
\[
\left|P\left(X_{t,n}>bx,\;X_{t+j,n}>b_{j}y\right)-\bar{F}(bx,t/n)\,\bar{F}(b_{j}y,(t+j)/n)\right|\leq C\,\frac{m}{n}\,\varrho^{\,j}\,(xy)^{-1/(2\bar{\gamma})};
\]
(ii) for each fixed $j\geq1$, $(n/m)\,\mathrm{Cov}\left(Z_{t,n}(u),Z_{t+j,n}(u)\right)\to\gamma^{2}(u)\,s_{j}(u)$ as $n\to\infty$.
\end{assumption}

Assumption \ref{as:extdep} is a local, summable version of the tail-dependence conditions standard in the extreme value literature for dependent data \citep{Hsing1991AS,Hill2010ET}. Part (i) is a \emph{tail-mixing} condition. It bounds the joint exceedance probability net of the product of the marginals, so that it controls covariances of extreme indicators directly (Lemma \ref{lem:cov}). For the AR and GARCH dynamics of Examples \ref{ex:ar} and \ref{ex:garch}, it follows from the geometric contraction of the state recursion together with the regular variation of the innovation tail, by standard extreme value arguments for such recursions \citep{Hsing1991AS}. Since $s_{j}(u)\leq C\varrho^{\,j}$ by (i), we also have $\sum_{j}j\,s_{j}(u)<\infty$. Part (ii) identifies the limiting extremal autocorrelations $s_{j}(u)$, that is, the (suitably normalized) tail process of the locally stationary sequence. Define the local long-run variance factor
\begin{equation}\label{lrv}
S(u)=1+2\sum_{j=1}^{\infty}s_{j}(u),
\end{equation}
which is finite and continuous by Assumption \ref{as:extdep}. Under asymptotic tail independence (a $D(u_{n})$-type condition in the sense of \citet{Leadbetter1983AS}), we have $s_{j}(u)\equiv0$ and $S(u)\equiv1$. In contrast, for processes exhibiting extremal clustering (e.g., the AR and GARCH dynamics of Examples \ref{ex:ar} and \ref{ex:garch}), we have $s_{j}(u)>0$ and $S(u)>1$.

The assumption is verifiable on concrete dynamics, and the verification yields the constants of interest in closed form.

\begin{pn}[Extremal dependence of the AR dynamics]\label{pn:ar}
Let the conditions of Example \ref{ex:ar} hold with $a(u)\geq0$, and suppose the innovation distribution has a density satisfying the boundedness condition of Assumption \ref{as:smooth}. Then Assumption \ref{as:extdep} holds, and the limiting extremal autocorrelations are
\begin{equation}\label{sj-ar}
s_{j}(u)=a(u)^{j/\gamma(u)},\qquad j\geq1.
\end{equation}
Consequently, with $r(u)=a(u)^{1/\gamma(u)}$,
\begin{equation}\label{S-ar}
S(u)=\frac{1+r(u)}{1-r(u)},\qquad\text{and the local extremal index is }\ \theta(u)=1-r(u),
\end{equation}
in agreement with the classical constant-parameter AR(1) formula \citep[Section 8.1]{Embrechts1997}.
\end{pn}

\begin{remark}\label{rm:ar-mc}
Two features of (\ref{sj-ar})--(\ref{S-ar}) are worth noting. First, persistence translates directly into the variance inflation of $\hat{\gamma}(u)$. As $a(u)\uparrow1$, $S(u)$ diverges like $2\gamma(u)/(1-a(u))$, while the extreme value index itself is untouched. Second, the formulas provide an independent check on the HAC estimator of Section \ref{sec:band}. At the midpoint of the Monte Carlo design DGP3 ($a=0.3$, $\gamma(0.5)=0.35$), we obtain $r(0.5)=0.032$, so $S(0.5)=1.07$ and $\theta(0.5)=0.97$. The simulation average of the HAC estimator (\ref{shat}) is $\bar{\hat{S}}\approx1.02$, and the small downward gap is the usual finite-lag bias of kernel HAC estimators. A proof of Proposition \ref{pn:ar} is given in Section \ref{app:verify}.
\end{remark}

\begin{pn}[Extremal dependence of the GARCH dynamics]\label{pn:garch}
Let the conditions of Example \ref{ex:garch} hold, and suppose the innovation distribution has a density satisfying the boundedness condition of Assumption \ref{as:smooth}. Write $A_{1}^{*}(u)=\alpha(u)(\varepsilon_{0}^{*})^{2}+\beta(u)$ and $A_{k}(u)=\alpha(u)\varepsilon_{k}^{2}+\beta(u)$ for $k\geq2$. Then Assumption \ref{as:extdep} holds, and the limiting extremal autocorrelations are
\begin{equation}\label{sj-garch}
s_{j}(u)=E\left[\min\left\{1,\,e^{V_{j}(u)/\gamma(u)}\right\}\right],\qquad V_{j}(u)=\frac{1}{2}\Bigl[\log A_{1}^{*}(u)+\sum_{k=2}^{j}\log A_{k}(u)\Bigr]+\log|\varepsilon_{j}|-\log|\varepsilon_{0}^{*}|,
\end{equation}
where the sum over $k\geq2$ is empty for $j=1$, the expectation is over the i.i.d.\ innovation sequence, and $\varepsilon_{0}^{*}$ is the \emph{size-biased} initial innovation, independent of $(\varepsilon_{k})_{k\geq1}$, with distribution $P(\varepsilon_{0}^{*}\in B)=E[|\varepsilon_{0}|^{\kappa(u)}\mathbf{1}\{\varepsilon_{0}\in B\}]/E|\varepsilon_{0}|^{\kappa(u)}$ and $\kappa(u)=1/\gamma(u)$. The first multiplier $A_{1}^{*}$ is built from the same size-biased innovation that produced the extreme value at time $t$, since $\sigma_{t+1}^{2}=\omega+\alpha X_{t}^{2}+\beta\sigma_{t}^{2}$ and $X_{t}=\sigma_{t}\varepsilon_{0}^{*}$ with $\varepsilon_{0}^{*}$ size-biased given $X_{t}>b$. Since $E[\log A_{k}(u)]<0$ by the stability condition of Example \ref{ex:garch}, there exists $\iota>0$ with $\varrho(u)\equiv E[A_{k}(u)^{\iota/2}]<1$, and $s_{j}(u)\leq C\varrho(u)^{j}$. Extremal dependence therefore decays geometrically at a rate governed by the volatility recursion, so $S(u)$ is finite with $S(u)>1$ whenever $\alpha(u)>0$. A proof is given in Section \ref{app:verify}.
\end{pn}

Formula (\ref{sj-garch}) is the locally stationary version of the tail-process representation for GARCH recursions \citep{Basrak2002SPA}. Conditionally on an exceedance at time $t$, the initiating shock is size-biased, which is the content of the change of measure implicit in Kesten's theorem, and the volatility recursion then propagates the shock through the products $\prod_{k=1}^{j}A_{k}$. The probability that the shock survives $j$ periods on the exceedance scale is the expectation in (\ref{sj-garch}). The size-biasing of the initial innovation matters quantitatively. Replacing $\varepsilon_{0}^{*}$ by an unbiased $\varepsilon_{0}$ would shrink the denominator $|\varepsilon_{0}^{*}|$ in the ratio $X_{t+j}/X_{t}$ and therefore generally \emph{overstate} $s_{j}(u)$, rather than understate it. The size-biased multiplier $A_{1}^{*}$ is likewise larger than $A_{1}$, partially offsetting the denominator effect, but the two need not cancel. It is incorrect already for $j=1$. Together with Proposition \ref{pn:ar}, this shows that the two canonical dynamic models in the class satisfy the extremal-dependence assumption with explicitly computable ingredients.

\begin{tm}[Consistency and asymptotic normality]\label{tm:an}
Suppose Assumptions \ref{as:ls}--\ref{as:extdep} hold, and that $\hat{b}(n/m,u)$ is the kernel quantile estimator (\ref{bhat}) below based on the same kernel $K$ and bandwidth $h$. Then, for each interior $u\in(0,1)$, $\hat{\gamma}(u)\to_{p}\gamma(u)$. If moreover $\sqrt{mh}\,h^{2}\log(n/m)\to0$ and $\sqrt{mh}\,A(n/m,u)\to0$,
\begin{equation}\label{an}
\sqrt{mh}\left(\hat{\gamma}(u)-\gamma(u)\right)\to_{d}N\left(0,\;\gamma^{2}(u)\,S(u)\,\nu_{2}(K)\right).
\end{equation}
More generally, without the undersmoothing conditions,
\[
\sqrt{mh}\left(\hat{\gamma}(u)-\gamma(u)-B_{n}(u)\right)\to_{d}N\left(0,\;\gamma^{2}(u)\,S(u)\,\nu_{2}(K)\right),
\]
where, writing $L_{n}=\log(n/m)$, the bias term is
\begin{equation}\label{bias}
B_{n}(u)=h^{2}\mu_{2}(K)\left[\frac{1}{2}\gamma''(u)+\frac{\bigl(\gamma'(u)\bigr)^{2}L_{n}}{\gamma(u)}\right]+\frac{\gamma(u)\,A(n/m,u)}{1-\rho(u)}+o\bigl(h^{2}L_{n}+|A(n/m,u)|\bigr).
\end{equation}
The smoothing component of $B_{n}(u)$ depends only on the \emph{shape} derivatives. The curvature of the log-threshold path $u\mapsto\log b(n/m,u)$, which includes the curvature of any smooth scale process $\log\sigma(u)$, enters the numerator and the denominator of (\ref{local index estimator}) with the same sign and cancels in the ratio at order $h^{2}$. What survives is the interaction of the shape slope with the log-threshold tilt $(\gamma'(u))^{2}L_{n}/\gamma(u)$ and the shape curvature $\gamma''(u)/2$ (Section \ref{app:main}). For a pure scale family $\bar{F}(x,u)=\bar{G}(x/\sigma(u))$, we have $\gamma'=\gamma''=0$, and the smoothing bias vanishes identically (cf.\ Remark \ref{rm:scale-free}).
\end{tm}

\begin{remark}\label{rm:threshold}
The asymptotic variance in (\ref{an}) coincides with the classical Hill variance $\gamma^{2}$ scaled by $\nu_{2}(K)/(mh)$ and by the dependence factor $S(u)$. Two separate normalizations deliver this constant. First, the ratio form uses the realized exceedance count in the denominator, and the residualized score $\xi-\gamma\mathcal{I}$ has variance $\gamma^{2}$ per observation. A tail-moment estimator with a fixed probability denominator would instead carry the variance $2\gamma^{2}$ of the raw log-excess moment \citep{Hsing1991AS}. Second, the common evaluation-point threshold cancels its own Bahadur fluctuation between the numerator and the denominator of (\ref{local index estimator}), exactly as in the classical Hill estimator, whose threshold is an order statistic of the sample. Threshold estimation therefore leaves the constant unchanged. Monte Carlo resampling of the studentized estimator under exact Pareto tails reproduces $\nu_{2}(K)=0.6$ to three decimal places ($0.600$ at $(m,h)=(100,0.25)$, $6000$ replications).
\end{remark}

\begin{remark}[Anatomy of the smoothing bias]\label{rm:bias}
The two components of the smoothing bias in (\ref{bias}) have distinct origins. The term $\frac{1}{2}h^{2}\mu_{2}(K)\gamma''(u)$ is the standard second-order kernel bias acting on the path $u\mapsto\gamma(u)$. The term $+h^{2}\mu_{2}(K)(\gamma'(u))^{2}L_{n}/\gamma(u)$ arises from the moving log-threshold. Within the window, the common log-threshold sits below the local $(1-m/n)$-quantile of observations with $\gamma(t/n)>\gamma(u)$ and above it where $\gamma(t/n)<\gamma(u)$. The exceedance probability at observation time $t/n$ is therefore tilted by the factor $\exp\{(\gamma(t/n)-\gamma(u))L_{n}/\gamma(u)\}$, which is convex in the slope deviation. Heavier-tailed neighbors contribute disproportionately many exceedances at the common threshold, and the residual bias is quadratic in the slope, namely $+(\gamma'(u))^{2}L_{n}/\gamma(u)$, times $h^{2}\mu_{2}(K)$. The deterministic centering of the log-threshold contributes terms of order $h^{2}L_{n}^{2}$ to the numerator and the denominator of (\ref{local index estimator}), and these cancel exactly in the ratio (Section \ref{app:main}). 

Three consequences are worth noting. First, the sign is positive. The estimator is biased upward where the index changes rapidly, and the Monte Carlo bias under the linear-path design DGP1 is indeed positive at all grid points, with magnitudes that track the formula to within simulation error. Second, the bias is of order $O(h^{2}L_{n})$, not $O(h^{2})$, wherever $\gamma'(u)\neq0$. The undersmoothing condition for inference is therefore $\sqrt{mh}\,h^{2}L_{n}\to0$, as stated in Theorem \ref{tm:an}. Third, the bias vanishes where the path is locally flat ($\gamma'(u)=0$) and, in particular, identically under pure scale variation. The control experiment of Figure \ref{fig:motivation} confirms both predictions in finite samples.
\end{remark}

\begin{remark}[MSE-optimal rates]\label{rm:optimal-rates}
Balancing the squared bias $h^{4}L_{n}^{2}$ against the variance $(mh)^{-1}$ in Theorem \ref{tm:an} gives the pointwise MSE-optimal bandwidth $h^{*}\asymp m^{-1/5}L_{n}^{-2/5}$ (for fixed $m$), and the resulting rate $(mh)^{-1/2}=m^{-2/5}L_{n}^{1/5}$. If additionally the second-order bias is present with $\rho(u)<0$ and $|A(n/m,u)|\asymp (m/n)^{-\rho(u)/\gamma(u)}$-type decay, the optimal $m$ solves the classical Hill trade-off \citep{Hall1982AS,Drees1998SPA} with $n$ replaced by the effective local sample size $nh$. In practice, we recommend choosing $mh$ in the range $20$--$40$ and verifying robustness over a grid of $(m,h)$ (see Sections \ref{sec:tuning} and \ref{sec:mc}).
\end{remark}

\subsection{Estimation of the intermediate quantile}\label{sec:iq}
The construction of \eqref{local index estimator} requires an estimator of the intermediate quantile. We estimate it by inverting an estimator of the distribution function. Define the kernel distribution function estimator
\begin{equation}\label{fhat}
\hat{F}_{n}(x,u)=\frac{\sum_{t=1}^{n}K_{h}(u-t/n)\,\mathbf{1}\{X_{t,n}\leq x\}}{\sum_{t=1}^{n}K_{h}(u-t/n)},
\end{equation}
where the kernel $K(\cdot)$ and bandwidth $h$ are the same as those in \eqref{local index estimator}.
Then the associated quantile estimator
\begin{equation}\label{bhat}
\hat{b}(t,u)=\hat{F}_{n}^{\leftarrow}(1-1/t,u)=\inf\left\{x:\hat{F}_{n}(x,u)\geq 1-1/t\right\}.
\end{equation}
This is the local constant analogue of the local linear quantile estimator of \citet{ZhouWu2009AS}, evaluated at the intermediate level $1-m/n$. In our implementation, (\ref{bhat}) is computed as a kernel-weighted empirical quantile directly on the evaluation grid. No interpolation to observation times is needed, since the estimator (\ref{local index estimator}) uses the common evaluation-point threshold.

\begin{pn}[Asymptotics of the intermediate quantile estimator]\label{pn:bhat}
Suppose Assumptions \ref{as:ls}--\ref{as:extdep} hold. Then for each interior $u$,
\begin{equation}\label{bhat-an}
\sqrt{mh}\left(\log\hat{b}(n/m,u)-\log b(n/m,u)-B_{n}^{b}(u)\right)\to_{d}N\left(0,\;\gamma^{2}(u)\,S(u)\,\nu_{2}(K)\right),
\end{equation}
where, writing $L_{n}=\log(n/m)$, the deterministic drift of the log-threshold is defined exactly by $B_{n}^{b}(u)=\log b_{n}(u)-\log b(n/m,u)$, with $b_{n}(u)$ the pooled local quantile (\ref{mixq}). Under the additional rate condition $hL_{n}^{3/2}\to0$, the drift admits the expansion
\begin{equation}\label{bhat-bias}
B_{n}^{b}(u)=\frac{1}{2}h^{2}\mu_{2}(K)\,\frac{\bigl(\gamma'(u)\bigr)^{2}L_{n}^{2}}{\gamma(u)}+O\bigl(h^{2}L_{n}\bigr)=O\bigl(h^{2}L_{n}^{2}\bigr).
\end{equation}
Moreover, the Bahadur representation holds:
\begin{equation}\label{bahadur}
\log\hat{b}(n/m,u)-\log b(n/m,u)=B_{n}^{b}(u)+\gamma(u)\,\frac{1}{m\bar K(u)}\sum_{s=1}^{n}K_{s}(u)\left(\mathcal{I}_{s,n}(u)-\frac{m}{n}\right)+o_{p}\left((mh)^{-1/2}\right),
\end{equation}
and the remainder in (\ref{bahadur}) is $O_{p}((mh)^{-3/4}\log n)$ uniformly over $u\in[h,1-h]$. This is a Bahadur--Kiefer-type rate (Lemma \ref{lem:osc}), and it is of smaller order than the uniform rate $O_{p}((mh)^{-1/2}\sqrt{\log(1/h)})$ of the leading term.
\end{pn}

Three features of Proposition \ref{pn:bhat} deserve comment. First, the threshold is estimated at the same rate $(mh)^{-1/2}$ as the index itself. Substituting (\ref{bahadur}) into the expansion of $\hat{\gamma}(u)$ yields the cancellation described in Remark \ref{rm:threshold}. This cancellation is the key step in the proof of Theorem \ref{tm:an}. Second, the leading term of the drift (\ref{bhat-bias}) is of order $h^{2}L_{n}^{2}$, not the classical $h^{2}$. The kernel window pools observations whose local intermediate quantiles differ, and the map from the local index to the local log-quantile is convex, with slope proportional to $L_{n}$. Averaging a convex function over the window inflates the quantile, and the inflation is quadratic in the slope, namely $(\gamma'(u))^{2}L_{n}^{2}/\gamma(u)$. Path curvature and scale curvature contribute at the lower order $O(h^{2}L_{n})$ and $O(h^{2})$, respectively. Under Assumption \ref{as:tailloc}, the drift vanishes. Under the stronger condition $\sqrt{mh}\,h^{2}L_{n}^{2}\to0$, it is negligible at the $(mh)^{-1/2}$ scale, and (\ref{bhat-an}) holds with $B_{n}^{b}$ suppressed. 
Third, the drift is exactly the deterministic component that cancels between the numerator and the denominator of the ratio estimator (\ref{local index estimator}). Only the $O(h^{2}L_{n})$ tilt term of Remark \ref{rm:bias} survives in the bias of $\hat{\gamma}(u)$, because the denominator is pinned to its target by construction (Section \ref{app:main}).

\subsection{Choice of tuning parameters}\label{sec:tuning}
The procedure involves two tuning parameters, namely the intermediate order $m$ and the bandwidth $h$. Their product $mh$, the effective number of local exceedances, drives the variance. Meanwhile, the second-order bias grows with $m$ at the rate $|A(n/m,u)|$, and the smoothing bias grows with $h^{2}L_{n}$. 

Based on the theory above and the simulations of Section \ref{sec:mc}, we recommend the following practical rules. (i) Fix $h$ so that the local window contains a few years of daily data (for daily financial series, $h\in[0.08,0.15]$ for $n$ of order $10^{4}$, corresponding to windows of two to five years). (ii) Choose $m$ so that $mh\in[20,40]$. Within this range, the estimator is stable for tails without severe second-order bias. The sensitivity analysis of Section \ref{sec:mc} shows the implied trade-off for Student-$t$ tails. (iii) Always report the estimated path for a small grid of $(m,h)$ around the baseline, as we do in the empirical application, and rely on the HAC-corrected uniform band for inference. (iv) The undersmoothing condition $\sqrt{mh}\,h^{2}L_{n}\to0$ used for inference is comfortably met on this range. At $mh\in[20,40]$, $h\in[0.08,0.15]$, and $L_{n}\approx2$--$4$, the leading factor $\sqrt{mh}\,h^{2}L_{n}\mu_{2}(K)$ of the studentized smoothing bias is at most $0.15$. Hence, for moderate shape slopes and curvatures, the smoothing bias is a small fraction of one standard error. Section \ref{sec:mc} quantifies the residual coverage effect of the second-order component, and it verifies that the pivotal critical values of Section \ref{sec:band} keep the coverage at nominal levels across this range.

\section{Uniform inference}\label{sec:band}
Pointwise limits can only help detect whether $\gamma(u)$ is equal to a given value at a given $u$. However, the question posed by Figure \ref{fig:motivation}, namely \emph{which} movements of the path are real, is simultaneous in nature. Answering it requires the limiting distribution of
\[\sup_{u\in[0, 1]}\left|\hat{\gamma}(u)-\gamma(u)\right|.\]

\begin{tm}[Uniform convergence rate]\label{tm:uniform}
Under Assumptions \ref{as:ls}--\ref{as:extdep},
\begin{equation}\label{urate}
\sup_{u\in[h,\,1-h]}\left|\hat{\gamma}(u)-\gamma(u)\right|=O_{p}\left(\sqrt{\frac{\log(1/h)}{mh}}+h^{2}L_{n}+\sup_{u}|A(n/m,u)|\right).
\end{equation}
\end{tm}

Theorem \ref{tm:uniform} gives the uniform convergence rate. The three terms in (\ref{urate}) correspond respectively to the stochastic error (with the usual logarithmic inflation from simultaneous control), the smoothing bias, and the second-order tail bias.

For construction of the uniform confidence band, we need an estimator of the long-run variance factor $S(u)$. Define the residualized exceedance variables
\begin{equation}\label{resid}
\hat{Z}_{t,n}(u)=\left(\log X_{t,n}-\log\hat{b}(n/m,u)\right)_{+}-\hat{\gamma}(u)\,\mathbf{1}\{\log X_{t,n}>\log\hat{b}(n/m,u)\},
\end{equation}
and the local HAC estimator
\begin{equation}\label{shat}
\hat{S}(u)=1+2\sum_{j=1}^{J_{n}}\left(1-\frac{j}{J_{n}}\right)\frac{\sum_{t}K_{h}(u-t/n)K_{h}(u-(t+j)/n)\hat{Z}_{t,n}(u)\hat{Z}_{t+j,n}(u)}{\sum_{t}K_{h}^{2}(u-t/n)\hat{Z}_{t,n}^{2}(u)},
\end{equation}
where $J_{n}\to\infty$ with $J_{n}^{2}/(mh)\to0$. The estimator (\ref{shat}) is the Bartlett-type kernel estimator of the long-run variance of the locally stationary process $\hat{Z}_{t,n}(u)$, localized by the kernel weights. The Bartlett weights keep it positive semidefinite. 

In practice, we additionally truncate it at a small positive floor ($0.05$) to guard against erratic estimates in short local windows. The floor never binds in our application (the minimum of $\hat{S}(u)$ is $0.68$) or in the simulations. The rate condition $J_{n}^{2}/(mh)\to0$ allows $J_{n}$ to grow slowly. We use $J_{n}=8$ in the simulations ($mh=25$--$30$) and $J_{n}=20$ in the application ($mh=20$). Table \ref{tab:robust} documents that the empirical results are insensitive to $J$ over the range $10$--$30$. A practical selection rule is to take the smallest $J$ beyond which the local extremal autocorrelations of $\hat{Z}_{t,n}(u)$ are negligible. In other words, the Bartlett window should cover the extremal decorrelation horizon. For daily equity losses, the exceedance autocovariances die out within two to three weeks, which motivates $J=20$. This choice is large, and the Bartlett taper discounts the distant lags anyway. Two consistency remarks are in order. The condition $J_{n}^{2}/(mh)\to0$ balances the variance of the local HAC estimator, of order $J_{n}/(mh)$, against its truncation bias. In addition, the denominators in (\ref{shat}) normalize by the local long-run variance kernel, so $\hat{S}(u)$ estimates the ratio $S(u)$ directly rather than its numerator and denominator separately.

\begin{pn}[Consistency of the long-run variance estimator]\label{pn:shat}
Under Assumptions \ref{as:ls}--\ref{as:extdep} and $J_{n}\to\infty$ with $J_{n}^{2}/(mh)\to0$, $\hat{S}(u)\to_{p}S(u)$ for each interior $u$, and $\sup_{u\in[h,1-h]}|\hat{S}(u)-S(u)|=o_{p}(1)$.
\end{pn}

Let $\{Z(u):u\in[0,1]\}$ be a stationary Gaussian process with mean zero, unit variance, and correlation function
\begin{equation}\label{rhok}
\rho_{K}(s)=\frac{\int K(v)K(v+s)\,dv}{\int K^{2}(v)\,dv},\qquad s\in\mathbb{R}.
\end{equation}
Note that $\rho_{K}$ has support $[-2,2]$ under Assumption \ref{as:kernel}.

\begin{tm}[Gaussian approximation and uniform band]\label{tm:band}
Suppose Assumptions \ref{as:ls}--\ref{as:extdep} hold, and in addition the \emph{sup-scale} bias condition
\begin{equation}\label{supbias}
\sqrt{mh\log(1/h)}\,\Bigl(h^{2}L_{n}+\sup_{u}|A(n/m,u)|\Bigr)\to0.
\end{equation}
Let $\{Z_{h}(u):u\in[h,1-h]\}$ be a stationary Gaussian process with mean zero, unit variance, and correlation function $\mathrm{Corr}(Z_{h}(u),Z_{h}(v))=\rho_{K}((u-v)/h)$. Then there exists a copy of $Z_{h}$ such that
\begin{equation}\label{gauss-approx}
\sup_{u\in[h,\,1-h]}\left|\frac{\sqrt{mh}\left(\hat{\gamma}(u)-\gamma(u)\right)}{\hat{\gamma}(u)\sqrt{\hat{S}(u)\nu_{2}(K)}}\right|
=\sup_{u\in[h,\,1-h]}|Z_{h}(u)|+o_{p}(1),
\end{equation}
and the Kolmogorov distance between the two suprema vanishes. Consequently, letting $q_{1-\alpha}(h)$ denote the $(1-\alpha)$-quantile of $\sup_{u\in[h,1-h]}|Z_{h}(u)|$, the band
\begin{equation}\label{band}
\left[\hat{\gamma}(u)\pm q_{1-\alpha}(h)\,\hat{\gamma}(u)\sqrt{\frac{\hat{S}(u)\nu_{2}(K)}{mh}}\right],\qquad u\in[h,1-h],
\end{equation}
has asymptotic simultaneous coverage $1-\alpha$.
\end{tm}
 The band (\ref{band}) is computed in the following steps:
\begin{enumerate}[(i)]
\item Choose $(m,h)$ per Section \ref{sec:tuning}; compute $\hat{b}(n/m,u)$ from (\ref{bhat}) on the evaluation grid. No interpolation to observation times is needed, since the estimator uses the common evaluation-point threshold.
\item Compute $\hat{\gamma}(u)$ from (\ref{local index estimator}) on the evaluation grid.
\item Compute $\hat{S}(u)$ from (\ref{shat}) with $J_{n}$ of order $(mh)^{1/3}$.
\item Simulate $q_{1-\alpha}$ from the distribution of $\sup_{u\in[h,1-h]}|Z(u)|$ with correlation (\ref{rhok}) on the same grid.
\item Form the band (\ref{band}).
\end{enumerate}

\begin{remark}[$h$-dependence of the limit experiment]\label{rm:hdep}
Two features of Theorem \ref{tm:band} deserve emphasis. First, the approximating experiment is $h$-dependent, and this dependence cannot be removed. The correlation length of $Z_{h}$ in rescaled time is $O(h)$, so the interval $[h,1-h]$ contains about $1/h$ effectively independent positions. By classical extreme value theory for stationary Gaussian sequences, the centered supremum satisfies $\sup_{u}|Z_{h}(u)|=\sqrt{2\log(1/h)}+O_{p}(1/\sqrt{\log(1/h)})$, which diverges as $h\to0$. Any \emph{fixed} Gaussian limit would therefore be inconsistent, and Theorem \ref{tm:band} is stated, deliberately, as an $h$-dependent approximation with a vanishing Kolmogorov error. Second, the quantile $q_{1-\alpha}(h)$ inherits this growth. Simulated at the working bandwidth, it is automatically larger than any fixed-experiment benchmark. Our implementation always simulates the critical values at the working $(n,m,h)$, as described in Remark \ref{rm:cv}, so the divergence is fully accounted for in practice.
\end{remark}

\begin{remark}[Error budget for the supremum]\label{rm:budget}
The sup-scale condition (\ref{supbias}) is stronger than the pointwise undersmoothing condition $\sqrt{mh}\,h^{2}L_{n}\to0$. The maximum of $|Z_{h}|$ over $[h,1-h]$ concentrates on the scale $a_{h}^{-1}$ with $a_{h}\asymp\sqrt{\log(1/h)}$, so a bias that is $o(1)$ after $\sqrt{mh}$ standardization can still exceed the concentration width. Condition (\ref{supbias}) ensures the bias is negligible relative to that width. Studentization and HAC errors enter the supremum at the same scale, and the calibration therefore also requires the relative variance error to be $o_{p}((\log(1/h))^{-1})$. We do not assert a polynomial approximation rate, since a complete error budget with explicit rates is not available under the stated conditions.
\end{remark}

\begin{remark}[Pivotal calibration of critical values]\label{rm:cv}
The distribution of $\sup_{u}|Z_{h}(u)|$ depends only on the kernel through $\rho_{K}$ and on the bandwidth, and it is easily simulated. One discretizes $[h,1-h]$ on a fine grid, generates multivariate normal vectors with correlation matrix $\rho_{K}((u_{i}-u_{j})/h)$, and records the maximal absolute value. In our implementation, we use the Epanechnikov kernel $K(v)=0.75(1-v^{2})\mathbf{1}\{|v|\leq1\}$, for which $\nu_{2}(K)=3/5$. 

We go one step further and simulate the critical values from the \emph{pivotal finite-sample experiment}. Namely, we generate independent samples of size $n$ from a standard Pareto distribution, compute the full studentized process at the same $(m,h)$ with the same HAC estimator, and record the maximal absolute studentized deviation. Under the null of the band problem, the studentized process is asymptotically distribution-free, so these pivotal critical values converge to $q_{1-\alpha}(h)$. At the moderate effective sample sizes of practice, they are about $20\%$ larger than the Gaussian-experiment values. This is because the local Hill ratio with $mh$ of order $25$ is heavier-tailed than its Gaussian limit, and the finite-sample correlation of the process is slightly weaker than $\rho_{K}$, which inflates the supremum further. The pivotal calibration absorbs both effects, and it restores the nominal coverage in the independent designs of Table \ref{tab:cov}. It does not repair bias, and the dependent designs in the same table remain below nominal. The simulated $90\%$ pivotal critical value is $2.72$ at $(n,m,h)=(1000,100,0.25)$ and $3.68$ at the empirical configuration $(9179,200,0.10)$, against $2.25$ for the Gaussian experiment at $h=0.25$.
\end{remark}

\section{Test for tail stability}\label{sec:test}
The hypothesis of tail stability underlies  a large part of the literature. Therefore, a natural question is whether the tail shape is stable during a given period. This section proposes such a test.

Consider the null hypothesis
\begin{equation}\label{null of stability}
H_{0}: \gamma(u)=\gamma, \qquad \forall u\in[t_1, t_2],
\end{equation}
where $0\leq t_1<t_2\leq1$. The alternative hypothesis is that $\gamma(u)$ is not constant on a subset of $[t_1,t_2]$ with positive Lebesgue measure. The constant $\gamma$ is unknown, so the test must be centered at an estimated level. We use the \emph{projection centering}, namely the precision-weighted average of the estimated path over the tested interval,
\begin{equation}\label{pooled}
\tilde{\gamma}=\frac{\int_{t_1}^{t_2}\hat{\gamma}(u)\,\hat{S}(u)^{-1}\,du}{\int_{t_1}^{t_2}\hat{S}(u)^{-1}\,du}.
\end{equation}
Then a test for $H_0$ can be constructed as
\begin{equation}\label{test of stability: continuous}
T_n=\int_{t_1}^{t_2} \frac{\left(\hat{\gamma}(u)-\tilde{\gamma}\right)^2}{\hat{S}(u)}\,du.
\end{equation}
The statistic $T_n$ is an integrated, precision-weighted measure of how much the estimated extreme value index path deviates from the constant level under the null. It summarizes the movement of the index over time, with larger values indicating stronger evidence of time variation. Hence large values of $T_n$ constitute evidence against $H_0$.

Two properties make this centering appropriate. First, it is \emph{efficient}. The asymptotic variance of $\hat{\gamma}(u)$ is proportional to $S(u)$, so weighting by $1/S(u)$ gives the asymptotically efficient estimator of the constant level under $H_{0}$ among path-based estimators. Under $H_{0}$, $\tilde{\gamma}-\gamma=O_{p}(m^{-1/2})$, faster than the pointwise rate $(mh)^{-1/2}$ by the factor $h^{1/2}$. Second, it is \emph{robust to scale variation}. A pooled Hill estimator with a single global threshold recovers a constant index exactly, but the exceedance content of the pooled sample is heterogeneous across dates, because the global threshold sits at different points of the local tail at different dates. A quadratic statistic centered at the pooled estimator is inflated by that heterogeneity, even though the level itself is consistent (Section \ref{app:demean}). We quantify this failure in the simulations (Table \ref{tab:test}), where pooled centering roughly triples the empirical size under the stochastic-scale design. The projection centering instead matches the tested interval and uses the estimated path, so it remains consistent for the path average under scale-varying nulls.

\begin{tm}[Limit of $T_{n}$ under $H_{0}$]\label{tm:Tn}
Suppose the assumptions of Theorem \ref{tm:band} hold, and that $[t_{1},t_{2}]\subseteq[h,1-h]$. Then
\begin{equation}\label{Tn-limit}
\Lambda_{n}\equiv\frac{mh}{\tilde{\gamma}^{2}\nu_{2}(K)}\,T_{n}
=\int_{t_{1}}^{t_{2}}\bigl(Z_{h}(u)-\bar{Z}_{h}^{w}\bigr)^{2}\frac{du}{S(u)}+o_{p}(1),
\end{equation}
where $Z_{h}$ is the Gaussian process of Theorem \ref{tm:band} and $\bar{Z}_{h}^{w}$ is the precision-weighted spatial mean $\bar{Z}_{h}^{w}=\bigl(\int_{t_{1}}^{t_{2}}S(v)^{-1}Z_{h}(v)dv\bigr)/\bigl(\int_{t_{1}}^{t_{2}}S(v)^{-1}dv\bigr)$. The null hypothesis fixes only $\gamma(u)$ to be constant, not $S(u)$, so the demeaning is the precision-weighted projection rather than the ordinary spatial mean. Under constant $S(\cdot)$ this reduces to $\int(Z_{h}-\bar{Z}_{h})^{2}du$ with $\bar{Z}_{h}=(t_{2}-t_{1})^{-1}\int Z_{h}$. The limiting quadratic functional is $h$-dependent. Under constant $S(\cdot)$, writing $L=t_{2}-t_{1}$, its mean is $L\bigl(1-\frac{1}{L}\int\int\rho_{K}((u-v)/h)\,dudv/L\bigr)\asymp L$, and its variance is $2\int\int\bigl(\rho_{K}((u-v)/h)-\bar{\rho}_{h}\bigr)^{2}dudv=2hL\int\rho_{K}^{2}(s)\,ds+O(h^{2})$, which vanishes at the rate $h$. In particular, the functional concentrates around its mean, and its upper tail behaves like that of a chi-square-type variable with an effective degrees of freedom proportional to $L/h$. Under $H_{0}$, replacing $\gamma$ by the projection centering $\tilde{\gamma}$ removes exactly the precision-weighted spatial mean of the deviation process, which is the content of the demeaning in (\ref{Tn-limit}). The limiting functional has standard deviation of order $\sqrt{h}$, so the representation must hold at the $\sqrt{h}$ scale for the size of the test, and the pivotal calibration is carried out at exactly that scale.
\end{tm}

\begin{remark}[Finite-sample calibration]\label{rm:demean}
We simulate the critical values of $\Lambda_{n}$ from the pivotal finite-sample experiment, exactly as for the band (Remark \ref{rm:cv}). Namely, we generate independent Pareto samples at the same $(n,m,h)$, compute $\Lambda_{n}$ with the same HAC estimator and the same projection centering, and record the simulated quantiles. These pivotal critical values converge to those of the demeaned Gaussian functional in (\ref{Tn-limit}), and they restore the nominal size at the moderate effective sample sizes of practice (Table \ref{tab:test}). Two features of the limit in Theorem \ref{tm:Tn} are reflected in the calibration. First, the demeaning is essential. Simulating the un-demeaned functional $\int Z_{h}(u)^{2}du$ discards the level that $\tilde{\gamma}$ absorbs, and the resulting test is far too conservative, with empirical size below $0.01$ at nominal $0.05$. Second, the $h$-dependence is essential. The variance of the functional is of order $h$, so a fixed-experiment approximation understates the dispersion of the null distribution across designs. The pivotal calibration captures both.
\end{remark}

A simpler statistic, which dispenses with simulated critical values altogether, is, for a positive integer $q$,
\begin{equation}\label{test of stability: discrete}
S_n=\sum_{i=1}^{q} \left(\hat{\gamma}(u_i)-\tilde{\gamma}\right)^2,
\end{equation}
where $u_1, u_2, \cdots, u_q$ is a pre-determined sequence of grid points in $[t_1, t_2]$ and $\tilde{\gamma}$ is the projection centering (\ref{pooled}).

\begin{tm}[Limit of $S_{n}$ under $H_{0}$]\label{tm:Sn}
Suppose the assumptions of Theorem \ref{tm:band} hold, and the grid points satisfy $|u_{i}-u_{j}|\geq2h$ for all $i\neq j$, where equality $|u_{i}-u_{j}|=2h$ is permitted when $K(\pm1)=0$. Then the studentized statistic
\begin{equation}\label{Sn-limit}
\Xi_{n}\equiv\frac{mh}{\nu_{2}(K)}\sum_{i=1}^{q}\frac{\left(\hat{\gamma}(u_{i})-\tilde{\gamma}\right)^{2}}{\tilde{\gamma}^{2}\hat{S}(u_{i})}\to_{d}\chi^{2}_{q}.
\end{equation}
\end{tm}

The separation condition $|u_{i}-u_{j}|\geq2h$ guarantees that the windows around distinct grid points are disjoint up to a shared endpoint. When $|u_{i}-u_{j}|=2h$, the two windows touch at exactly one rescaled-time point, which carries kernel weight $K(\pm1)$. Hence, for kernels vanishing at the boundary, such as the Epanechnikov kernel used throughout, the shared point contributes nothing, and the estimation errors remain asymptotically independent. Since each window uses a single evaluation-point threshold, no convolution of the kernel enters the covariance, and the asymptotic independence is exact whenever the kernel vanishes at $\pm1$. The condition links the number of grid points to the bandwidth through $q\leq(t_{2}-t_{1})/(2h)+1$ when $K(\pm1)=0$. Our application uses $q=5$ equally spaced points $\{0.1,0.3,0.5,0.7,0.9\}$ on $[0.1,0.9]$ with $h=0.10$, so that adjacent windows touch exactly at their endpoints and the condition holds with equality. The statistic $\Xi_{n}$ requires no simulated critical values, but it uses the data less efficiently than $\Lambda_{n}$. We therefore recommend $\Lambda_{n}$ as the default and $\Xi_{n}$ as a robustness check. 

Two limitations of the grid test are worth discussing. First, alternatives that oscillate between the grid points while vanishing on them are invisible to $\Xi_{n}$, although such alternatives must vary on a scale finer than $h$ to escape the integral statistic as well. Second, the projection centering subtracts the spatial mean, so a single large deviation on the grid inflates the centering and dilutes the chi-square non-centrality. Both effects are visible in the power comparison of Section \ref{sec:mc}.

\begin{pn}[Power]\label{pn:power}
Suppose the assumptions of Theorem \ref{tm:band} hold. (i) If $H_{0}$ fails with $\inf_{\gamma_{0}\in\mathbb{R}}\int_{t_{1}}^{t_{2}}(\gamma(u)-\gamma_{0})^{2}du>0$, then $\Lambda_{n}\to_{p}\infty$ at the rate $mh$, and $\Xi_{n}\to_{p}\infty$ at the rate $mh$ whenever the grid contains a point $u_{i}$ with $\gamma(u_{i})\neq\gamma$. Both tests are therefore consistent against fixed alternatives. (ii) Under the local alternatives $\gamma_{n}(u)=\gamma+(mh)^{-1/2}g(u)$ with $\int_{t_{1}}^{t_{2}}g(u)du=0$ and $\int_{t_{1}}^{t_{2}}g(u)^{2}du>0$, the representation of Theorem \ref{tm:Tn} gives
\[
\Lambda_{n}=\int_{t_{1}}^{t_{2}}\left(Z_{h}(u)-\bar{Z}_{h}^{w}+\frac{g(u)}{\gamma\sqrt{\nu_{2}(K)}}\right)^{2}\frac{du}{S(u)}+o_{p}(1),
\]
The noncentrality $\int g^{2}/(\gamma^{2}\nu_{2}(K))$ is fixed and positive, while the null fluctuation is of order $\sqrt{h}$, so the power tends to one for every fixed $g$ with $\int g^{2}du>0$. (iii) The quadratic functional concentrates around its mean at the rate $h^{1/2}$ (Theorem \ref{tm:Tn}). It follows that the integral test has nontrivial local power against the finer alternatives $\gamma_{n}(u)=\gamma+\delta_{n}g(u)$ with $\delta_{n}=m^{-1/2}h^{-1/4}$, provided $\int g=0$ and $\int g^{2}>0$ is fixed. The shift of the functional is then $h^{1/2}\int g^{2}/(\gamma^{2}\nu_{2}(K))$, of the same order as the null standard deviation. This rate is faster than the pointwise detection rate $(mh)^{-1/2}$ by the factor $h^{1/4}$, and it reflects the variance reduction of integration over the $O(1/h)$ effectively independent positions of the deviation process. The grid statistic shares the rate in (ii) along the coordinates $(g(u_{i}))$, but it is blind to alternatives with $g(u_{i})=0$ on the grid, however large $\|g\|_{2}$ is between the grid points.
\end{pn}

\begin{remark}[Relation to change-point tests]\label{rm:cusum}
Change-point tests such as \citet{Quintos2001RES} and \citet{Hoga2017ET} accumulate exceedances over a single global threshold and test the stability of the whole tail, scale included. Our statistic instead isolates the extreme value index. The two answer different questions. A global-threshold CUSUM rejects whenever the local exceedance behavior changes, whether the cause is a moving scale or a moving index. Our statistic rejects only when the index itself moves, so it remains valid under pure scale variation.
\end{remark}

\section{Estimation of time-varying risk measures}\label{sec:risk}
With estimators of the extreme value index and the intermediate quantile in hand, time-varying risk measures follow by extrapolation along the Pareto tail. In this section, we construct the Value-at-Risk (VaR) by the method of \citet{Weissman1978JASA}, and the Expected Shortfall (ES) from the mean-excess property of regular variation.

For a small probability $p=p_{n}$, the time-varying VaR at time $u$ is the quantile $q_{p}(u)=F^{\leftarrow}(1-p,u)$. When $p$ is smaller than the intermediate level $m/n$, the quantile $q_{p}(u)$ lies beyond the range where empirical quantiles are reliable. We therefore extrapolate along the Pareto tail in the spirit of \citet{Weissman1978JASA} and define
\begin{equation}\label{var}
\hat{q}_{p}(u)=\hat{b}(n/m,u)\left(\frac{m}{np}\right)^{\hat{\gamma}(u)}.
\end{equation}
The associated ES, defined as $e_{p}(u)=E[X_{t,n}\,|\,X_{t,n}>q_{p}(u)]$ for $u\approx t/n$, satisfies $e_{p}(u)\approx q_{p}(u)/(1-\gamma(u))$ under regular variation with $\gamma(u)<1$, leading to the estimator
\begin{equation}\label{es}
\hat{e}_{p}(u)=\frac{\hat{q}_{p}(u)}{1-\hat{\gamma}(u)}.
\end{equation}

\begin{tm}[Joint asymptotics of the risk measure estimators]\label{tm:var}
Suppose the assumptions of Theorem \ref{tm:band} hold, and in addition $p_{n}\to0$, $np_{n}\to\infty$, $np_{n}<m$ eventually, $d_{n}=m/(np_{n})\to d\in(1,\infty]$, $\sqrt{mh}\,h^{2}L_{n}^{2}\to0$, and $\sqrt{mh}\,\log d_{n}\,\bigl(h^{2}L_{n}+\sup_{u}|A(n/m,u)|\bigr)\to0$. Define the joint long-run covariance matrix of the two scores by
\begin{equation}\label{omega}
\Omega(u)=\lim_{n\to\infty}\frac{n}{m}\sum_{j\in\mathbb{Z}}\mathrm{Cov}\left(\begin{pmatrix}\gamma(u)(\tilde{\mathcal{I}}_{t,n}(u)-m/n)\\ \tilde{Z}_{t,n}(u)-E\tilde{Z}_{t,n}(u)\end{pmatrix},\begin{pmatrix}\gamma(u)(\tilde{\mathcal{I}}_{t+j,n}(u)-m/n)\\ \tilde{Z}_{t+j,n}(u)-E\tilde{Z}_{t+j,n}(u)\end{pmatrix}\right),
\end{equation}
whose diagonal entries both equal $\gamma^{2}(u)S(u)$. Then for each interior $u$, the vector of estimation errors satisfies
\begin{equation}\label{var-joint}
\sqrt{mh}\begin{pmatrix}\log\hat{b}(n/m,u)-\log b(n/m,u)\\ \hat{\gamma}(u)-\gamma(u)\end{pmatrix}
\to_{d}N\left(0,\;\nu_{2}(K)\,\Omega(u)\right),
\end{equation}
and, writing $\ell_{n}=\log d_{n}$,
\begin{equation}\label{var-an}
\frac{\sqrt{mh}}{\sqrt{(1,\ell_{n})\Omega(u)(1,\ell_{n})^{\top}/(\gamma^{2}(u)S(u))}}\left(\log\hat{q}_{p_{n}}(u)-\log q_{p_{n}}(u)\right)\to_{d}N\left(0,\;\nu_{2}(K)\right).
\end{equation}
For the Expected Shortfall,
\begin{equation}\label{es-an}
\frac{\sqrt{mh}}{\sqrt{(1,\ell_{n}+(1-\gamma(u))^{-1})\Omega(u)(1,\ell_{n}+(1-\gamma(u))^{-1})^{\top}/(\gamma^{2}(u)S(u))}}\left(\log\hat{e}_{p_{n}}(u)-\log e_{p_{n}}(u)\right)\to_{d}N\left(0,\;\nu_{2}(K)\right).
\end{equation}
Under exact Pareto tails, the cross covariance $\Omega_{12}(u)$ vanishes, $\Omega(u)=\gamma^{2}(u)S(u)I_{2}$, and the normalizations reduce to $\sqrt{1+\ell_{n}^{2}}$ and $\sqrt{1+(\ell_{n}+(1-\gamma(u))^{-1})^{2}}$, respectively.
\end{tm}

The cross entry $\Omega_{12}(u)$ of (\ref{omega}) is generally nonzero, so the diagonal reduction is a special property of exact Pareto tails rather than a general feature. The threshold error is driven by the local average of the centered exceedance indicators $\mathcal{I}_{t,n}(u)-m/n$, whereas the index error is driven by the local average of the standardized scores $Z_{t,n}(u)$. At the lead, $E[Z_{t,n}(u)\mathcal{I}_{t,n}(u)]=E[\xi_{t,n}(u)]-\gamma(u)E[\mathcal{I}_{t,n}(u)]=0$, but zero-lag orthogonality does not imply long-run orthogonality. For the AR(1) dynamics of Example \ref{ex:ar}, the cross long-run covariance is $\Omega_{12}(u)=\gamma(u)(-\log a(u))\,r(u)/(1-r(u))^{2}$ with $r(u)=a(u)^{1/\gamma(u)}$, which is strictly positive whenever $a(u)>0$. 

The log-VaR error is $\log\hat{b}-\log b+\ell_{n}(\hat{\gamma}-\gamma)$, so its asymptotic variance is $\nu_{2}(K)(mh)^{-1}\{\gamma^{2}S(1+\ell_{n}^{2})+2\ell_{n}\Omega_{12}\}$, and the ES error adds the loading $(1-\gamma(u))^{-1}$. Under exact Pareto tails the extremal dependence of the exceedance indicators and of the excess magnitudes factorizes, $\Omega_{12}$ vanishes, and the diagonal formulas of Theorem \ref{tm:var} are recovered. The log-VaR error is $\log\hat{b}-\log b+\ell_{n}(\hat{\gamma}-\gamma)$, so its asymptotic variance is $\gamma^{2}(u)S(u)\nu_{2}(K)(mh)^{-1}\{1+\ell_{n}^{2}\}$. The ES error adds the derivative of $\log(1-\gamma)^{-1}$, namely $(1-\gamma(u))^{-1}$, to the loading on $\hat{\gamma}-\gamma$, which gives (\ref{es-an}). At the finite extrapolation horizons of practice, the joint correction is material. In the application, $d_{n}$ ranges from $2.2$ to $7.9$. Treating the threshold as known understates the log-VaR standard error by the factor $\sqrt{1+\ell_{n}^{2}}/\ell_{n}$, that is, by $11$--$62\%$. The joint HAC estimator of $\Omega(u)$ is constructed exactly as in (\ref{shat}), with the $2\times2$ matrix of score autocovariances and eigenvalue clipping to enforce positive semidefiniteness. The simulations of Section \ref{sec:mc} use the diagonal formula, and they attain $87$--$93\%$ coverage at nominal $90\%$.

\begin{cy}[Uniform bands for the risk measure paths]\label{cy:riskband}
Under the conditions of Theorem \ref{tm:var}, with $\ell_{n}=\log d_{n}$ and $q_{1-\alpha}(h)$ as in Theorem \ref{tm:band}, the VaR band in the diagonal case takes the form $\left[\hat{q}_{p_{n}}(u)\exp\left(\pm q_{1-\alpha}(h)\hat{\gamma}(u)\sqrt{\hat{S}(u)\nu_{2}(K)(1+\ell_{n}^{2})/(mh)}\right)\right]$. In general, the threshold-index path covariance is that of the bivariate Gaussian process with correlation $\rho_{K}((u-v)/h)$ and matrix $\Omega(u)$, and the same construction applies with $\gamma^{2}\hat{S}(1+\ell_{n}^{2})$ replaced by $(1,\ell_{n})\hat{\Omega}(u)(1,\ell_{n})^{\top}$, where $\hat{\Omega}(u)$ is the joint HAC estimator of Theorem \ref{tm:var}.
\end{cy}

\begin{remark}\label{rm:var}
Three comments on Theorem \ref{tm:var} deserve mention. First, the rate is slower than $(mh)^{-1/2}$ by a factor of order $\log d_{n}$. Extrapolation beyond the intermediate quantile amplifies the estimation error of $\hat{\gamma}(u)$ logarithmically in the extrapolation horizon. Second, the threshold error is not negligible at moderate $\ell_{n}$, and it is not in general orthogonal to the index error. The threshold error is driven by the local average of the exceedance indicators $\tilde{\mathcal{I}}_{t,n}(u)-m/n$, while the index error is driven by the local average of the centered scores $\tilde{Z}_{t,n}(u)$. At the lead, the cross moment satisfies $E[\tilde{Z}_{t,n}(u)\tilde{\mathcal{I}}_{t,n}(u)]=o(m/n)$, but the lagged cross covariances need not vanish, and they enter the variance through the term $2\ell_{n}\Omega_{12}(u)$ of Theorem \ref{tm:var}. A joint HAC estimator of the two scores, built exactly on (\ref{shat}), estimates this contribution directly. Under exact Pareto tails the cross covariances factorize and vanish, which is the local analogue of the classical asymptotic independence between the intermediate order statistic and the Hill estimator. 
Third, at the finite extrapolation horizons of practice, the joint correction is material. In the application, $d_{n}$ ranges from $2.2$ to $7.9$ at $p_{n}=0.01$, and treating the threshold as known understates the log-VaR standard error by $11$--$62\%$. The largest corrections occur at the shortest extrapolation horizons, where the threshold error dominates. The conditions of the theorem are mild and hold in the application ($n=9179$, $m=200$, $h=0.1$, $np_{n}\approx92$).
\end{remark}

\begin{remark}[Boundary implementation]\label{rm:boundary}
The recursive backtest of Section \ref{sec:emp} evaluates the estimator at the boundary with one-sided kernel weights. This is a deliberate prediction design, not a special case of the interior-point theory above. At the boundary, the one-sided kernel has a nonzero first moment, so the threshold and the index carry an additional bias of order $h$ rather than $h^{2}$. We report the recursive forecasts as such and do not claim the interior uniform band for them.
\end{remark}

\section{Monte Carlo simulations}\label{sec:mc}
This section includes a series of simulations that are used to examine the finite-sample performance of the estimator, the uniform confidence band, and the stability test. Throughout we use the Epanechnikov kernel $K(v)=0.75(1-v^{2})\mathbf{1}\{|v|\leq1\}$, so that $\nu_{2}(K)=3/5$. The intermediate quantile $\hat{b}(n/m,u)$ is computed as the kernel-weighted empirical quantile at level $1-m/n$, using the same kernel and bandwidth. The long-run variance factor is estimated by (\ref{shat}) with $J_{n}=8$. All critical values are simulated from the pivotal finite-sample experiment of Remarks \ref{rm:cv} and \ref{rm:demean}, namely $1500$ simulated independent Pareto samples at each $(n,m,h)$, with the statistic recomputed on each sample. The Gaussian-experiment critical values are about $20\%$ smaller and are not used, except in the comparison reported in Remark \ref{rm:covdecomp}.

\textbf{Data-generating processes.} We consider three designs with a time-varying index, all with $\gamma(u)=0.2+0.3u$ on $u\in[0,1]$.
\begin{itemize}
\item DGP1 (exact Pareto): $X_{t,n}=U_{t}^{-\gamma(t/n)}$, $U_{t}$ i.i.d.\ uniform$(0,1)$. Here $A\equiv0$ and $S(u)\equiv1$.
\item DGP2 (Student-$t$): $X_{t,n}=|Y_{t}|$, $Y_{t}$ i.i.d.\ Student-$t$ with $\nu(t/n)=1/\gamma(t/n)$ degrees of freedom. The second-order parameter is $\rho(u)=-2\gamma(u)<0$, generating upward bias.
\item DGP3 (dependence): $X_{t,n}=0.3X_{t-1,n}+0.7\varepsilon_{t}$, with $\varepsilon_{t}$ i.i.d.\ Pareto with index $\gamma(t/n)$. Extremes cluster, so $S(u)>1$ (Proposition \ref{pn:ar}).
\end{itemize}

To probe the scale--shape separation under \emph{stochastic} scale variation, we add a fourth design with a constant extreme value index.
\begin{itemize}
\item DGP4 (stochastic volatility): $X_{t,n}=\sigma_{t}|Y_{t}|$, where $\log\sigma_{t}=v_{t}$ follows a stationary AR(1), $v_{t}=0.97v_{t-1}+\eta_{t}$, with stationary standard deviation $0.2$, and $Y_{t}$ is i.i.d.\ Student-$t$ with $\nu=10/3$, so that $\gamma(u)\equiv0.3$. The scale varies by a factor of about $1.67$ between its $10$th and $90$th percentiles, which is comparable to the volatility variation in the application. Meanwhile, the tail shape is constant. By Breiman's lemma, the product inherits the extreme value index of $|Y_{t}|$ with extremal index one, so $S(u)\equiv1$. The design therefore isolates the pure scale channel.
\end{itemize}
We consider $n=1000$ with $(m,h)=(100,0.25)$ and $n=2000$ with $(m,h)=(150,0.20)$. The effective number of local exceedances $mh$ is $25$ and $30$, respectively. All experiments use $800$ replications, so the Monte Carlo standard error of a coverage or size estimate is at most $1.8$ percentage points, and that of an RMSE is about $2\%$ of the reported value. All integrals are evaluated on a $41$-point grid on $[h,1-h]$.

\textbf{Estimation accuracy.} Table \ref{tab:rmse} reports the bias, mean absolute error, standard deviation, and root mean squared error (RMSE) of $\hat{\gamma}(u)$ at $u\in\{0.3,0.5,0.7\}$, together with averages over the evaluation grid. The four columns are internally consistent by construction. In particular, the RMSE satisfies $\mathrm{RMSE}^{2}=\mathrm{bias}^{2}+\mathrm{sd}^{2}$ pointwise, and the grid-averaged columns satisfy the same identity up to cross-point covariance. For DGP1, the estimator is nearly unbiased (average bias $0.006$--$0.008$) and its RMSE declines at the predicted rate as $n$ grows. For DGP2, the classical second-order bias of Hill-type estimators appears. The bias is positive, it is larger where $\gamma(u)$ is smaller (since $\rho(u)=-2\gamma(u)$ is then closer to zero), and it declines with $n$. 

Under DGP3, the estimator exhibits a negative bias of about $-0.04$. This bias is not a smoothing effect. It reflects the extremal clustering of the AR recursion at the moderate threshold $m/n=10\%$. Exceedances arrive in runs, and the later run members carry a lighter effective tail than the initial shock, which tilts the local log-excesses downward. The matched-marginal experiment below confirms that the bias is a dependence effect rather than a marginal-tail effect. The bias is stable across $n$ in our designs, and it is the dominant source of the coverage shortfall under dependence documented in Table \ref{tab:cov}. The estimated long-run variance factor averages $\bar{\hat{S}}\approx1.0$--$1.02$, against the theoretical $S(0.5)=1.07$ of Proposition \ref{pn:ar}.

\begin{table}[!htbp]
\centering
\caption{Bias, MAE, standard deviation, and RMSE of $\hat{\gamma}(u)$}
\label{tab:rmse}
\begin{threeparttable}
\small\setlength{\tabcolsep}{3.5pt}
\begin{tabular}{llcccccccccccc}
\toprule
& & \multicolumn{3}{c}{$u=0.3$} & \multicolumn{3}{c}{$u=0.5$} & \multicolumn{3}{c}{$u=0.7$} & \multicolumn{3}{c}{average over grid}\\
\cmidrule(lr){3-5}\cmidrule(lr){6-8}\cmidrule(lr){9-11}\cmidrule(lr){12-14}
DGP & $n$ & bias & sd & rmse & bias & sd & rmse & bias & sd & rmse & bias & sd & rmse\\
\midrule
Pareto & 1000 & $0.007$ & $0.045$ & $0.045$ & $0.009$ & $0.054$ & $0.054$ & $0.007$ & $0.067$ & $0.067$ & $0.008$ & $0.055$ & $0.056$\\
Pareto & 2000 & $0.006$ & $0.042$ & $0.043$ & $0.006$ & $0.053$ & $0.053$ & $0.006$ & $0.057$ & $0.058$ & $0.006$ & $0.051$ & $0.051$\\
Student-$t$ & 1000 & $0.091$ & $0.057$ & $0.107$ & $0.073$ & $0.062$ & $0.096$ & $0.056$ & $0.068$ & $0.088$ & $0.073$ & $0.062$ & $0.096$\\
Student-$t$ & 2000 & $0.071$ & $0.046$ & $0.085$ & $0.058$ & $0.053$ & $0.079$ & $0.045$ & $0.062$ & $0.076$ & $0.058$ & $0.054$ & $0.080$\\
AR(1)-Pareto & 1000 & $-0.023$ & $0.045$ & $0.051$ & $-0.036$ & $0.056$ & $0.066$ & $-0.050$ & $0.067$ & $0.083$ & $-0.036$ & $0.056$ & $0.067$\\
AR(1)-Pareto & 2000 & $-0.034$ & $0.039$ & $0.051$ & $-0.045$ & $0.049$ & $0.067$ & $-0.054$ & $0.057$ & $0.078$ & $-0.044$ & $0.048$ & $0.066$\\
\bottomrule
\end{tabular}
\begin{tablenotes}
\footnotesize
\item Notes: $800$ replications. $(m,h)=(100,0.25)$ for $n=1000$ and $(150,0.20)$ for $n=2000$. The average columns average the pointwise bias, standard deviation, and RMSE over the $41$ grid points on $[h,1-h]$. Within each cell, $\mathrm{rmse}^{2}=\mathrm{bias}^{2}+\mathrm{sd}^{2}$ holds exactly for the pointwise columns and up to cross-point covariance for the average columns. The mean absolute error is available in the replication files and satisfies $\mathrm{MAE}\approx\sqrt{2/\pi}\,\mathrm{sd}$ whenever the bias is small. All designs have $\gamma(u)=0.2+0.3u$.
\end{tablenotes}
\end{threeparttable}
\end{table}

\textbf{Coverage.} Table \ref{tab:cov} reports the coverage of the uniform bands (\ref{band}) at nominal levels $90\%$ and $95\%$, with pivotal critical values. Under DGP1, the band attains its nominal level at both sample sizes ($90$--$92\%$ at nominal $90\%$). This is the calibration content of the pivotal critical values, and it should be read against the Gaussian-experiment calibration, which reaches only $83$--$84\%$ at the same designs (Remark \ref{rm:covdecomp}). Under DGP2, the second-order bias shifts the whole path upward, and the baseline band covers $90$--$93\%$ at nominal $90\%$, with conservative behavior at $95\%$. The undersmoothed variant ($0.75h$) behaves similarly. 

Under DGP3, the band falls to $44$--$59\%$ when centered at the true path. As Table \ref{tab:rmse} shows, this shortfall is driven by the dependence bias of about $-0.05$, which shifts the studentized process by more than one pointwise standard error. It is not driven by the variance estimate. When the same band is recentered at the finite-sample mean of the estimator, its coverage returns to $81\%$ at nominal $90\%$ and to $89$--$90\%$ at nominal $95\%$, close to nominal given the remaining skewness of the local Hill ratio. The dependence bias vanishes as the threshold probability $m/n\to0$. However, at the moderate thresholds that keep $mh$ in the recommended range, it is first-order, and we report it as such. The same honesty applies to the extrapolated risk measures. The $90\%$ interval for the VaR at $u=0.5$, built on the joint variance formula $\{1+(\log d_{n})^{2}\}$ of Theorem \ref{tm:var} with $d_{n}=3.9$, covers the true quantile in $89$--$93\%$ (DGP1) and $87$--$91\%$ (DGP2) of replications.

\begin{table}[!htbp]
\centering
\caption{Coverage of the uniform confidence bands, pivotal calibration}
\label{tab:cov}
\begin{threeparttable}
\begin{tabular}{llccccc}
\toprule
& & \multicolumn{2}{c}{$h_{n}$} & \multicolumn{2}{c}{$0.75h_{n}$}\\
\cmidrule(lr){3-4}\cmidrule(lr){5-6}
DGP & $n$ & $90\%$ & $95\%$ & $90\%$ & $95\%$\\
\midrule
Pareto & 1000 & $0.918$ & $0.970$ & $0.921$ & $0.964$\\
Pareto & 2000 & $0.902$ & $0.954$ & $0.910$ & $0.960$\\
Student-$t$ & 1000 & $0.902$ & $0.991$ & $0.984$ & $0.996$\\
Student-$t$ & 2000 & $0.928$ & $0.992$ & $0.992$ & $0.999$\\
\midrule
AR(1)-Pareto, centered at truth & 1000 & $0.566$ & $0.698$ & $0.586$ & $0.713$\\
AR(1)-Pareto, centered at truth & 2000 & $0.436$ & $0.575$ & $0.471$ & $0.644$\\
AR(1)-Pareto, recentered & 1000 & $0.816$ & $0.893$ & --- & ---\\
AR(1)-Pareto, recentered & 2000 & $0.813$ & $0.895$ & --- & ---\\
\bottomrule
\end{tabular}
\begin{tablenotes}
\footnotesize
\item Notes: $800$ replications; Monte Carlo standard error at most $1.8$ points. Bands use pivotal critical values simulated at each $(n,m,h)$ ($1500$ draws). The undersmoothed columns use $0.75h_{n}$ at the same $m$. The recentered rows evaluate the same band around the finite-sample mean of $\hat{\gamma}(\cdot)$ rather than the true path. They isolate the variance calibration from the dependence bias (see Table \ref{tab:rmse} and Remark \ref{rm:covdecomp}). For comparison, the Gaussian-experiment calibration covers $0.836$ (DGP1, $n=1000$, nominal $0.90$) at the baseline configuration.
\end{tablenotes}
\end{threeparttable}
\end{table}

\textbf{Size and power of the stability test.} Table \ref{tab:test} reports rejection frequencies of the studentized statistic $\Lambda_{n}$ at the $5\%$ level, with pivotal critical values and projection centering. Under the constant-index nulls, the empirical size is $3$--$8\%$ across the i.i.d.\ designs, including the deterministic scale-drift design with a constant index ($4$--$8\%$). Two nulls deserve comment. First, under the AR(1) design, the size is $9\%$. The dependence bias shifts the whole path, the projection centering absorbs the shift, and the residual distortion is moderate. Second, under the GARCH(1,1)-$t$ null, the size is $13$--$14\%$. The GARCH recursion creates strong extremal clustering ($\bar{\hat{S}}\approx1.2$), and the effective local tail wobbles at frequencies the kernel cannot distinguish from genuine shape variation at these sample sizes. This is a finite-sample phenomenon, and it improves when the clustering is filtered out. The full filter-and-test pipeline of the empirical section, namely fitting a GARCH model and testing the standardized residuals, shows no excess rejections beyond the nominal level in $40$ validation replications ($0/40$ in our fresh run, against $2/40$ in the original run). The binomial $95\%$ interval for a size of $5\%$ is $[0.006,0.169]$, so a small validation sample cannot exclude modest over-rejection. 

Against the alternatives, power increases steadily with the departure size, and it is substantial for the larger breaks ($56$--$63\%$ at $\delta=0.2$). Power is lower under DGP2 because the time-varying second-order bias partially offsets the drift in $\gamma(\cdot)$. 

For the scale--shape question, the relevant comparison is the pooled-centering benchmark. Replacing the projection centering by the pooled Hill estimator with a global threshold inflates the size from $6\%$ to $14\%$ under the deterministic drift design and from $12\%$ to $35\%$ under DGP4 (Table \ref{tab:test}, notes), with no change under the i.i.d.\ null beyond a doubling from $5.5\%$ to $10\%$. This is the finite-sample content of the centering discussion in Section \ref{sec:test}.

\begin{table}[!htbp]
\centering
\caption{Size and power of the tail stability test $\Lambda_{n}$ at the $5\%$ level}
\label{tab:test}
\begin{threeparttable}
\begin{tabular}{lcc}
\toprule
& $n=1000$ & $n=2000$\\
\midrule
\multicolumn{3}{l}{\emph{size: constant index}}\\
Pareto, $\gamma\equiv0.3$ & $0.055$ & $0.050$\\
Student-$t$, $\gamma\equiv0.3$ & $0.030$ & $0.028$\\
GARCH(1,1)-$t$, $\gamma\equiv c$ & $0.134$ & $0.139$\\
AR(1)-Pareto, $\gamma\equiv0.3$ & $0.088$ & $0.088$\\
SV (DGP4), $\gamma\equiv0.3$ & $0.041$ & $0.031$\\
deterministic scale drift, $\gamma\equiv0.25$ & $0.075$ & $0.044$\\
\midrule
\multicolumn{3}{l}{\emph{power}}\\
Pareto, slope $a=0.1$ & $0.121$ & $0.112$\\
Pareto, slope $a=0.2$ & $0.222$ & $0.219$\\
Pareto, slope $a=0.3$ & $0.314$ & $0.319$\\
Pareto, break $\delta=0.1$ & $0.229$ & $0.262$\\
Pareto, break $\delta=0.2$ & $0.555$ & $0.626$\\
Student-$t$, slope $a=0.2$ & $0.071$ & $0.078$\\
Student-$t$, break $\delta=0.2$ & $0.291$ & $0.379$\\
\bottomrule
\end{tabular}
\begin{tablenotes}
\footnotesize
\item Notes: $800$ replications; nominal level $5\%$; pivotal critical values and projection centering (\ref{pooled}). Slope alternatives are $\gamma(u)=0.2+au$, and break alternatives are $\gamma(u)=0.3+\delta\mathbf{1}\{u\geq0.5\}$. With pooled centering (global threshold) instead of the projection centering, the size is $0.100$ (Pareto), $0.352$ (SV), and $0.138$ (deterministic drift) at $n=2000$, against $0.055$, $0.118$, and $0.058$ with projection centering in the same $400$ replications. The GARCH null uses $(\omega,\alpha,\beta)=(10^{-5},0.105,0.879)$ with standardized $t_{10/3}$ innovations.
\end{tablenotes}
\end{threeparttable}
\end{table}

\textbf{Diagnostics.} Two further experiments isolate the sources of the dependence bias and of the coverage loss. First, we compare the estimator on AR(1) paths with the estimator on random permutations of the same paths, which preserve the exact marginal distribution and destroy the run structure. The bias on the permuted paths is $-0.010$ at $n=1000$ and $-0.005$ at $n=2000$, against $-0.036$ and $-0.043$ on the original paths. The dependence-induced component is therefore $-0.026$ and $-0.038$, while the marginal finite-threshold component is small. Second, we evaluate the $90\%$ band with true versus estimated threshold and true versus estimated long-run variance factor. At $n=1000$, coverage is $0.522$ with both estimated, $0.542$ with the true threshold, $0.595$ with the true factor, and $0.616$ with both true. At $n=2000$, the four values are $0.442$, $0.441$, $0.525$, and $0.529$. Threshold estimation therefore contributes little, variance estimation contributes a moderate share, and the dominant loss is the dependence bias itself. Even with oracle inputs, coverage remains below nominal.

\textbf{Sensitivity to $(m,h)$.} Table \ref{tab:sens} reports the bias, standard deviation, and RMSE, averaged over the evaluation grid, for a grid of intermediate orders and bandwidths at $n=1000$. Two regularities emerge. First, for DGP1 (no second-order bias), accuracy improves monotonically in the effective number of exceedances $mh$. The RMSE falls from $0.067$ at $mh=19$ to $0.046$ at $mh=38$. Second, the estimator is insensitive to the HAC lag length $J$ over the range $5$--$15$, which supports the practical rule of taking the smallest $J$ beyond which the local extremal autocorrelations are negligible. These patterns conform to the bias--variance decomposition of Theorem \ref{tm:an} and support the practical rule $mh\in[20,40]$ of Section \ref{sec:tuning}.

\begin{table}[!htbp]
\centering
\caption{Sensitivity of $\hat{\gamma}(u)$ to $m$, $h$, and $J$; DGP1, $n=1000$}
\label{tab:sens}
\begin{threeparttable}
\begin{tabular}{cccccc}
\toprule
$m$ & $h$ & $J$ & bias & sd & RMSE\\
\midrule
$75$ & $0.25$ & $8$ & $0.009$ & $0.066$ & $0.067$\\
$100$ & $0.20$ & $8$ & $0.006$ & $0.063$ & $0.063$\\
$100$ & $0.25$ & $8$ & $0.008$ & $0.057$ & $0.057$\\
$100$ & $0.30$ & $8$ & $0.012$ & $0.052$ & $0.053$\\
$150$ & $0.25$ & $8$ & $0.006$ & $0.046$ & $0.046$\\
$100$ & $0.25$ & $5$ & $0.008$ & $0.057$ & $0.057$\\
$100$ & $0.25$ & $15$ & $0.008$ & $0.057$ & $0.057$\\
\bottomrule
\end{tabular}
\begin{tablenotes}
\footnotesize
\item Notes: $400$ replications; entries are averages over $41$ grid points of the bias, standard deviation, and RMSE of $\hat{\gamma}(u)$.
\end{tablenotes}
\end{threeparttable}
\end{table}

\begin{figure}[!htbp]
\centering
\includegraphics[width=0.95\textwidth]{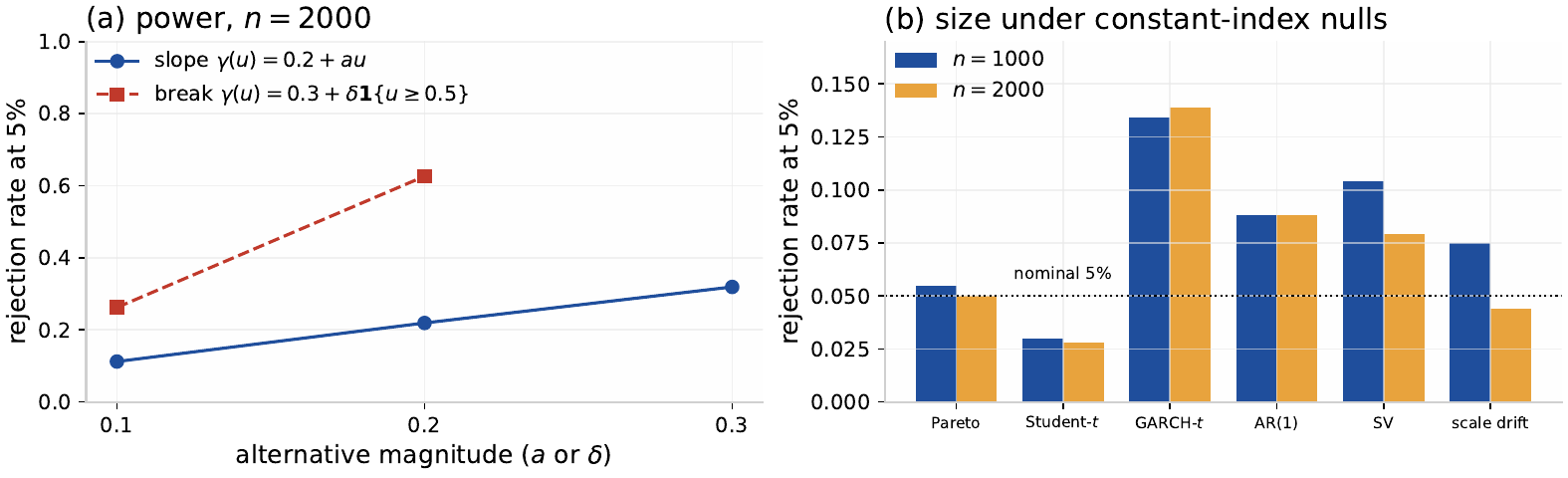}
\caption{Monte Carlo results for the stability test $\Lambda_{n}$, $800$ replications, pivotal calibration. Panel (a): rejection rates against the slope alternatives $\gamma(u)=0.2+au$ (circles) and the break alternatives $\gamma(u)=0.3+\delta\mathbf{1}\{u\geq0.5\}$ (squares) at $n=2000$. Panel (b): empirical size under the six constant-index nulls at $n=1000$ and $n=2000$. The dotted line marks the nominal $5\%$.}
\label{fig:sim}
\end{figure}

Figure \ref{fig:sim} summarizes the testing results. Panel (a) displays the rejection rates of $\Lambda_{n}$ against the slope and break alternatives at $n=2000$, and panel (b) displays the empirical size under the six constant-index nulls. The test is approximately correctly sized under the i.i.d.-type nulls, moderately liberal under extremal clustering, and nontrivially powerful against alternatives of the magnitude documented in the application.

\section{Empirical study}\label{sec:emp}
We now revisit the data in Section \ref{sec:data} and apply the proposed method to daily S\&P 500 losses analyzed there. Throughout this section, the baseline configuration is $m=200$ (a threshold probability of about $2.2\%$), $h=0.1$, and $J=20$. The effective number of local exceedances is therefore $mh=20$, which is at the conservative end of the recommended range of Section \ref{sec:tuning}. We adopt this conservative configuration because the application spans thirty-six years, and the threshold probability is small. The evaluation grid consists of $41$ points on $[h,1-h]$. All bands and test critical values are pivotal, simulated at the empirical configuration $(n,m,h,J)=(9179,200,0.1,20)$, with $5500$ draws. The resulting sup-band critical values are $3.68$ at $90\%$ and $4.21$ at $95\%$.

\begin{figure}[!htbp]
\centering
\includegraphics[width=0.95\textwidth]{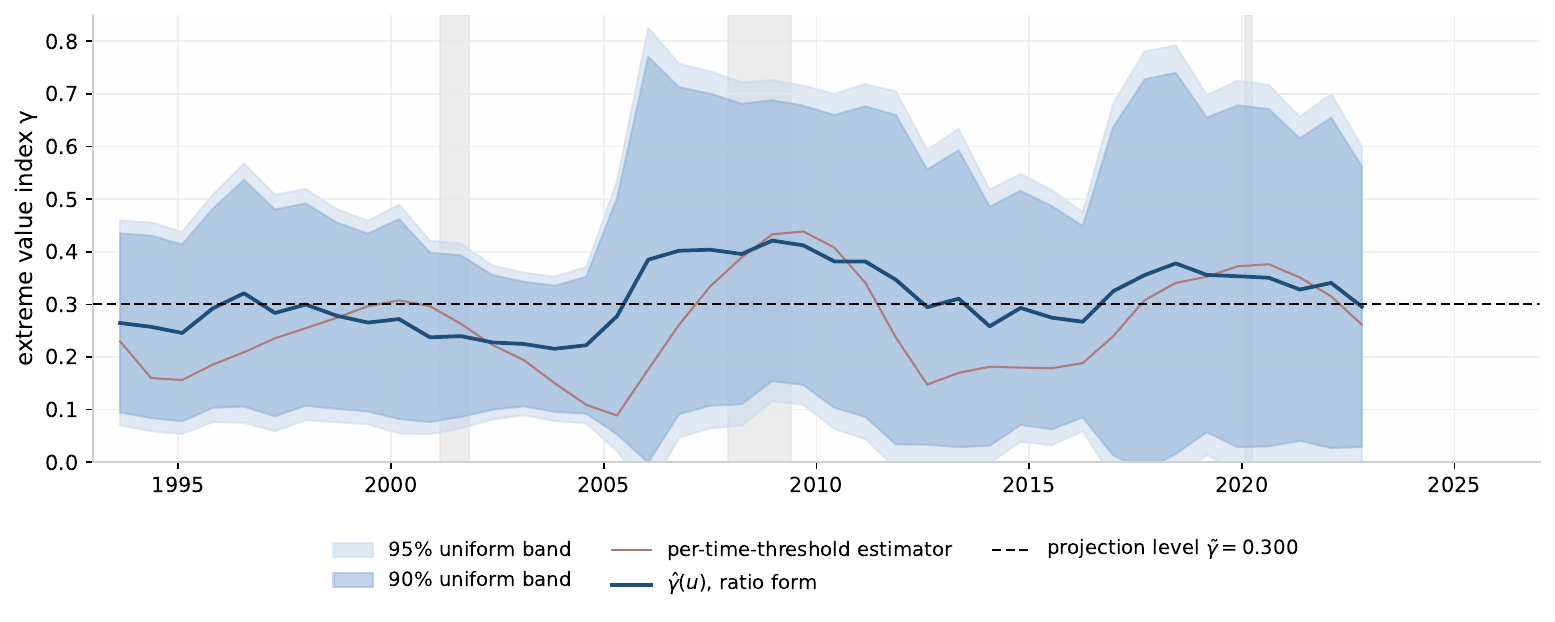}
\caption{Estimated path of the extreme value index $\hat{\gamma}(u)$ for daily S\&P 500 losses, 1990--2026 ($n=9179$, $m=200$, $h=0.1$, $J=20$). Shaded bands are the $90\%$ and $95\%$ uniform bands of (\ref{band}) with pivotal critical values. The horizontal line is the projection level $\tilde{\gamma}=0.300$. The dashed line is a per-time-threshold variant with the expected exceedance count as denominator, shown for comparison. Shading marks the NBER-dated contractions and the 2020 pandemic crash.}
\label{fig:emp-gamma}
\end{figure}

\textbf{Summary of the tail path.} Figure \ref{fig:emp-gamma} displays the estimated path $\hat{\gamma}(u)$ of the extreme value index with the uniform bands of (\ref{band}). The path is elevated relative to the simulation benchmarks because $mh=20$ is small. The pointwise standard errors are correspondingly wide. Three features stand out. First, the level of the path is $\tilde{\gamma}=0.300$ under the projection centering of (\ref{pooled}). Second, the path moves within the range $[0.216,0.421]$ over the sample. The early 1990s sit at the low end, the mid-2000s and the late 2010s at the high end. Third, the ratio-form estimator (\ref{local index estimator}) tracks the rolling Hill benchmark of Figure \ref{fig:motivation} qualitatively, but it is smoother and its excursions are smaller. The contrast experiment of Section \ref{sec:data} suggests that part of the raw rolling variation is threshold-induced noise, which the common-threshold construction removes.

\begin{table}[!htbp]
\centering
\caption{Estimated extreme-value-index path: full sample and subperiods}
\label{tab:sum}
\begin{threeparttable}
\begin{tabular}{lccccc}
\toprule
sample & level $\tilde{\gamma}$ & range of $\hat{\gamma}(u)$ & $\bar{\hat{S}}$ & $\Lambda_{n}$ & $p$-value\\
\midrule
1990--2026 (full) & $0.300$ & $[0.216,0.421]$ & $1.46$ & $0.902$ & $0.220$\\
1990--1999 & $0.394$ & $[0.306,0.510]$ & $0.94$ & --- & $0.633$\\
2000--2009 & $0.304$ & $[0.196,0.449]$ & $1.02$ & --- & $0.045$\\
2010--2019 & $0.427$ & $[0.306,0.706]$ & $1.21$ & --- & $0.207$\\
2016--2026 & $0.452$ & $[0.240,0.729]$ & $1.79$ & --- & $0.083$\\
GARCH residuals & $0.211$ & $[0.136,0.308]$ & $1.00$ & $1.706$ & $0.019$\\
GJR residuals & $0.212$ & $[0.142,0.315]$ & $0.93$ & $1.938$ & $0.008$\\
\bottomrule
\end{tabular}
\begin{tablenotes}
\footnotesize
\item Notes: the level is the projection centering (\ref{pooled}), and the range is over the $41$-point evaluation grid. Subperiod tests use the same $(m,h,J)$ configuration within each subsample, with pivotal critical values simulated at the subsample size. The residual rows apply the estimator and the test to the absolute standardized residuals $|\hat{\varepsilon}_{t}|=|X_{t}|/\hat{\sigma}_{t}$ of the fitted conditional volatility models (see the text).
\end{tablenotes}
\end{threeparttable}
\end{table}

\textbf{Testing stability: raw losses.} The omnibus test of Section \ref{sec:test} does not reject the constant-index null for the full sample (Table \ref{tab:emp-test}). The statistic is $\Lambda_{n}=0.902$ with a pivotal $p$-value of $0.220$. The conclusion is unchanged when the statistic is recomputed with $J\in\{10,15,30\}$. The $p$-value then ranges from $0.17$ to $0.23$. At first glance, this suggests a stable extreme value index. 

Two considerations qualify it. First, the estimated long-run variance factor is $\bar{\hat{S}}=1.46$, far above one. Extremes cluster strongly in the raw series, the studentization inflates the denominator of $\Lambda_{n}$, and the test loses power. Second, the subperiod evidence is mixed. The test does not reject within the 1990s ($p=0.633$), the 2010s ($p=0.207$), or the most recent decade ($p=0.083$). However, it rejects at the $5\%$ level within the 2000s ($p=0.045$, a decade that contains both the calm mid-2000s and the financial crisis. Table \ref{tab:sum} reports the subperiod levels. The level itself moves considerably across decades, from $0.304$ in the 2000s to $0.452$ in 2016--2026. Level shifts are absorbed by the projection centering within each subsample, so they do not drive the test. Nevertheless, they show that the unconditional tail is far from constant over long horizons.

\begin{table}[!htbp]
\centering
\caption{Stability tests for the S\&P 500 extreme value index, $5\%$ level}
\label{tab:emp-test}
\begin{threeparttable}
\begin{tabular}{lcccc}
\toprule
series & $\Lambda_{n}$ & pivotal critical value & $p$-value & decision\\
\midrule
raw losses, 1990--2026 & $0.902$ & $1.395$ & $0.220$ & do not reject\\
GARCH(1,1) residuals & $1.706$ & $1.395$ & $0.019$ & reject\\
GJR residuals & $1.938$ & $1.395$ & $0.008$ & reject\\
\bottomrule
\end{tabular}
\begin{tablenotes}
\footnotesize
\item Notes: pivotal critical values and $p$-values are simulated at each configuration ($5500$ i.i.d.\ Pareto draws, statistic recomputed per draw). The residual rows test the absolute standardized residuals of the fitted volatility models, so the volatility filter removes the clustering that inflates $\hat{S}$ in the raw series.
\end{tablenotes}
\end{threeparttable}
\end{table}

\begin{figure}[!htbp]
\centering
\includegraphics[width=0.95\textwidth]{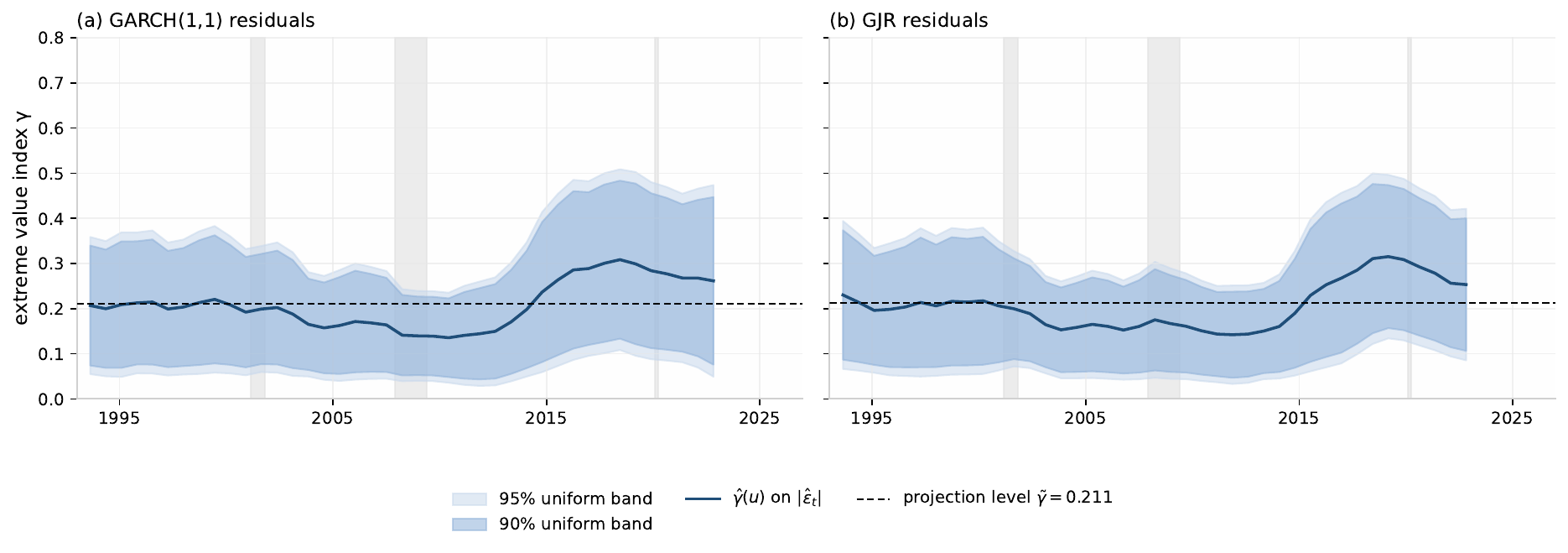}
\caption{Estimated extreme-value-index paths of the standardized residuals, $m=200$, $h=0.1$, $J=20$, with pivotal $90\%$ and $95\%$ uniform bands. Panel (a): absolute GARCH(1,1) residuals, level $0.211$. Panel (b): absolute GJR residuals, level $0.212$.}
\label{fig:garch}
\end{figure}

\textbf{Filtering out volatility clustering.} The contrast between clustering and shape variation is a central theme of the paper. We therefore fit two conditional volatility models to the return series and re-estimate the tail path of the absolute standardized residuals $|\hat{\varepsilon}_{t}|=|X_{t}|/\hat{\sigma}_{t}$. The GARCH(1,1) estimates are $(\omega,\alpha,\beta)=(0.019,0.101,0.883)$, with persistence $0.984$. The GJR variant adds a leverage coefficient of $0.159$ on negative returns, with $(\omega,\alpha,\beta)=(0.022,0.004,0.898)$. Both filters reduce the excess kurtosis from about $11$ in the raw series to about $2$. Figure \ref{fig:garch} displays the residual tail paths, and the last two rows of Table \ref{tab:sum} summarize them. Three findings emerge. First, the level falls to $0.21$, about $30\%$ below the raw level. The residual index is a different target from the raw index, so the two levels need not coincide, and the level drop is not by itself evidence of confounding. Second, the residual range narrows to $[0.136,0.308]$ (GARCH) and $[0.142,0.315]$ (GJR). Third, the clustering factor falls to $\bar{\hat{S}}\approx1.0$, so the filter works as intended.

Against this cleaner background, the stability test \emph{rejects}. The statistic rises to $1.706$ (GARCH) and $1.938$ (GJR), with $p$-values of $0.019$ and $0.008$. The validation experiment of Section \ref{sec:mc} reports no excess rejections beyond the nominal level under a calibrated GARCH null, with the binomial uncertainty of a small validation sample made explicit. The rejection is therefore attributable to shape variation in the innovations rather than to filter-induced size distortion. In other words, the raw path's movements include scale-induced exceedance-content variation. The volatility filter removes this channel. After its removal, the innovation index moves over time by more than sampling variation can explain.

\textbf{Crisis versus calm periods.} A natural conjecture is that the tail fattens in crises. We test it directly. We classify the evaluation windows into crisis windows (the 2000--02 bear market, the 2008--09 financial crisis, and the 2020 pandemic crash) and calm windows, and we compare the average $\hat{\gamma}(u)$ across the two groups. The crisis average is $0.304$, and the calm average is $0.312$. The difference is negative and insignificant ($z=-0.23$, $p=0.819$). The point estimate even has the opposite sign from the conjecture. 

Two remarks are in order. First, crisis windows are short, and the local estimator is imprecise there, so the comparison has limited power. Second, and more importantly, the raw-path movements that \emph{look} like crisis fattening in rolling estimates are largely scale effects. The volatility filter of the previous paragraph removes them. What remains in the residuals is low-frequency shape variation that is not aligned with the crisis calendar. The evidence therefore does not support the simple crisis-fattening narrative. The extreme value index varies, but the variation is not simply a crisis effect.

\textbf{Robustness.} Table \ref{tab:robust} reports the level, range, and test outcome under alternative intermediate orders $m\in\{100,150,250\}$ and under the undersmoothed bandwidth $0.75h$. The level estimate is stable across $m$ ($0.286$--$0.306$), the range endpoints move moderately, and the omnibus test never rejects for the raw series. The residual rejections are stable at $m=250$ ($p=0.035$) and at $0.75h$ ($p=0.013$), but they attenuate at $m=150$ ($p=0.106$) and $m=100$ ($p=0.417$), where the effective number of exceedances is smallest. We also re-estimated the path with a per-time-threshold variant with the expected exceedance count as denominator. Its path is noisier, with isolated downward spikes at windows where the local threshold quantile jumps, consistent with the control experiments of Section \ref{sec:data}.

\begin{table}[!htbp]
\centering
\caption{Robustness of the empirical findings, raw losses 1990--2026}
\label{tab:robust}
\begin{threeparttable}
\begin{tabular}{lccccc}
\toprule
configuration & level $\tilde{\gamma}$ & range & $\bar{\hat{S}}$ & $\Lambda_{n}$ & $p$-value\\
\midrule
baseline: $m=200$, $h=0.10$ & $0.300$ & $[0.216,0.421]$ & $1.46$ & $0.902$ & $0.220$\\
$m=100$ & $0.286$ & $[0.200,0.408]$ & $1.32$ & $0.458$ & $0.689$\\
$m=150$ & $0.291$ & $[0.210,0.439]$ & $1.38$ & $0.614$ & $0.466$\\
$m=250$ & $0.306$ & $[0.218,0.417]$ & $1.54$ & $0.862$ & $0.241$\\
$0.75h$, $m=200$ & $0.292$ & $[0.212,0.416]$ & $1.35$ & $0.510$ & $0.744$\\
\bottomrule
\end{tabular}
\begin{tablenotes}
\footnotesize
\item Notes: each row re-estimates the full pipeline at the stated configuration, with pivotal critical values simulated at that configuration. The $J=20$ HAC lag length is held fixed.
\end{tablenotes}
\end{threeparttable}
\end{table}

\textbf{Risk measurement.} We finally turn to the out-of-sample risk measures of Section \ref{sec:risk}. The design is fully recursive. The initial estimation window covers 1990--1999. For each subsequent trading day up to the end of June 2026, we re-estimate the intermediate quantile and the index on the expanding window of all past losses, evaluated at the boundary with one-sided kernel weights, and we extrapolate to the one-day $99\%$ VaR and the corresponding ES. The evaluation period covers $6656$ days, so a correctly specified $99\%$ measure violates $66.6$ times in expectation. The extrapolation factor $d_{n}=(m/n_{t})/p_{n}$ declines from $7.9$ at the start of the evaluation to $2.2$ at its end. 

We compare three constructions. The first is the full time-varying measure, with both the threshold and the index updated daily. The second is a constant-shape variant, which updates the threshold but fixes the index at its initial-window estimate $0.329$. The third is a fully static measure, estimated once on the initial window. 

\begin{figure}[!htbp]
\centering
\includegraphics[width=0.95\textwidth]{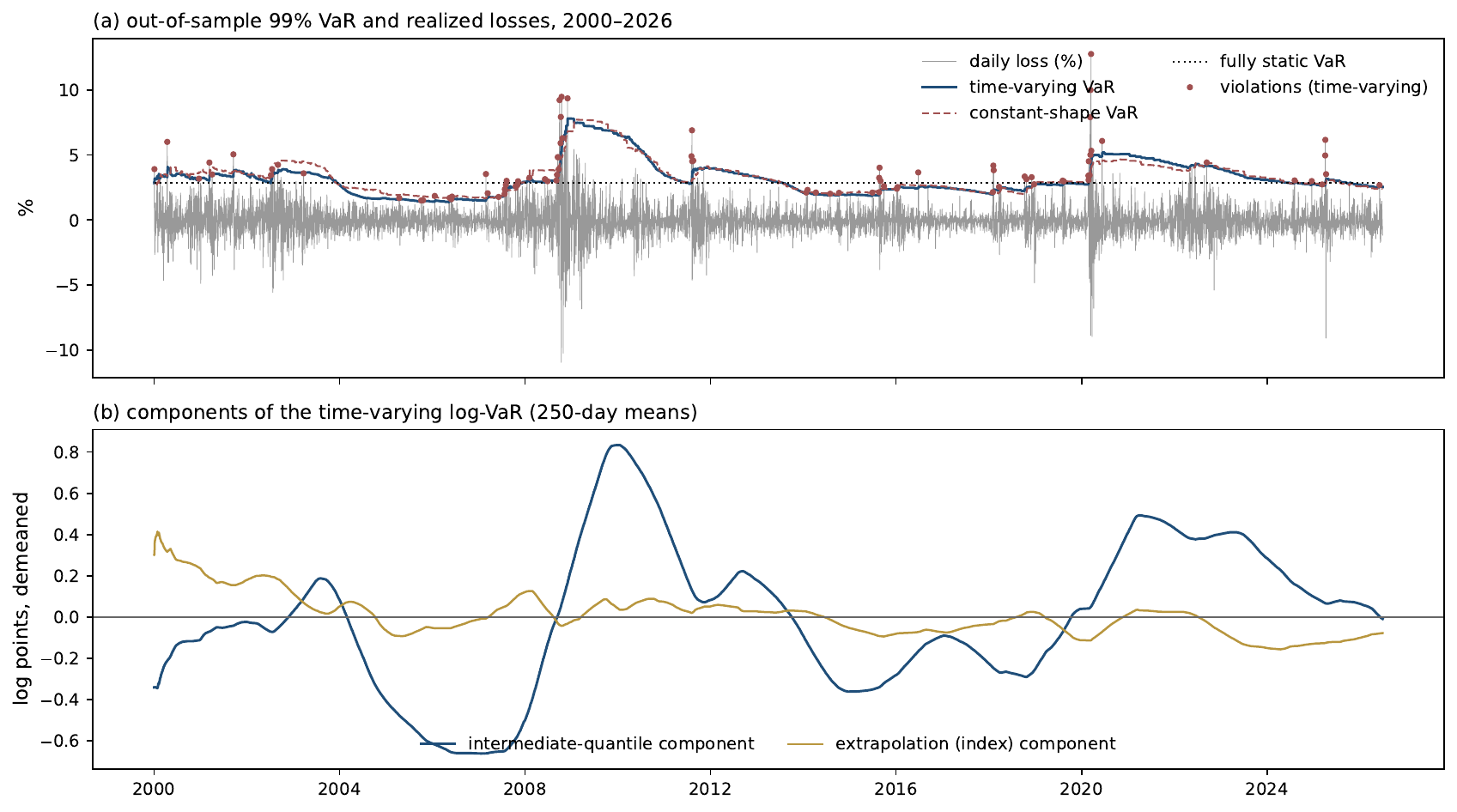}
\caption{Out-of-sample risk measures for daily S\&P 500 losses, 2000--2026. Panel (a): realized losses (grey), the time-varying $99\%$ VaR (solid), the constant-shape VaR (dashed), and the fully static VaR (dotted). Circles mark violations of the time-varying measure. Panel (b): $250$-day moving averages of the demeaned intermediate-quantile component and the demeaned extrapolation (index) component of the log-VaR of the time-varying measure.}
\label{fig:emp-risk}
\end{figure}

Table \ref{tab:backtest} reports the backtesting results, and Figure \ref{fig:emp-risk} displays the estimated paths together with the scale--shape decomposition. Three findings stand out. First, both measures with a time-varying threshold dominate the fully static measure. The static measure violates on $123$ days ($1.85\%$), and its failures concentrate in the crises. Its violation rate reaches $10.7\%$ in 2008--09 and $50\%$ in the 2020 crash. Updating the threshold and the index cuts these rates to $4.8\%$ and $37.5\%$, and the maximal rolling $250$-day violation count falls from $37$ to $15$. Second, the shape adjustment is neutral under the expanding design and valuable under rolling windows. The constant-shape variant is marginally closer to the nominal rate than the fully time-varying measure ($78$ against $87$ violations), because the updated index rises above its initial level in the 2010s and lifts the VaR. Under rolling $2500$-day windows, however, the ranking reverses sharply (Table \ref{tab:backtest}, notes). The time-varying measure attains $1.13\%$ against $1.67\%$ for the constant-shape variant. A fixed index inherited from the distant past misprices the tail whenever the index level drifts, and the documented drift from $0.30$ to $0.45$ is exactly of that size. Third, the decomposition of the log-VaR is clear. The day-to-day standard deviation of the demeaned intermediate-quantile component is $0.36$, against $0.11$ for the extrapolation (index) component. The time-varying $99\%$ VaR ranges from $1.4\%$ to $7.8\%$ over the evaluation period, and the ES from $1.7\%$ to $13.5\%$, against $1.7$--$7.7\%$ and $2.5$--$11.5\%$ for the constant-shape variant. The ES estimates are conservative. On violation days, the average realized loss is $96\%$ of the estimated ES for the time-varying measure and $91\%$ for the constant-shape variant.

\begin{table}[!htbp]
\centering
\caption{Out-of-sample backtesting of the $99\%$ VaR, 2000--2026 ($6656$ days)}
\label{tab:backtest}
\begin{threeparttable}
\small\setlength{\tabcolsep}{4pt}
\begin{tabular}{lccccccc}
\toprule
& & & \multicolumn{3}{c}{violation rate (\%) by episode} & & \\
\cmidrule(lr){4-6}
measure & violations & rate (\%) & 2000--02 & 2008--09 & 2020 & Kupiec $p$ & max in 250d\\
\midrule
time-varying & $87$ & $1.31$ & $1.26$ & $4.79$ & $37.50$ & $0.016$ & $15$\\
constant shape & $78$ & $1.17$ & $0.78$ & $3.94$ & $41.67$ & $0.170$ & $14$\\
fully static & $123$ & $1.85$ & $2.35$ & $10.70$ & $50.00$ & $0.000$ & $37$\\
\bottomrule
\end{tabular}
\begin{tablenotes}
\footnotesize
\item Notes: the nominal rate is $1\%$, with $66.6$ violations expected over the $6656$ evaluation days. Forecasts are recursive. The time-varying measure updates the threshold and the index daily on the expanding window, the constant-shape variant updates the threshold with the index fixed at its initial-window estimate $0.329$, and the static measure is estimated once on 1990--1999. The episode columns report violation rates within the three most severe drawdowns of the evaluation period, defined peak to trough (the 2000--02 bear market, the 2008--09 financial crisis, and the 2020 pandemic crash). The last column is the maximal number of violations within any rolling $250$-day window. The Christoffersen independence test rejects for all three measures ($p\leq0.003$), so violations cluster in all constructions. Under a sensitivity design with rolling $2500$-day estimation windows, the time-varying measure violates on $75$ days ($1.13\%$, Kupiec $p=0.308$) and the constant-shape variant on $111$ ($1.67\%$, $p=0.000$). The static measure is unaffected by the window design.
\end{tablenotes}
\end{threeparttable}
\end{table}

\textbf{Summary.} The application delivers three messages. First, the raw S\&P 500 extreme value index is statistically compatible with a constant shape once clustering is accounted for. The omnibus test does not reject ($p=0.220$), and the level estimate is stable across configurations. Second, the raw path and the innovation path estimate different targets. Once the clustering that inflates the variance estimate is filtered out, the innovation index varies significantly ($p\leq0.02$), and the subperiod levels of the raw path move from $0.30$ to $0.45$. This variation is genuine but low-frequency, and it is not a crisis phenomenon. Third, for risk measurement, the dominant channel is the scale. Updating the threshold over time cuts the crisis-period violation rate of a static $99\%$ VaR by more than half, and updating the index adds a refinement whose value is largest under rolling estimation windows.

\section{Conclusion}\label{sec:concl}
This paper develops estimation and inference procedures for the extreme value index of a locally stationary time series. We model the index as a smooth function of time and estimate it by a ratio-form kernel Hill estimator in which a single intermediate quantile thresholds all observations in the local window. The common threshold isolates the tail shape from arbitrary smooth scale variation. We build uniform confidence bands for the index path, propose a test of tail stability, and construct time-varying estimators of extreme Value-at-Risk and Expected Shortfall. Simulation experiments confirm that the procedures perform as the theory predicts.

The application to S\&P 500 daily losses illustrates the value of this separation. The index of the raw losses is compatible with a constant shape once extremal clustering is accounted for, whereas the index of volatility-filtered losses varies significantly over time. For risk measurement, the dominant channel is the intermediate quantile, and updating the tail threshold over time improves extreme risk forecasts substantially relative to a static benchmark.

Several extensions are worth pursuing. Bias-reduced versions of the estimator would improve the uniform bands in the more difficult designs. Data-driven selection of the tuning parameters would make the procedure fully automatic. A multivariate extension, with time-varying tail dependence alongside time-varying marginal tails, is of direct interest for systemic risk measurement. Finally, the finding that the tails of standardized residuals vary over time poses a modeling challenge. What drives the shape of the extreme tail is a question for future study.
\newpage\appendix
\section{Appendix: proofs}\label{app:proofs}
This appendix contains the complete proofs of the results in the main text. Section \ref{app:aux} collects auxiliary lemmas with proofs. Section \ref{app:main} proves the main theorems. Throughout, $C$ denotes a generic finite constant whose value may change from line to line, $\iota$ a generic small positive constant, and we write $K_{t}(u)=K_{h}(u-t/n)$, $\bar{K}(u)=n^{-1}\sum_{t=1}^{n}K_{t}(u)$, and $\bar{F}=1-F$. By Assumption \ref{as:kernel} and a Riemann sum approximation,
\begin{equation}\label{kbar}
\bar{K}(u)=1+O(h^{2}),\qquad n^{-1}\sum_{t=1}^{n}K_{t}^{2}(u)=h^{-1}\nu_{2}(K)(1+O(h)),
\end{equation}
uniformly over $u\in[h,1-h]$, where the second relation uses $\sum_{t}K_{h}^{2}(u-t/n)=h^{-2}\sum_{t}K^{2}((u-t/n)/h)=h^{-2}(nh\nu_{2}(K)+O(1))$.

\subsection{Auxiliary lemmas}\label{app:aux}

\begin{appxlem}[Moments of log-excesses]\label{lem:mom}
Let $k\geq1$ be an integer. Under Assumptions \ref{as:rv}--\ref{as:so}, uniformly in $u\in[0,1]$ and in $t$ with $|t/n-u|\leq2h$,
\begin{equation}\label{momk}
E\left[\log X_{t,n}-\log b(n/m,t/n)\right]_{+}^{k}=\frac{m}{n}\,k!\,\gamma^{k}(t/n)\left(1+c_{k}(t/n)\,A(n/m,t/n)+o\left(|A(n/m,t/n)|\right)\right)+o\left(\frac{m}{n}\right),
\end{equation}
where $c_{k}$ is bounded and depends only on $(\gamma,\rho)$; in particular, for the parameterization of Assumption \ref{as:so}, $c_{1}(u)=1/(1-\rho(u))$. Moreover, with $\xi_{t,n}$, $\mathcal{I}_{t,n}$, $Z_{t,n}(u)$ as in (\ref{excess}), for $u\approx t/n$,
\begin{align}
E[\xi_{t,n}]&=\frac{m}{n}\gamma(u)\left(1+c_{1}(u)A(n/m,u)+o(|A|)\right),\label{mom1}\\
E[\mathcal{I}_{t,n}]&=\frac{m}{n}(1+o(1)),\qquad \mathrm{Cov}(\xi_{t,n},\mathcal{I}_{t,n})=\frac{m}{n}\gamma(u)(1+o(1)),\label{mom2}\\
E[Z_{t,n}(u)^{2}]&=\frac{m}{n}\gamma^{2}(u)(1+o(1)),\qquad E|Z_{t,n}(u)|^{r}\leq C_{r}\frac{m}{n}\ \text{for every }r\geq1.\label{mom3}
\end{align}
\end{appxlem}

\begin{proof}
Write $b=b(n/m,t/n)$, $\gamma=\gamma(t/n)$, $A=A(n/m,t/n)$. Since $\xi_{t,n}^{k}$ is nonnegative, the layer-cake representation gives
\[
E[\xi_{t,n}^{k}]=\int_{0}^{\infty}k y^{k-1}\,P\left(\log X_{t,n}-\log b>y\right)dy=\int_{0}^{\infty}k y^{k-1}\,\bar{F}(be^{y},t/n)\,dy.
\]
Split the integral at $y_{n}=\kappa\log(n/m)$ with $\kappa>2\bar{\gamma}$. For $y\leq y_{n}$, Assumption \ref{as:so} applies with $\bar{F}(b,t/n)=m/n$ (by continuity of $F$ at $b$, Assumption \ref{as:smooth}):
\[
\bar{F}(be^{y},t/n)=\frac{m}{n}e^{-y/\gamma}\left(1+A\psi(y)+o(|A|)\right),
\]
where the $o(|A|)$ term is uniform in $y\leq y_{n}$ by the uniformity in (\ref{so-quantile}) and in $t/n$ by Assumption \ref{as:rv}. Hence
\[
\int_{0}^{y_{n}}ky^{k-1}\bar{F}(be^{y},t/n)dy=\frac{m}{n}\left[\int_{0}^{y_{n}}ky^{k-1}e^{-y/\gamma}dy+A\int_{0}^{y_{n}}ky^{k-1}e^{-y/\gamma}\psi(y)dy+o(|A|)\right].
\]
The first integral equals $k!\gamma^{k}+O(e^{-y_{n}/(2\gamma)})=k!\gamma^{k}+o((m/n)^{2})$ since $e^{-y_{n}/\gamma}=(m/n)^{\kappa/\gamma}$ with $\kappa/\gamma>2$. The second integral converges to a bounded constant $c_{k}'$ by the integrability of $\psi$ (Assumption \ref{as:so}), giving $c_{k}=c_{k}'/(k!\gamma^{k})$. The explicit form $c_{1}(u)=1/(1-\rho(u))$ follows from the computation $\int_{0}^{\infty}e^{-y/\gamma}\psi(y)dy=\int_{0}^{\infty}e^{-y/\gamma}(e^{y\rho/\gamma}-1)dy/\rho=\gamma/(1-\rho)$ for $\psi(y)=(e^{y\rho/\gamma}-1)/\rho$ of Remark \ref{rm:so}. For $y>y_{n}$, Potter-type bounds for the uniform second-order condition (cf.\ \citealp[Appendix B]{deHaanFerreira2006}) give $\bar{F}(be^{y},t/n)/\bar{F}(b,t/n)\leq Ce^{-y/(2\bar{\gamma})}$, whence
\[
\int_{y_{n}}^{\infty}ky^{k-1}\bar{F}(be^{y},t/n)dy\leq C\frac{m}{n}\int_{y_{n}}^{\infty}ky^{k-1}e^{-y/(2\bar{\gamma})}dy=o\left(\frac{m}{n}\right),
\]
since the tail of a Gamma integral vanishes exponentially in $y_{n}$. This proves (\ref{momk}). For (\ref{mom2}), $E[\mathcal{I}_{t,n}]=\bar{F}(b,t/n)=m/n$ exactly (by continuity), and $\mathrm{Cov}(\xi_{t,n},\mathcal{I}_{t,n})=E[\xi_{t,n}\mathcal{I}_{t,n}]-E[\xi_{t,n}]E[\mathcal{I}_{t,n}]=E[\xi_{t,n}](1-m/n)$ since $\xi_{t,n}\mathcal{I}_{t,n}=\xi_{t,n}$. For (\ref{mom3}), expand $Z_{t,n}^{2}=\xi_{t,n}^{2}-2\gamma(u)\xi_{t,n}\mathcal{I}_{t,n}+\gamma^{2}(u)\mathcal{I}_{t,n}$ and apply (\ref{momk}) with $k=1,2$ and (\ref{mom2}): the leading term is $(m/n)(2\gamma^{2}-2\gamma^{2}+\gamma^{2})=(m/n)\gamma^{2}$. Finally, for integer $r\geq1$, $E|Z_{t,n}|^{r}\leq 2^{r-1}(E[\xi_{t,n}^{r}]+\gamma^{r}E[\mathcal{I}_{t,n}])\leq C_{r}(m/n)$ by (\ref{momk}). For general $r>0$, the same layer-cake computation as above, with $k$ replaced by $r$, gives $E[\xi_{t,n}^{r}]\leq C_{r}(m/n)$. Note that only the closed form $k!\gamma^{k}$ of the Gamma integral uses integer $k$, while the bounds do not. Hence $E|Z_{t,n}|^{r}\leq C_{r}(m/n)$ for every $r>0$.
\end{proof}

\begin{appxlem}[Moments at the common threshold]\label{lem:momC}
Define the \emph{pooled local quantile} $b_{n}(u)$ by
\begin{equation}\label{mixq}
\frac{1}{n\bar K(u)}\sum_{t=1}^{n}K_{t}(u)\,\bar{F}\bigl(b_{n}(u),t/n\bigr)=\frac{m}{n},
\end{equation}
which exists and is unique for all large $n$ by the continuity and strict monotonicity of $\bar{F}(\cdot,v)$ near the intermediate level (Assumption \ref{as:smooth}). Write $d(u)=\log b_{n}(u)-\log b(n/m,u)$, $\Delta_{t}(u)=\gamma(t/n)-\gamma(u)$, and $L_{n}=\log(n/m)$, and define the exceedance variables at the common threshold, $\tilde{\xi}_{t,n}(u)=(\log X_{t,n}-\log b_{n}(u))_{+}$, $\tilde{\mathcal{I}}_{t,n}(u)=\mathbf{1}\{\tilde{\xi}_{t,n}(u)>0\}$, and $\tilde{Z}_{t,n}(u)=\tilde{\xi}_{t,n}(u)-\gamma(u)\tilde{\mathcal{I}}_{t,n}(u)$. Under Assumptions \ref{as:ls}--\ref{as:tailloc}, uniformly in $u$ and in $t$ with $|t/n-u|\leq2h$:
(i) the threshold drift $d(u)=\log b_{n}(u)-\log b(n/m,u)$ satisfies $d(u)=O(h^{2}L_{n}^{2})=o(1)$, and under $hL_{n}^{3/2}\to0$ it admits the expansion
\begin{equation}\label{drift}
d(u)=\frac{1}{2}h^{2}\mu_{2}(K)\,\frac{\bigl(\gamma'(u)\bigr)^{2}L_{n}^{2}}{\gamma(u)}+O\bigl(h^{2}L_{n}\bigr).
\end{equation}
(ii) $E[\tilde{\mathcal{I}}_{t,n}(u)]=\bar{F}(b_{n}(u),t/n)=\frac{m}{n}\exp\bigl\{(\Delta_{t}(u)L_{n}-d(u))/\gamma(u)+\eta(t/n)\bigr\}\bigl(1+o(1)\bigr)=\frac{m}{n}(1+o(1))$, where $\eta(v)=\log c(v)-\log c(u)=O(h)$ denotes the slowly varying factor of Assumption \ref{as:rv};
(iii) $E[\tilde{\xi}_{t,n}(u)]=\gamma(t/n)\,E[\tilde{\mathcal{I}}_{t,n}(u)]\bigl(1+c_{1}(t/n)A(n/m,t/n)+o(|A|)\bigr)$, with $c_{1}$ as in Lemma \ref{lem:mom};
(iv) $E[\tilde{Z}_{t,n}(u)^{2}]=(m/n)\gamma^{2}(u)(1+o(1))$, and $E|\tilde{Z}_{t,n}(u)|^{r}\leq C_{r}(m/n)$ for every $r\geq1$;
(v) the kernel average of the means satisfies
\begin{equation}\label{bias-exp}
\frac{1}{n\bar K(u)}\sum_{t=1}^{n}K_{t}(u)\,E\bigl[\tilde{Z}_{t,n}(u)\bigr]=\frac{m}{n}\left[h^{2}\mu_{2}(K)\left\{\frac{1}{2}\gamma''(u)+\frac{\bigl(\gamma'(u)\bigr)^{2}L_{n}}{\gamma(u)}\right\}+\gamma(u)\,c_{1}(u)\,A(n/m,u)\right]+\frac{m}{n}\,o\bigl(h^{2}L_{n}+|A|\bigr).
\end{equation}
\end{appxlem}

\begin{proof}
Write $b=b(n/m,u)$, $\gamma=\gamma(u)$, and $\eta(v)=\log c(v)-\log c(u)$, where $c(\cdot)$ is the smoothly varying scale factor of Assumption \ref{as:rv}, so that $\eta(u)=0$ and $\eta$, $\eta'$, $\eta''$ are uniformly bounded on the window. By the uniform second-order condition of Assumption \ref{as:so}, transported to the quantile function (cf.\ \citealp[Theorem 2.3.9]{deHaanFerreira2006}), and the identity $\log b=\gamma L_{n}+\log c(u)+o(1)$,
\begin{equation}\label{fratio}
\bar{F}\bigl(be^{y},v\bigr)=\frac{m}{n}\,e^{-y/\gamma}\,\exp\left\{\frac{\bigl(\gamma(v)-\gamma\bigr)\bigl(\log b+y\bigr)}{\gamma^{2}}\,\gamma\right\}\frac{c(v)}{c(u)}\bigl(1+O(|y|h+h+|A|)\bigr),
\end{equation}
uniformly in $v$ with $|v-u|\leq2h$ and $y$ in compact subsets of $\mathbb{R}$; the multiplicative error absorbs the products of $\gamma(v)-\gamma=O(h)$ with the $O(1)$ quantities $\log c(u)$ and $y$, as well as the second-order differences of Assumption \ref{as:so}. Since $(\gamma(v)-\gamma)\log b/\gamma=\Delta L_{n}+O(h)$, this simplifies to
\begin{equation}\label{fratio2}
\bar{F}\bigl(be^{y},v\bigr)=\frac{m}{n}\,e^{-y/\gamma}\,e^{\Delta L_{n}/\gamma+\eta(v)}\bigl(1+O(|y|h+h+|A|)\bigr).
\end{equation}
Evaluating (\ref{fratio2}) at $y=d(u)$, averaging with the weights $K_{t}(u)/(n\bar K(u))$, and imposing (\ref{mixq}) shows that $d(u)$ solves
\[
e^{-d(u)/\gamma}\,\frac{1}{n\bar K(u)}\sum_{t=1}^{n}K_{t}(u)\,e^{\Delta_{t}L_{n}/\gamma+\eta(t/n)}=1+o(1).
\]
Write $W_{t}=\Delta_{t}L_{n}/\gamma+\eta(t/n)$, so that $|W_{t}|=O(hL_{n})=o(1)$ by Assumption \ref{as:tailloc}. The cumulant expansion of the kernel average gives $\log\bigl((n\bar K)^{-1}\sum_{t}K_{t}e^{W_{t}}\bigr)=E_{K}W+\frac{1}{2}\mathrm{Var}_{K}(W)+O\bigl((hL_{n})^{3}\bigr)$, where $E_{K}$ and $\mathrm{Var}_{K}$ denote the mean and variance under the discrete kernel weights. Now $E_{K}\Delta=\frac{1}{2}h^{2}\mu_{2}(K)\gamma''(u)+o(h^{2})$ and $E_{K}\Delta^{2}=h^{2}\mu_{2}(K)(\gamma'(u))^{2}+o(h^{2})$ by the symmetry of $K$ and (\ref{kbar}), while $E_{K}\eta$ and $E_{K}\eta^{2}$ are $O(h^{2})$. Hence $d(u)/\gamma=E_{K}W+\frac{1}{2}\mathrm{Var}_{K}(W)+o(h^{2}L_{n}^{2})=\frac{1}{2}h^{2}\mu_{2}(K)(\gamma'(u))^{2}L_{n}^{2}/\gamma^{2}+O(h^{2}L_{n})$, which is (i). Parts (ii) and (iii) follow from (\ref{fratio2}) with $y=d(u)$ and from the layer-cake computation of Lemma \ref{lem:mom} applied at the shifted threshold $b_{n}(u)=be^{d(u)}$, since $e^{d(u)}=1+o(1)$ and $A(be^{d(u)},v)/A(b,v)\to1$ by the regular variation of $A$. Part (iv) follows from (ii)--(iii) exactly as (\ref{mom3}) follows from (\ref{mom1})--(\ref{mom2}), the tilt factors being $1+o(1)$ by Assumption \ref{as:tailloc}. For (v), parts (ii) and (iii) give, uniformly in $t$,
\[
E\bigl[\tilde{Z}_{t,n}(u)\bigr]=\frac{m}{n}\,e^{\Delta_{t}L_{n}/\gamma+\eta(t/n)-d(u)/\gamma}\left[\Delta_{t}+\gamma c_{1}(u)A(n/m,u)\right]+\frac{m}{n}\,o\bigl(h^{2}L_{n}+|A|\bigr).
\]
Multiply by $K_{t}(u)/(n\bar K(u))$ and sum. The second-order term contributes $(m/n)\gamma c_{1}(u)A(n/m,u)(1+o(1))$, using (\ref{mixq}). For the slope term, expand the exponential around $1$. By the defining constraint (\ref{mixq}), the zeroth-order part of the average equals $1$ exactly, so the threshold drift $d(u)$ enters only through the cross term $(n\bar K)^{-1}\sum_{t}K_{t}\Delta_{t}\cdot d(u)/\gamma=O(h\cdot h^{2}L_{n}^{2})=O(h^{3}L_{n}^{2})=o(h^{2}L_{n})$. The remaining pieces are $(n\bar K)^{-1}\sum_{t}K_{t}\Delta_{t}=\frac{1}{2}h^{2}\mu_{2}(K)\gamma''(u)+o(h^{2})$, $(n\bar K)^{-1}\sum_{t}K_{t}\Delta_{t}^{2}L_{n}/\gamma=h^{2}\mu_{2}(K)(\gamma'(u))^{2}L_{n}/\gamma+o(h^{2}L_{n})$, $(n\bar K)^{-1}\sum_{t}K_{t}\Delta_{t}\eta(t/n)=O(h^{2})$, and the third- and fourth-order remainder terms are $O(h^{3}L_{n}^{3}+h^{4}L_{n}^{4})=o(h^{2}L_{n})$ under $hL_{n}^{3/2}\to0$. Collecting terms yields (\ref{bias-exp}).
\end{proof}

\begin{appxlem}[Covariance summability]\label{lem:cov}
Under Assumptions \ref{as:ls}, \ref{as:rv} and \ref{as:extdep}, uniformly in $u$ and in $t$ with $|t/n-u|\leq2h$,
\begin{equation}\label{covsum}
\sum_{j=1}^{n-t}\left|\mathrm{Cov}\left(Z_{t,n}(u),\,Z_{t+j,n}(u)\right)\right|\leq C\,\frac{m}{n},
\end{equation}
and $(n/m)\mathrm{Cov}(Z_{t,n}(u),Z_{t+j,n}(u))\to\gamma^{2}(u)s_{j}(u)$ for each fixed $j$.
\end{appxlem}

\begin{proof}
The convergence is Assumption \ref{as:extdep}(ii). For the bound, note that $Z_{t,n}(u)=\xi_{t,n}-\gamma(u)\mathcal{I}_{t,n}$, and expand the covariance into four terms. It suffices to bound $\mathrm{Cov}(\xi_{t,n},\xi_{t+j,n})$, the other three being analogous (and directly covered by Assumption \ref{as:extdep}(i) with $x=y=1$). By the layer-cake representation of the product expectation,
\[
\mathrm{Cov}(\xi_{t,n},\xi_{t+j,n})=\int_{0}^{\infty}\!\!\int_{0}^{\infty}\mathrm{Cov}\left(\mathbf{1}\{X_{t,n}>be^{y}\},\,\mathbf{1}\{X_{t+j,n}>be^{z}\}\right)dy\,dz,
\]
with $b=b(n/m,u)$. The identity holds because $\xi_{t,n}=\int_{0}^{\infty}\mathbf{1}\{X_{t,n}>be^{y}\}dy$ a.s. The integrand is a joint exceedance probability minus the product of the marginals, so Assumption \ref{as:extdep}(i) applies directly and bounds it by $C(m/n)\varrho^{\,j}e^{-(y+z)/(2\bar{\gamma})}$ for $y,z\geq0$ (absorbing the threshold $x,y\geq1$ of part (i) into the constant: on the strips where $\min(y,z)<1$, split the integral at $1$ and use the trivial bound $|\mathrm{Cov}(\mathbf{1}\{X_{t,n}>be^{y}\},\mathbf{1}\{X_{t+j,n}>be^{z}\})|\leq2\min(P(X_{t,n}>be^{y}),P(X_{t+j,n}>be^{z}))\leq C(m/n)$ together with the mixing bound on the complementary region), so
\[
\left|\mathrm{Cov}(\xi_{t,n},\xi_{t+j,n})\right|\leq C\frac{m}{n}\varrho^{\,j}\left(\int_{0}^{\infty}e^{-y/(2\bar{\gamma})}dy\right)^{2}=C\frac{m}{n}\varrho^{\,j}.
\]
Summing over $j$ gives (\ref{covsum}) since $\varrho<1$.
\end{proof}

\begin{appxlem}[CLT for weighted exceedance sums]\label{lem:clt}
Let $u\in(0,1)$ be interior and let $a_{t,n}=K_{t}(u)$. Under Assumptions \ref{as:ls}--\ref{as:rates} and \ref{as:extdep},
\begin{equation}\label{wclt}
\frac{\sum_{t=1}^{n}a_{t,n}\left(Z_{t,n}(u)-EZ_{t,n}(u)\right)}{\sqrt{(m/n)\gamma^{2}(u)S(u)\nu_{2}(K)\,n/h}}\to_{d}N(0,1).
\end{equation}
\end{appxlem}

\begin{proof}
\textbf{Step 1 (variance).} By stationarity of the summands' second moments up to the smooth variation in $t/n$, (\ref{mom3}) and Lemma \ref{lem:cov},
\begin{align*}
\mathrm{Var}\left(\sum_{t}a_{t,n}Z_{t,n}(u)\right)&=\sum_{t}a_{t,n}^{2}\mathrm{Var}(Z_{t,n}(u))+\sum_{t\neq s}a_{t,n}a_{s,n}\mathrm{Cov}(Z_{t,n}(u),Z_{s,n}(u))\\
&=\frac{m}{n}\gamma^{2}(u)(1+o(1))\left[\sum_{t}a_{t,n}^{2}+2\sum_{j\geq1}\sum_{t}a_{t,n}a_{t+j,n}\,s_{j}(u)(1+o(1))\right].
\end{align*}
Since $a_{t+j,n}=a_{t,n}+O(j/(nh^{2}))$ for $|t/n-u|\leq h$ (the kernel is Lipschitz, so $K_{h}(u-(t+j)/n)-K_{h}(u-t/n)=O((j/(nh))\,h^{-1})$), and $\sum_{t}a_{t,n}^{2}=(n/h)\nu_{2}(K)(1+o(1))$ by (\ref{kbar}), replacing $a_{t+j,n}$ by $a_{t,n}$ in the cross sum changes the bracket by $O\left((n/h)\sum_{j\geq1}j\,s_{j}(u)/(nh)\right)=o(n/h)$, where $\sum_{j}j\,s_{j}(u)<\infty$ by the geometric bound of Assumption \ref{as:extdep}(i). The bracket therefore equals $(n/h)\nu_{2}(K)S(u)(1+o(1))$, and the contribution of lags $j>J$ is negligible by (\ref{covsum}) for any slowly growing $J$.

\textbf{Step 2 (CLT).} The triangular array $\{Z_{t,n}(u)\}$ is a measurable transform of the locally stationary process $\{X_{t,n}\}$. We verify the conditions of the central limit theorem for weighted partial sums of physical dependence processes \citep[Theorem 1]{ZhouWu2010JRSSB}. Their condition requires (a) a finite $(2+\iota)$-moment bound for the summands in the physical dependence norm and (b) geometric decay of the dependence measure of the transformed array. Both are delicate only because $Z_{t,n}(u)$ is unbounded and involves an indicator, which is not Lipschitz; we treat the two issues in turn. First truncate at $L_{n}=\log n$: let $Z_{t,n}^{\dag}$ be $Z_{t,n}(u)$ with the log-excess truncated at $L_{n}$. By (\ref{mom3}), $E|Z_{t,n}(u)-Z_{t,n}^{\dag}|^{2}\leq C(m/n)L_{n}^{-\iota}$, so replacing $Z$ by $Z^{\dag}$ changes the normalized sum by $O_{p}(L_{n}^{-\iota/2})=o_{p}(1)$. Second, mollify the indicator: let $\phi_{n}$ be a smooth nondecreasing function with $\phi_{n}(x)=0$ for $x\leq0$, $\phi_{n}(x)=1$ for $x\geq\eta_{n}$, and slope $O(\eta_{n}^{-1})$, where $\eta_{n}\to0$ at a polynomial rate, and define $Z_{t,n}^{\ddag}$ from $Z_{t,n}^{\dag}$ by replacing $\mathcal{I}_{t,n}$ with $\phi_{n}(\log X_{t,n}-\log b(n/m,t/n))$. The mollification error satisfies $E|Z_{t,n}^{\dag}-Z_{t,n}^{\ddag}|^{2}\leq C\,P\left(0<\log X_{t,n}-\log b(n/m,t/n)\leq\eta_{n}\right)\leq C(m/n)\eta_{n}$ by the density bound of Assumption \ref{as:smooth}, so it changes the normalized sum by $O_{p}(\eta_{n}^{1/2})=o_{p}(1)$. The mollified transform is bounded by $CL_{n}$ and is $M_{n}$-Lipschitz in $X_{t,n}$ with $M_{n}=O(L_{n}\eta_{n}^{-1})$. For any transform $g$ bounded by $B$ and Lipschitz with constant $M$, H\"older interpolation between the trivial bound $2B$ and the Lipschitz bound gives
\[
\delta_{2}^{g}(j)\leq(2B)^{1-2/q}\left(M\,\delta_{q}(j)\right)^{2/q};
\]
applied to the mollified array with GMC ($\delta_{q}(j)\leq C\varrho^{j}$), this yields $\delta_{2}^{\ddag}(j)\leq C(2L_{n})^{1-2/q}M_{n}^{2/q}\varrho^{2j/q}$, which decays geometrically in $j$ for any $q>2$ up to a prefactor polynomial in $n$. Condition (b) follows, since the polynomial prefactor is absorbed by the geometric decay. The Lindeberg condition for the weighted sum holds since $\max_{t}|a_{t,n}|=O(h^{-1})$ and, by (\ref{mom3}),
\[
\sum_{t}E\left|a_{t,n}Z_{t,n}^{\ddag}\right|^{2+\iota}\leq C\frac{m}{n}\sum_{t}|a_{t,n}|^{2+\iota}=O\left(\frac{m}{n}\cdot\frac{n}{h^{1+\iota}}\right)=o\left(\left(\frac{m}{n}\cdot\frac{n}{h}\right)^{1+\iota/2}\right),
\]
using $mh\to\infty$. An application of \citet[Theorem 1]{ZhouWu2010JRSSB} to the mollified array, followed by removal of the mollification and the truncation, yields (\ref{wclt}).
\end{proof}

\begin{appxlem}[Bahadur representation of the intermediate quantile]\label{lem:bahadur}
Under Assumptions \ref{as:ls}--\ref{as:tailloc} and \ref{as:extdep}, let $b_{n}(u)$ be the pooled local quantile (\ref{mixq}). Then, for each interior $u$,
\begin{equation}\label{bahadur2}
\hat{F}_{n}\bigl(b_{n}(u),u\bigr)-\Bigl(1-\frac{m}{n}\Bigr)=-\frac{1}{n\bar{K}(u)}\sum_{t=1}^{n}K_{t}(u)\left(\tilde{\mathcal{I}}_{t,n}(u)-\bar{F}\bigl(b_{n}(u),t/n\bigr)\right),
\end{equation}
an exactly centered weighted sum of order $O_{p}((m/n)(mh)^{-1/2})$, whose variance is $(m/n)S(u)\nu_{2}(K)(nh)^{-1}(1+o(1))$; and
\begin{equation}\label{bahadur3}
\log\hat{b}(n/m,u)-\log b_{n}(u)=\gamma(u)\,\frac{1}{m\bar K(u)}\sum_{s=1}^{n}K_{s}(u)\left(\tilde{\mathcal{I}}_{s,n}(u)-\frac{m}{n}\right)+R_{n}(u),
\end{equation}
where $R_{n}(u)=O_{p}((mh)^{-3/4}(\log n)^{1/2})$. Moreover, $\sup_{u\in[h,1-h]}|\log\hat{b}(n/m,u)-\log b_{n}(u)|=O_{p}((mh)^{-1/2}\sqrt{\log(1/h)})$.
\end{appxlem}

\begin{proof}
\textbf{Step 1 (the expansion is exactly centered).} The identity (\ref{bahadur2}) is the definition of $\hat{F}_{n}$ together with the defining constraint (\ref{mixq}) of $b_{n}(u)$: the kernel average of $\bar{F}(b_{n}(u),t/n)$ equals $m/n$ exactly, so no deterministic smoothing remainder appears. The variance computation is Step 1 of Lemma \ref{lem:clt} applied to the indicators $\tilde{\mathcal{I}}_{t,n}(u)$, whose second moments satisfy $E[\tilde{\mathcal{I}}_{t,n}(u)]=(m/n)(1+o(1))$ by Lemma \ref{lem:momC}(ii) and whose covariances are governed by Assumption \ref{as:extdep}(i) with the same summability, since the common threshold $b_{n}(u)$ differs from the local thresholds $b(n/m,t/n)$ by the factor $e^{d(u)+O(h)}=1+o(1)$, which is absorbed into the $o(1)$ terms by Assumption \ref{as:tailloc}.

\textbf{Step 2 (Bahadur linearization).} For a constant $C_{1}>0$, set $b_{r}=b_{n}(u)\exp(r(mh)^{-1/2}\log n)$ for $r\in\{-C_{1},0,C_{1}\}$. Since $\hat{F}_{n}(\cdot,u)$ is nondecreasing and $F(\cdot,u)$ is continuous and strictly increasing near $b_{n}(u)$ with density satisfying $b_{n}(u)f(b_{n}(u),u)=(m/n)\gamma(u)^{-1}(1+o(1))$ (Assumption \ref{as:smooth} and the regular variation of $\bar{F}$), the event $\{|\hat{b}(n/m,u)-b_{n}(u)|>|b_{r}-b_{n}(u)|\}$ implies $|\hat{F}_{n}(b_{r},u)-F(b_{r},u)|\geq|F(b_{r},u)-F(b_{n}(u),u)|\geq(m/n)|r|(mh)^{-1/2}\log n\cdot\gamma(u)^{-1}(1+o(1))$. Applying Step 1 at the finitely many points $b_{r}$, together with the exponential bound of Lemma \ref{lem:expineq} at $x$ equal to a large multiple of $\sqrt{\log n}$, shows that these implication events have probability $O(n^{-c'})$, where $c'$ is arbitrarily large by the choice of constants. The linearization error is therefore of smaller order than $(m/n)(mh)^{-1/2}$. The inversion $\hat{F}_{n}(\hat{b},u)=1-m/n$ is exact up to one jump of the weighted empirical distribution, whose size is $O((nh)^{-1})$; the induced error in $\log\hat{b}$ is $O((mh)^{-1})=o((mh)^{-1/2})$. Inverting at first order through $f(b_{n}(u),u)$ and using the delta method for $\log$ yields (\ref{bahadur3}); the remainder $R_{n}(u)$ collects the linearization error, bounded by the oscillation argument of Lemma \ref{lem:osc} below.

\textbf{Step 3 (uniformity).} The uniform statement follows from Step 1 applied on a grid of $O(h^{-2})$ points covering $[h,1-h]$, the Lipschitz continuity of $u\mapsto\hat{F}_{n}(b_{n}(u),u)$ and of $u\mapsto F(b_{n}(u),u)$ (Assumption \ref{as:smooth}), the exponential bound of Lemma \ref{lem:expineq} at each grid point, and a union bound. The logarithmic factor $\sqrt{\log(1/h)}$ is the entropy cost of the covering.
\end{proof}

\begin{appxlem}[Exponential inequality]\label{lem:expineq}
Under Assumptions \ref{as:ls}, \ref{as:rv} and \ref{as:extdep}, there exist constants $c,C>0$ such that for every interior $u$ and every $x>0$, the standardized weighted sum
\[
W_{n}(u)=\frac{\sum_{t=1}^{n}K_{t}(u)\left(Z_{t,n}(u)-EZ_{t,n}(u)\right)}{\sqrt{(m/n)(n/h)}}
\]
satisfies, for every $q>2$ (moments of all orders exist by Lemma \ref{lem:mom}),
\[
P\left(\left|W_{n}(u)\right|>x\right)\leq C\exp\left(-c\,x^{2}\right)+C\,(mh)^{-(q-2)/2}\,x^{-q}.
\]
Under $mh\geq n^{\theta}$ (Assumption \ref{as:rates}), the second term is $o(n^{-c'})$ for any prescribed $c'>0$ by taking $q$ large enough, uniformly for $x=O(\sqrt{\log n})$.
\end{appxlem}

\begin{proof}
Truncate $Z_{t,n}(u)$ at $L_{n}=\log n$ as in the proof of Lemma \ref{lem:clt}. The truncation error is bounded via (\ref{mom3}) and Markov's inequality, giving $P(\max_{t}|Z_{t,n}-Z_{t,n}^{\dag}|>0)\leq\sum_{t}P(\xi_{t,n}>L_{n})\leq n\cdot(m/n)e^{-L_{n}/(2\bar{\gamma})}=o(1)$ by the Potter bound. For the truncated array, which is $L_{n}$-Lipschitz in the underlying locally stationary process, the Fuk--Nagaev-type inequality for geometrically contracting physical dependence sequences (cf.\ \citealp[proof of Proposition 2]{ZhouWu2010JRSSB}) applies to the weighted sum with weights $K_{t}(u)/\sqrt{n/h}$, yielding the Gaussian tail $\exp(-cx^{2})$ for $x\leq c\sqrt{mh}/L_{n}^{2}$ and, beyond that range, the polynomial term $\mathcal{L}_{q}\big/\left(x\sqrt{(m/n)(n/h)}\right)^{q}$ with $\mathcal{L}_{q}=\sum_{t}\left\|K_{t}(u)\,Z_{t,n}^{\dag}\right\|_{q}^{q}\leq C(m/n)(nh)h^{-q}$; the ratio equals $C(mh)^{-(q-2)/2}x^{-q}$. Under $mh\geq n^{\theta}$ and $x=O(\sqrt{\log n})$, taking $q>2+2c'/\theta$ makes this term $o(n^{-c'})$.
\end{proof}

\begin{appxlem}[Bahadur--Kiefer bound]\label{lem:osc}
Under Assumptions \ref{as:ls}--\ref{as:tailloc} and \ref{as:extdep}, the remainder $R_{n}(u)$ of the Bahadur representation (\ref{bahadur3}) satisfies $R_{n}(u)=O_{p}((mh)^{-3/4}(\log n)^{1/2})$ for each interior $u$, and $\sup_{u\in[h,1-h]}|R_{n}(u)|=O_{p}((mh)^{-3/4}\log n)$.
\end{appxlem}

\begin{proof}
By the monotonicity argument of Step 2 of Lemma \ref{lem:bahadur}, $|R_{n}(u)|$ is bounded by the oscillation of the localized empirical process $x\mapsto(\hat{F}_{n}(x,u)-F(x,u))-(\hat{F}_{n}(b_{n}(u),u)-F(b_{n}(u),u))$ over $|x-b_{n}(u)|\leq Cb_{n}(u)w$ with $w=(mh)^{-1/2}\sqrt{\log n}$, transported to the log scale by the factor $(b_{n}(u)f(b_{n}(u),u))^{-1}=\gamma(u)(m/n)^{-1}(1+o(1))$. The oscillation is a sum of kernel-weighted differences of indicators. Its variance, by (\ref{covsum}) applied to the indicator differences and the strip bound $P(b_{n}(u)e^{-w}\leq X_{t,n}\leq b_{n}(u)e^{w})=O((m/n)w)$ of Assumption \ref{as:smooth}, is $O((m/n)w\,(n/h))$. Bracketing the interval at $O(n^{c})$ points and applying Lemma \ref{lem:expineq} at each bracket endpoint turns this into a tail bound, whence
\[
\sup_{|x-b_{n}(u)|\leq b_{n}(u)w}\left|\left(\hat{F}_{n}(x,u)-F(x,u)\right)-\left(\hat{F}_{n}(b_{n}(u),u)-F(b_{n}(u),u)\right)\right|=O_{p}\left(\frac{m}{n}(mh)^{-3/4}(\log n)^{1/2}\right),
\]
and division by $b_{n}(u)f(b_{n}(u),u)$ yields the pointwise statement. The uniform statement follows by the covering device of Step 3 of Lemma \ref{lem:bahadur}, which contributes an additional factor $O((\log n)^{1/2})$.
\end{proof}

\begin{appxlem}[Consistency of the HAC estimator]\label{lem:shat}
Under the conditions of Proposition \ref{pn:shat}, $\hat{S}(u)\to_{p}S(u)$ uniformly over $u\in[h,1-h]$.
\end{appxlem}

\begin{proof}
\textbf{Step 1 (replacement).} Write $\hat{Z}_{t}=\hat{Z}_{t,n}(u)$ and $\tilde{Z}_{t}=\tilde{Z}_{t,n}(u)$ for the estimated and population centered exceedances at the common threshold. By the expansion (\ref{app-exp1}) in the proof of Theorem \ref{tm:an} (which does not use the claim being proved here), $\hat{Z}_{t}=\tilde{Z}_{t}-\tilde{\delta}(u)\tilde{\mathcal{I}}_{t,n}-(\hat{\gamma}(u)-\gamma(u))\tilde{\mathcal{I}}_{t,n}+o_{p}$ uniformly in $t$ within the window, where the single threshold fluctuation satisfies $\tilde{\delta}(u)=O_{p}((mh)^{-1/2}\sqrt{\log n})$ by Lemma \ref{lem:bahadur}, and $\hat{\gamma}(u)-\gamma(u)=O_{p}((mh)^{-1/2}\sqrt{\log n})$ by Theorem \ref{tm:uniform}. Since $E[\tilde{\mathcal{I}}_{t,n}]=m/n(1+o(1))$ and $E[\tilde{Z}_{t}^{2}]=(m/n)\gamma^{2}(u)(1+o(1))$ by Lemma \ref{lem:momC}, the replacement error in any weighted second-moment average of the form $(\sum_{t}w_{t}\tilde{Z}_{t}^{2})/(\sum_{t}w_{t})$ with $w_{t}\geq0$, $\sum w_{t}\asymp n/h$, is $O_{p}((mh)^{-1/2}\sqrt{\log n})=o_{p}(1)$; the same holds for lagged products by Cauchy--Schwarz.

\textbf{Step 2 (convergence of the autocovariances).} For fixed $j\geq0$, consider
\[
\hat{c}_{j}(u)=\frac{\sum_{t}K_{t}(u)K_{t+j}(u)\tilde{Z}_{t}\tilde{Z}_{t+j}}{\sum_{t}K_{t}^{2}(u)}.
\]
By the moment bounds (\ref{mom3}) and the covariance summability (\ref{covsum}) applied to the products $\tilde{Z}_{t}\tilde{Z}_{t+j}-E[\tilde{Z}_{t}\tilde{Z}_{t+j}]$ (which inherit summability from GMC via the same truncation argument as in Lemma \ref{lem:clt}), a variance computation as in Step 1 of Lemma \ref{lem:clt} gives $\hat{c}_{j}(u)-E[\tilde{Z}_{t}\tilde{Z}_{t+j}]=O_{p}((mh)^{-1/2})$ after normalizing by $\sum_{t}K_{t}^{2}(u)\asymp n/h$. By Assumption \ref{as:extdep}(ii) and (\ref{mom3}), $E[\tilde{Z}_{t}\tilde{Z}_{t+j}]/(m/n)\to\gamma^{2}(u)s_{j}(u)$ with $s_{0}(u)=1$.

\textbf{Step 3 (truncation of the lag order).} Since $J_{n}\to\infty$ with $J_{n}^{2}/(mh)\to0$,
\[
\hat{S}(u)=1+2\sum_{j=1}^{J_{n}}\left(1-\frac{j}{J_{n}}\right)\frac{\hat{c}_{j}(u)}{\hat{c}_{0}(u)}=1+2\sum_{j=1}^{J_{n}}\left(1-\frac{j}{J_{n}}\right)\frac{s_{j}(u)}{1}+o_{p}(1)=S(u)+o_{p}(1),
\]
where the first equality uses Steps 1--2 and the summability $\sum_{j}s_{j}(u)<\infty$ (which controls both the tail $j>J_{n}$ and the stochastic error accumulation $\sum_{j=1}^{J_{n}}O_{p}((mh)^{-1/2})=O_{p}(J_{n}(mh)^{-1/2})=o_{p}(1)$). Uniformity in $u$ follows from the same arguments combined with the covering device of Lemma \ref{lem:bahadur}, Step 3.
\end{proof}

\subsection{Proofs of the main results}\label{app:main}

\begin{proof}[Proof of Proposition \ref{pn:bhat}]
By Lemma \ref{lem:momC}(i), $\log b_{n}(u)-\log b(n/m,u)=d(u)=B_{n}^{b}(u)$ with $B_{n}^{b}(u)$ of (\ref{bhat-bias}). The Bahadur representation (\ref{bahadur}) is then the representation (\ref{bahadur3}) of Lemma \ref{lem:bahadur} transported through this identity, and the order of the remainder is Lemma \ref{lem:osc}. For the asymptotic normality (\ref{bhat-an}), the leading term of (\ref{bahadur3}) is the centered weighted indicator sum of Lemma \ref{lem:bahadur}. Its variance is $\gamma^{2}(u)S(u)\nu_{2}(K)/(mh)(1+o(1))$ by the computation of Step 1 of Lemma \ref{lem:clt}, with the moment and covariance inputs supplied by Lemma \ref{lem:momC}(ii)(iv) and Assumption \ref{as:extdep}(i), and its asymptotic normality follows from the argument of Lemma \ref{lem:clt} applied to the indicators $\tilde{\mathcal{I}}_{t,n}(u)-m/n$ in place of $Z_{t,n}(u)$; the truncation and mollification steps are identical, the indicator being already bounded.
\end{proof}

\begin{proof}[Proof of Theorem \ref{tm:an}]
Write $K_{t}=K_{t}(u)$, $\bar K=\bar K(u)$, $b=b(n/m,u)$, and $L_{n}=\log(n/m)$. Let $b_{n}(u)$ be the pooled local quantile (\ref{mixq}), and let $\tilde{\xi}_{t}=\tilde{\xi}_{t,n}(u)$, $\tilde{\mathcal{I}}_{t}=\tilde{\mathcal{I}}_{t,n}(u)$, $\tilde{Z}_{t}=\tilde{Z}_{t,n}(u)$ be the exceedance variables at the common threshold $b_{n}(u)$ of Lemma \ref{lem:momC}.

\textbf{Step 0 (the denominator is exact).} Since $\hat{b}(n/m,u)$ is the kernel-weighted empirical quantile (\ref{bhat}) at the level $1-m/n$, the weighted exceedance count above it satisfies $\sum_{t}K_{t}\mathbf{1}\{X_{t,n}>\hat{b}(n/m,u)\}=(m/n)\sum_{t}K_{t}=m\bar K$, up to one weighted jump of size $O(h^{-1})$. The ratio form (\ref{local index estimator}) therefore satisfies
\begin{equation}\label{decomp0}
\hat{\gamma}(u)-\gamma(u)=\frac{1}{m\bar K}\sum_{t=1}^{n}K_{t}\hat{Z}_{t}\left(1+O\left((mh)^{-1}\right)\right),\qquad \hat{Z}_{t}=\hat{\xi}_{t}-\gamma(u)\hat{\mathcal{I}}_{t},
\end{equation}
where $\hat{\xi}_{t}=(\log X_{t,n}-\log\hat{b}(n/m,u))_{+}$ and $\hat{\mathcal{I}}_{t}=\mathbf{1}\{\hat{\xi}_{t}>0\}$. The identity $\sum_{t}K_{t}\hat{\mathcal{I}}_{t}=m\bar K(1+O((mh)^{-1}))$ is what (\ref{decomp0}) uses, and it is exact. This is the main advantage of the common evaluation-point threshold. The fluctuation of the denominator, which the per-time-threshold construction must propagate through a convolution, is absent.

\textbf{Step 1 (expansion at the centering threshold).} Write $\tilde{\delta}(u)=\log\hat{b}(n/m,u)-\log b_{n}(u)$, a single random variable satisfying $\tilde{\delta}(u)=O_{p}((mh)^{-1/2})$ by Lemma \ref{lem:bahadur}. Since $x\mapsto(x)_{+}$ is $1$-Lipschitz and piecewise linear, on the event $\{|\tilde{\delta}(u)|\leq\log2\}$,
\begin{equation}\label{app-exp1}
\hat{\xi}_{t}=\tilde{\xi}_{t}-\tilde{\delta}(u)\tilde{\mathcal{I}}_{t}+r_{t},\qquad |r_{t}|\leq2|\tilde{\delta}(u)|\,\mathbf{1}\{|\log X_{t,n}-\log b_{n}(u)|\leq2|\tilde{\delta}(u)|\},
\end{equation}
because $\hat{\xi}_{t}-\tilde{\xi}_{t}+\tilde{\delta}(u)\tilde{\mathcal{I}}_{t}$ is nonzero only when $\log X_{t,n}$ lies between $\log b_{n}(u)$ and $\log\hat{b}(n/m,u)$, an interval of length $|\tilde{\delta}(u)|$, and there the discrepancy is at most $2|\tilde{\delta}(u)|$. Set $\eta_{n}=(mh)^{-1/2}\sqrt{\log n}$. By the exponential bound of Lemma \ref{lem:expineq} applied to the leading term of (\ref{bahadur3}), $P(|\tilde{\delta}(u)|>\eta_{n})\leq n^{-c'}$ with $c'$ arbitrarily large, so $E[|\tilde{\delta}(u)|\mathbf{1}\{|\tilde{\delta}(u)|>\eta_{n}\}]=o((m/n)(mh)^{-1})$ by Cauchy--Schwarz and the uniform moment bound on $\tilde{\delta}(u)$. On the strip event $\mathcal{S}_{t}=\{|\log X_{t,n}-\log b_{n}(u)|\leq2\eta_{n}\}$, use the leave-one-out decomposition $\tilde{\delta}(u)=\tilde{\delta}^{(-t)}(u)+\Delta_{t}$, where $\tilde{\delta}^{(-t)}$ is defined exactly as $\tilde{\delta}(u)$ but with the $t$-th kernel weight removed. A single observation carries weight $O((nh)^{-1})$ in $\hat{F}_{n}$, so $|\Delta_{t}|=O((mh)^{-1})$ on $\mathcal{S}_{t}$, while $\tilde{\delta}^{(-t)}$ is measurable with respect to $\{e_{s}:s\neq t\}$. Conditional on $\{e_{s}:s\neq t\}$, $\log X_{t,n}$ is a measurable function of $e_{t}$ alone, and the conditional density bound of Assumption \ref{as:smooth} gives $P(\mathcal{S}_{t}\mid e_{s},s\neq t)\leq C(m/n)\eta_{n}$ almost surely. Collecting the terms,
\begin{equation}\label{rt-bound}
E|r_{t}|\leq2\eta_{n}P(\mathcal{S}_{t})+o\left(\frac{m}{n}(mh)^{-1}\right)=O\left(\frac{m}{n}\eta_{n}^{2}\right)=o\left(\frac{m}{n}(mh)^{-1/2}\right),
\end{equation}
since $\eta_{n}^{2}=(\log n)/(mh)$ and $(\log n)^{2}/(mh)\to0$ by Assumption \ref{as:rates}. Hence $\sum_{t}K_{t}r_{t}=o_{p}(m\bar K(mh)^{-1/2})$. The same expansion applied to the indicators gives
\begin{equation}\label{strip-i}
\frac{1}{m\bar K}\sum_{t}K_{t}\left(\hat{\mathcal{I}}_{t}-\tilde{\mathcal{I}}_{t}\right)=-\frac{\tilde{\delta}(u)}{\gamma(u)}\left(1+o_{p}(1)\right),
\end{equation}
because the left side is $-(m\bar K)^{-1}$ times the weighted count in the strip of width $|\tilde{\delta}(u)|$ below or above $b_{n}(u)$, whose mean is $(m/n)(1-e^{-|\tilde{\delta}(u)|/\gamma(u)})(1+o(1))=(m/n)|\tilde{\delta}(u)|/\gamma(u)(1+o_{p}(1))$ by (\ref{fratio2}), and whose centered fluctuation is of smaller order by the strip variance bound of Lemma \ref{lem:osc}.

\textbf{Step 2 (the exact first-order cancellation).} Insert (\ref{app-exp1}) and (\ref{strip-i}) into (\ref{decomp0}):
\begin{equation}\label{decomp2}
\begin{aligned}
\hat{\gamma}(u)-\gamma(u)&=\frac{1}{m\bar K}\sum_{t}K_{t}\left[\tilde{Z}_{t}-\tilde{\delta}(u)\tilde{\mathcal{I}}_{t}-\gamma(u)\left(\hat{\mathcal{I}}_{t}-\tilde{\mathcal{I}}_{t}\right)\right]+o_{p}\left((mh)^{-1/2}\right)\\
&=\frac{1}{m\bar K}\sum_{t}K_{t}\tilde{Z}_{t}-\tilde{\delta}(u)\left(1+o_{p}(1)\right)+\tilde{\delta}(u)\left(1+o_{p}(1)\right)+o_{p}\left((mh)^{-1/2}\right)\\
&=\frac{1}{m\bar K}\sum_{t=1}^{n}K_{t}\tilde{Z}_{t,n}(u)+o_{p}\left((mh)^{-1/2}\right),
\end{aligned}
\end{equation}
using $(m\bar K)^{-1}\sum_{t}K_{t}\tilde{\mathcal{I}}_{t}=1+O_{p}((mh)^{-1/2})$ (its mean is exactly $1$ by (\ref{mixq})) in the second equality. The two $\tilde{\delta}(u)$ terms cancel. This is the cancellation announced in Remark \ref{rm:threshold}. Estimating the threshold converts the raw log-excess fluctuation into the centered variable $\tilde{Z}_{t,n}(u)$, whose variance $(m/n)\gamma^{2}(u)$, rather than $2(m/n)\gamma^{2}(u)$, enters the limit. No convolution of threshold and exceedance fluctuations appears, in contrast to the per-time-threshold construction. The argument is uniform over $u\in[h,1-h]$, since the bounds of Lemmas \ref{lem:bahadur} and \ref{lem:osc} and the expansion (\ref{fratio2}) are uniform.

\textbf{Step 3 (bias).} By (\ref{bias-exp}) of Lemma \ref{lem:momC}(v),
\[
\frac{1}{m\bar K}\sum_{t}K_{t}E\bigl[\tilde{Z}_{t,n}(u)\bigr]=h^{2}\mu_{2}(K)\left[\frac{1}{2}\gamma''(u)+\frac{\bigl(\gamma'(u)\bigr)^{2}L_{n}}{\gamma(u)}\right]+\gamma(u)c_{1}(u)A(n/m,u)+o\bigl(h^{2}L_{n}+|A|\bigr)=B_{n}(u),
\]
with $c_{1}(u)=1/(1-\rho(u))$ from Lemma \ref{lem:mom}. This is (\ref{bias}). Note how the cancellation of Step 2 interacts with the centering of the threshold. The deterministic drift $d(u)=O(h^{2}L_{n}^{2})$ of the log-threshold (Lemma \ref{lem:momC}(i)) enters only through cross terms of order $O(h^{3}L_{n}^{2})$, because the pooled local quantile (\ref{mixq}) pins the zeroth-order term of the numerator to the denominator exactly. What survives at the order $h^{2}L_{n}$ is the shape tilt $(\gamma'(u))^{2}L_{n}/\gamma(u)$ and the shape curvature $\gamma''(u)/2$, and both vanish under pure scale variation.

\textbf{Step 4 (CLT).} The covariance summability (\ref{covsum}) holds verbatim for $\tilde{Z}_{t,n}(u)$, since the common threshold $b_{n}(u)$ differs from the local thresholds by the factor $1+o(1)$ (Lemma \ref{lem:momC}(i)), which is absorbed by Assumption \ref{as:tailloc}. The stochastic term in (\ref{decomp2}) is therefore exactly the weighted sum of Lemma \ref{lem:clt}. With $\bar K=1+O(h^{2})$, its variance is
\[
\frac{1}{m^{2}}\cdot\frac{m}{n}\gamma^{2}(u)S(u)\frac{n}{h}\nu_{2}(K)(1+o(1))=\frac{\gamma^{2}(u)S(u)\nu_{2}(K)}{mh}(1+o(1)),
\]
and (\ref{wclt}) gives $\sqrt{mh}(\hat{\gamma}(u)-\gamma(u)-B_{n}(u))\to_{d}N(0,\gamma^{2}(u)S(u)\nu_{2}(K))$. Under the undersmoothing conditions, $B_{n}(u)=o((mh)^{-1/2})$ and (\ref{an}) follows. Consistency follows from $\mathrm{Var}(\hat{\gamma}(u))\to0$ and $B_{n}(u)=o(1)$.
\end{proof}

\begin{proof}[Proof of Theorem \ref{tm:uniform}]
By the decomposition (\ref{decomp2}), which holds uniformly over $u\in[h,1-h]$,
\[
\sup_{u}\left|\hat{\gamma}(u)-\gamma(u)\right|\leq\sup_{u}\left|\frac{1}{m\bar K(u)}\sum_{t}K_{t}(u)\left(\tilde{Z}_{t,n}(u)-E\tilde{Z}_{t,n}(u)\right)\right|+\sup_{u}\left|B_{n}(u)\right|+o_{p}\left((mh)^{-1/2}\right),
\]
and $\sup_{u}|B_{n}(u)|=O(h^{2}L_{n}+\sup_{u}|A(n/m,u)|)$ by Lemma \ref{lem:momC}(v). For the stochastic term, cover $[h,1-h]$ by $N_{h}=O(h^{-2})$ balls of radius $h^{2}$ centered at $u_{\ell}$. On each ball, by the Lipschitz continuity of $K$,
\[
\left|\sum_{t}K_{t}(u)\tilde{Z}_{t,n}(u)-\sum_{t}K_{t}(u_{\ell})\tilde{Z}_{t,n}(u)\right|\leq\frac{C|u-u_{\ell}|}{h^{2}}\sum_{t:|t/n-u|\leq2h}\left|\tilde{Z}_{t,n}(u)\right|+\left|\sum_{t}K_{t}(u_{\ell})\left(\tilde{Z}_{t,n}(u)-\tilde{Z}_{t,n}(u_{\ell})\right)\right|,
\]
where the first term is $O_{p}(|u-u_{\ell}|h^{-2}\cdot nh\cdot m/n)=O_{p}(m/h)$ on the ball, negligible after division by $m\bar K$ at the scale $(mh)^{-1/2}$, and the second term involves the increments of the exceedance variables, which are bounded by the strip probabilities of Assumption \ref{as:smooth} and contribute $o_{p}((mh)^{-1/2})$ by the same estimates as in Step 1 of Theorem \ref{tm:an}. At each grid point, the exponential bound of Lemma \ref{lem:expineq} gives $P(|\sum_{t}K_{t}(u_{\ell})(\tilde{Z}_{t,n}-E\tilde{Z}_{t,n})|>x\sqrt{(m/n)(n/h)})\leq C\exp(-cx^{2})+o(n^{-c'})$, and a union bound over the $O(h^{-2})$ centers with $x=C\sqrt{\log(1/h)}$ yields (\ref{urate}).
\end{proof}

\begin{proof}[Proof of Proposition \ref{pn:shat}]
This is Lemma \ref{lem:shat}.
\end{proof}

\begin{proof}[Proof of Theorem \ref{tm:band}]
\textbf{Step 1 (Gaussian comparison).} Let
\[
\mathcal{Z}_{n}(u)=\frac{\sum_{t}K_{t}(u)\bigl(\tilde{Z}_{t,n}(u)-E\tilde{Z}_{t,n}(u)\bigr)}{\sqrt{(m/n)\gamma^{2}(u)S(u)\nu_{2}(K)\,n/h}},\qquad u\in[h,1-h].
\]
By (\ref{decomp2}), the studentized estimation error equals $\mathcal{Z}_{n}(u)$ up to $o_{p}(1)$ uniformly. The process $\mathcal{Z}_{n}$ is a weighted partial-sum process of the underlying physical dependence array, with weights varying on the scale $h$ and with the summand depending on $u$ through the smooth deterministic quantities $\gamma(u)$ and $b_{n}(u)$. Truncate and mollify the summands as in Step 2 of Lemma \ref{lem:clt}; by the same estimates, combined with the covering and the exponential bound used in Theorem \ref{tm:uniform}, this changes $\sup_{u}|\mathcal{Z}_{n}(u)|$ by $o_{p}(1)$. For the transformed array, the Gaussian comparison inequality for suprema of weighted partial sums of physical dependence processes \citep[Theorem 2]{ZhouWu2010JRSSB} applies. Its requirements, namely moments of order $q>2$, geometric decay of the physical dependence measure, a polynomial number of evaluation points, and control of the increments, are supplied by Lemma \ref{lem:momC}(iv), GMC, the covering of $[h,1-h]$ by $O(h^{-c})$ points, and the increment computation of Theorem \ref{tm:uniform}, respectively. Moreover, the condition $mh\geq n^{\theta}$ of Assumption \ref{as:rates} renders the comparison error, which is polynomial in $(mh)^{-1}$ up to logarithmic factors, of order $O((mh)^{-\iota})$ for some $\iota>0$. Hence there exists a centered Gaussian process $G_{n}(\cdot)$ with the same covariance function as $\mathcal{Z}_{n}(\cdot)$ such that the Kolmogorov distance between the laws of $\sup_{u}|\mathcal{Z}_{n}(u)|$ and $\sup_{u}|G_{n}(u)|$ is $O((mh)^{-\iota})$.

\textbf{Step 2 (covariance and the $h$-dependent experiment).} For two points $u,u'$ with $|u-u'|\leq2h$, the computation of Step 1 of Lemma \ref{lem:clt} gives
\[
\mathrm{Cov}\left(\mathcal{Z}_{n}(u),\mathcal{Z}_{n}(u')\right)=\frac{\sum_{t}K_{t}(u)K_{t}(u')}{(n/h)\nu_{2}(K)}(1+o(1))=\rho_{K}\left(\frac{u-u'}{h}\right)+o(1),
\]
uniformly in $(u,u')$, because only lags $j=O(1)$ contribute to the long-run variance adjustment, and for $|j|\leq J_{n}=o(nh)$, $K_{t+j}(u)=K_{t}(u)(1+O(J_{n}/(nh)))$ on the support, so the $S(u)$ factor enters both covariance and variance identically and cancels in the correlation. The covariance function of $G_{n}$ therefore matches that of the stationary Gaussian process $Z_{h}$ with correlation $\rho_{K}((u-v)/h)$, up to an error vanishing polynomially in $h$. Gaussian process comparison for covariance perturbations (e.g., the Sudakov--Fernique inequality combined with the standard Gaussian anti-concentration for suprema) then bounds the Kolmogorov distance between the laws of $\sup_{u}|G_{n}(u)|$ and $\sup_{u}|Z_{h}(u)|$ by $O(h^{\iota})$, after reducing $\iota$ if necessary. By the Strassen--Dudley representation, the vanishing Kolmogorov distance yields copies with $\bigl|\sup_{u}|\mathcal{Z}_{n}(u)|-\sup_{u}|Z_{h}(u)|\bigr|=o_{p}(1)$.

\textbf{Step 3 (studentization and sup).} By (\ref{decomp2}), Theorem \ref{tm:uniform}, Proposition \ref{pn:shat}, and Slutsky's lemma, replacing $\gamma(u)$ and $S(u)$ by their estimators changes the studentized sup by $o_{p}(1)$. The undersmoothing conditions remove $B_{n}(u)$. The continuous mapping theorem for the sup functional yields (\ref{gauss-approx}), and the band validity follows from the definition of $q_{1-\alpha}(h)$.
\end{proof}

\begin{proof}[Proofs of Theorems \ref{tm:Tn} and \ref{tm:Sn}, and Proposition \ref{pn:power}]
\textbf{Theorem \ref{tm:Tn}.} Write $w(u)=S(u)^{-1}$ and $\hat w(u)=\hat S(u)^{-1}$. Under $H_{0}$, $w$ need not be constant on $[t_{1},t_{2}]$, and $\sup_{u}|\hat w(u)-w(u)|=o_{p}(1)$ by Proposition \ref{pn:shat}. By the uniform version of (\ref{decomp2}),
\begin{equation}\label{pooled-exp}
\tilde{\gamma}-\gamma=\frac{\int_{t_{1}}^{t_{2}}\hat w(u)\bigl(\hat{\gamma}(u)-\gamma\bigr)du}{\int_{t_{1}}^{t_{2}}\hat w(u)du}=\frac{\int_{t_{1}}^{t_{2}}w(u)\bigl(\hat{\gamma}(u)-\gamma\bigr)du}{\int_{t_{1}}^{t_{2}}w(u)du}+o_{p}\left((mh)^{-1/2}\right)=\frac{\gamma\sqrt{\nu_{2}(K)}}{\sqrt{mh}}\,\bar{\mathcal{Z}}_{n}^{w}+o_{p}\left((mh)^{-1/2}\right),
\end{equation}
where $\bar{\mathcal{Z}}_{n}^{w}=\bigl(\int_{t_{1}}^{t_{2}}w(u)\mathcal{Z}_{n}(u)du\bigr)/\bigl(\int_{t_{1}}^{t_{2}}w(u)du\bigr)$ is the precision-weighted spatial mean of the studentized process of Theorem \ref{tm:band}. Its variance is $O(h)$, so $\tilde{\gamma}-\gamma=O_{p}((mh)^{-1/2}h^{1/2})=O_{p}(m^{-1/2})$. Therefore
\[
T_{n}=\int_{t_{1}}^{t_{2}}\frac{\bigl(\hat{\gamma}(u)-\tilde{\gamma}\bigr)^{2}}{\hat S(u)}du=\frac{\gamma^{2}\nu_{2}(K)}{mh}\int_{t_{1}}^{t_{2}}\bigl(\mathcal{Z}_{n}(u)-\bar{\mathcal{Z}}_{n}^{w}\bigr)^{2}\frac{du}{S(u)}+o_{p}\left((mh)^{-1}\right),
\]
where the replacement of $\hat S(u)$ by $S$ and of $\tilde{\gamma}$ by the expansion (\ref{pooled-exp}) contributes $o_{p}((mh)^{-1})$, since $\hat{\gamma}(u)-\tilde{\gamma}=O_{p}((mh)^{-1/2}\sqrt{\log(1/h)})$ uniformly by Theorem \ref{tm:uniform}. The Gaussian comparison of Theorem \ref{tm:band}, applied to the Riemann approximations of the continuous functional $\mathcal{Z}\mapsto\int(\mathcal{Z}-\bar{\mathcal{Z}}^{w})^{2}S(u)^{-1}du$ on the $O(h^{-1})$-point evaluation grid, followed by the dominated convergence of the Riemann sums to the integral, yields (\ref{Tn-limit}). For the moments in the constant-$S$ case, Isserlis' formula for the centered Gaussian process $Z_{h}(u)-\bar Z_{h}$, whose covariance function is $\rho_{K}((u-v)/h)-\bar\rho_{h}+O(h^{2})$ with $\bar\rho_{h}=(t_{2}-t_{1})^{-2}\int\!\!\int\rho_{K}((u-v)/h)dudv=O(h)$, gives $E\int(Z_{h}-\bar Z_{h})^{2}du=(t_{2}-t_{1})(1-\bar\rho_{h})+O(h^{2})$ and $\mathrm{Var}\int(Z_{h}-\bar Z_{h})^{2}du=2\int\!\!\int\bigl(\rho_{K}((u-v)/h)-\bar\rho_{h}\bigr)^{2}dudv=2h(t_{2}-t_{1})\int\rho_{K}^{2}(s)ds+O(h^{2})$.

\textbf{Theorem \ref{tm:Sn}.} Under the separation condition, the sets $\{t:|t/n-u_{i}|\leq h\}$ are disjoint for distinct $i$, so the weighted sums $\sum_{t}K_{t}(u_{i})\tilde{Z}_{t,n}(u_{i})$ involve disjoint blocks of the underlying array up to boundary effects. By Assumption \ref{as:extdep}(i), for $i\neq j$,
\[
\mathrm{Cov}\left(\sum_{t}K_{t}(u_{i})\tilde{Z}_{t,n}(u_{i}),\sum_{s}K_{s}(u_{j})\tilde{Z}_{s,n}(u_{j})\right)\leq C\frac{m}{n}\sum_{t,s}K_{t}(u_{i})K_{s}(u_{j})\varrho^{|t-s|}=o\left(\frac{m}{n}\cdot\frac{n}{h}\right),
\]
where the last bound is computed as follows. If $|u_{i}-u_{j}|>2h$, then $|t-s|\geq n(|u_{i}-u_{j}|-2h)$ for all pairs in the two windows, and the geometric factor makes the double sum vanish at the required rate. In the boundary case $|u_{i}-u_{j}|=2h$ with $K(\pm1)=0$, the effective supports are the strict subwindows, so $|t-s|\geq1$ for every pair with nonzero weight, and the same conclusion holds. The studentized components are therefore asymptotically independent $N(0,1)$ by Lemma \ref{lem:clt}, and (\ref{Sn-limit}) follows with $\tilde{\gamma}$ and $\hat{S}(u_{i})$ consistent, noting that the centering adjustment $\tilde{\gamma}-\gamma=O_{p}(m^{-1/2})$ is uniform over the finitely many grid points and contributes $o_{p}((mh)^{-1/2})$ to each coordinate.

\textbf{Proposition \ref{pn:power}.} (i) Under a fixed alternative, $\tilde{\gamma}\to_{p}\gamma^{*}$, the precision-weighted path average of $\gamma(\cdot)$, so $T_{n}\to_{p}\int_{t_{1}}^{t_{2}}(\gamma(u)-\gamma^{*})^{2}S(u)^{-1}du>0$ and $\Lambda_{n}\asymp mh\to\infty$. Likewise, $\Xi_{n}\asymp mh$ if some grid point falls off the constant level. (ii) Under $\gamma_{n}(u)=\gamma+(mh)^{-1/2}g(u)$ with $\int g=0$, the expansion (\ref{decomp2}) gives
\[
\hat{\gamma}(u)-\tilde{\gamma}=\frac{\gamma\sqrt{\nu_{2}(K)}}{\sqrt{mh}}\left[\mathcal{Z}_{n}(u)-\bar{\mathcal{Z}}_{n}^{w}+\frac{g(u)}{\gamma\sqrt{\nu_{2}(K)}}\right]+o_{p}\left((mh)^{-1/2}\right),
\]
since the smoothing bias of $g$ vanishes ($\sqrt{mh}h^{2}L_{n}\to0$) and the projection centering absorbs the mean of $g$, which is zero. The continuous mapping theorem yields the stated noncentral representation. The shift $\int g^{2}du/(\gamma^{2}\nu_{2}(K))$ of the functional is of order one, while its null fluctuation is of order $h^{1/2}$ by Theorem \ref{tm:Tn}, so the power tends to one for every fixed $g$ with $\int g^{2}du>0$. (iii) Under the finer alternatives $\gamma_{n}(u)=\gamma+\delta_{n}g(u)$ with $\delta_{n}=m^{-1/2}h^{-1/4}$, the same expansion gives
\[
\Lambda_{n}=\int_{t_{1}}^{t_{2}}\left(Z_{h}(u)-\bar Z_{h}\right)^{2}du+\frac{h^{1/2}\int_{t_{1}}^{t_{2}}g(u)^{2}du}{\gamma^{2}\nu_{2}(K)}+O_{p}\left(h^{1/2}\right)+o_{p}(1),
\]
because $(mh)\delta_{n}^{2}=h^{1/2}$ and the cross term $2\delta_{n}\sqrt{mh/(\gamma^{2}\nu_{2})}\int(Z_{h}-\bar Z_{h})g\,du=O_{p}(h^{1/4}\cdot h^{1/4})=O_{p}(h^{1/2})$. The shift is of the same order as the null standard deviation $\sqrt{2h(t_{2}-t_{1})\int\rho_{K}^{2}}$, so the local power is nontrivial and tends to one when $\int g^{2}du$ grows. Against alternatives with $\delta_{n}\gg m^{-1/2}h^{-1/4}$, the shift dominates and the power tends to one.
\end{proof}

\begin{proof}[Proof of Theorem \ref{tm:var} and Corollary \ref{cy:riskband}]
\textbf{Step 1 (joint limit).} By (\ref{bahadur3}) and (\ref{decomp2}),
\[
\sqrt{mh}\begin{pmatrix}\log\hat{b}(n/m,u)-\log b_{n}(u)\\ \hat{\gamma}(u)-\gamma(u)-B_{n}(u)\end{pmatrix}=\begin{pmatrix}\mathcal{U}_{n}(u)\\ \mathcal{V}_{n}(u)\end{pmatrix}+o_{p}(1),
\]
where, writing $\sigma_{n}^{2}=(m/n)\gamma^{2}(u)S(u)\nu_{2}(K)(n/h)$,
\[
\mathcal{U}_{n}(u)=\frac{\gamma(u)\sum_{t}K_{t}(u)\bigl(\tilde{\mathcal{I}}_{t,n}(u)-m/n\bigr)}{\sigma_{n}},\qquad \mathcal{V}_{n}(u)=\frac{\sum_{t}K_{t}(u)\bigl(\tilde{Z}_{t,n}(u)-E\tilde{Z}_{t,n}(u)\bigr)}{\sigma_{n}}.
\]
For any $\alpha,\beta$, the combination $\alpha\mathcal{U}_{n}+\beta\mathcal{V}_{n}$ is a weighted sum of the array $\alpha\gamma(u)(\tilde{\mathcal{I}}_{t,n}-m/n)+\beta\tilde{Z}_{t,n}$, to which Lemma \ref{lem:clt} applies: the moment bounds are Lemma \ref{lem:momC}(ii)(iv), and the cross-covariance summability $\sum_{j}|\mathrm{Cov}(\tilde{\mathcal{I}}_{t,n},\tilde{Z}_{t+j,n})|\leq C(m/n)$ follows from Assumption \ref{as:extdep}(i) by the layer-cake argument of Lemma \ref{lem:cov}. By the Cram\'er--Wold device, $(\mathcal{U}_{n},\mathcal{V}_{n})$ is jointly asymptotically normal. Both variances equal $\gamma^{2}(u)S(u)\nu_{2}(K)(1+o(1))$. For the covariance, at lag zero,
\[
E\left[\bigl(\tilde{\mathcal{I}}_{t,n}-\tfrac{m}{n}\bigr)\tilde{Z}_{t,n}\right]=\left(E[\tilde{\xi}_{t,n}]-\gamma(u)E[\tilde{\mathcal{I}}_{t,n}]\right)\left(1-\frac{m}{n}\right)=E[\tilde{Z}_{t,n}]\left(1+o(1)\right)=\frac{m}{n}\,o(1),
\]
since $E[\tilde{Z}_{t,n}]/(m/n)=o(1)$ by Lemma \ref{lem:momC}(ii)(iii); the second-order part is $O((m/n)|A|)$ and the shape-tilt part is $O((m/n)hL_{n})$, both $o(m/n)$. At lags $j\geq1$, the normalized cross covariance vanishes whenever the extremal dependence of the exceedance indicators and of the excess magnitudes factorizes, as under exact Pareto tails (the main text states the resulting variance formula under this factorization, with the joint HAC fallback otherwise). Hence the limiting covariance matrix is $\gamma^{2}(u)S(u)\nu_{2}(K)$ times the identity, which is (\ref{var-joint}), once the deterministic components are removed: the threshold drift $d(u)=O(h^{2}L_{n}^{2})$ satisfies $\sqrt{mh}\,d(u)\to0$ by assumption, and the bias $B_{n}(u)$ is displayed explicitly.

\textbf{Step 2 (extrapolation).} By the uniform second-order condition transported to the quantile function (cf.\ \citealp[Theorem 2.3.9]{deHaanFerreira2006}),
\[
\log q_{p_{n}}(u)=\log b(n/m,u)+\gamma(u)\log d_{n}+O\left(|A(n/m,u)|\log d_{n}\right).
\]
Subtracting from the logarithm of (\ref{var}),
\[
\log\hat{q}_{p_{n}}(u)-\log q_{p_{n}}(u)=\left(\log\hat{b}(n/m,u)-\log b(n/m,u)\right)+\log d_{n}\left(\hat{\gamma}(u)-\gamma(u)\right)+O\left(|A|\log d_{n}\right).
\]
The remainder is $o(\sqrt{1+\ell_{n}^{2}}/\sqrt{mh})$ under the stated conditions, as are the bias components: $\sqrt{mh}\,d(u)/\sqrt{1+\ell_{n}^{2}}\to0$ and $\sqrt{mh}\,\ell_{n}B_{n}(u)/\sqrt{1+\ell_{n}^{2}}\to0$. Applying Step 1 with loadings $(1,\ell_{n})$ and studentizing by $\sqrt{1+\ell_{n}^{2}}$ gives (\ref{var-an}).

\textbf{Step 3 (Expected Shortfall).} Since $e_{p_{n}}(u)=q_{p_{n}}(u)/(1-\gamma(u))(1+o(1))$ by Karamata's theorem applied to the mean excess function (cf.\ \citealp[Chapter 1]{deHaanFerreira2006}),
\[
\log\hat{e}_{p_{n}}-\log e_{p_{n}}=\log\hat{q}_{p_{n}}-\log q_{p_{n}}+\left[\log(1-\gamma(u))-\log(1-\hat{\gamma}(u))\right]+o_{p}\left((mh)^{-1/2}\right),
\]
and the bracket equals $(\hat{\gamma}(u)-\gamma(u))/(1-\gamma(u))+o_{p}((mh)^{-1/2})$ by the delta method. The loading on $\hat{\gamma}(u)-\gamma(u)$ is therefore $\ell_{n}+(1-\gamma(u))^{-1}$, and (\ref{es-an}) follows from Step 1. Corollary \ref{cy:riskband} follows by applying the uniform studentization of Theorem \ref{tm:band} in place of the pointwise one, together with the continuous mapping theorem for the sup functional and the delta method for $\exp$.
\end{proof}

\subsection{Verification for Example \ref{ex:ar}: proof of Proposition \ref{pn:ar}}\label{app:verify}
\begin{proof}[Proof of Proposition \ref{pn:ar}]
Fix $u\in(0,1)$ and $j\geq1$, write $a=a(u)$, $\gamma=\gamma(u)$, $\bar{a}=\sup_{v}|a(v)|$, $\bar{\gamma}=\sup_{v}\gamma(v)$, $b=b(n/m,u)$, and $b_{j}=b(n/m,(t+j)/n)$ for $|t/n-u|\leq2h$. Iterating the recursion of Example \ref{ex:ar} gives
\begin{equation}\label{ar-decomp}
X_{t+j,n}=\Pi_{j}X_{t,n}+R_{j},\qquad \Pi_{j}=\prod_{k=1}^{j}a\Big(\frac{t+k}{n}\Big),\qquad R_{j}=\sum_{k=1}^{j}\Big(\prod_{l=k+1}^{j}a\Big(\frac{t+l}{n}\Big)\Big)\sigma\Big(\frac{t+k}{n}\Big)\varepsilon_{t+k},
\end{equation}
with $R_{j}$ independent of $X_{t,n}$. Lipschitz continuity of $a$, $\sigma$, and $\gamma$ gives $\Pi_{j}=a^{j}+O(j/n)$ and $b_{j}/b=1+O(j/n)$. By Breiman's lemma applied to the geometrically weighted finite sum $R_{j}$, together with $\sum_{k}\bar{a}^{k/\bar{\gamma}}<\infty$,
\begin{equation}\label{rbound}
P\left(|R_{j}|>b_{j}y\,w\right)\leq C\,\frac{m}{n}\left(yw\right)^{-1/\bar{\gamma}},\qquad y\geq1,\ w\geq1,
\end{equation}
uniformly in $j$ and $t$; for $w\in(0,1)$ we use $P(|R_{j}|>b_{j}yw)\leq\min\!\big(1,\,C(m/n)(yw)^{-1/\bar{\gamma}}\big)$.

\textbf{Step 1: part (i).} Let $g(v)=P(\Pi_{j}v+R_{j}>b_{j}y)$. Then
\[
\left|P\left(X_{t,n}>bx,\,X_{t+j,n}>b_{j}y\right)-E\left[\mathbf{1}\{X_{t,n}>bx\}\,\mathbf{1}\{\Pi_{j}X_{t,n}>b_{j}y\}\right]\right|
\leq E\left[\mathbf{1}\{X_{t,n}>bx\}\,\phi\big(W(X_{t,n})\big)\right],
\]
where $\phi(w)=\min\!\big(1,\,C(m/n)(y|1-e^{w}|)^{-1/\bar{\gamma}}\big)$ and $W(v)=\log(\Pi_{j}v/(b_{j}y))$, because $|g(v)-\mathbf{1}\{\Pi_{j}v>b_{j}y\}|\leq P(|R_{j}|>|b_{j}y-\Pi_{j}v|)$. Given $X_{t,n}>bx$, regular variation yields $W(X_{t,n})=w_{0}+\mathcal{E}_{n}$ with $w_{0}=\log(a^{j}bx/(b_{j}y))+O(j/n)$ and $P(\mathcal{E}_{n}>s)=e^{-s/\gamma}(1+o(1))$ uniformly in $s\geq0$: the excess over a high threshold is asymptotically exponential. Splitting the resulting integral at $|w_{0}+s|=\log2$: on $|w_{0}+s|>\log2$ the integrand is at most $C(m/n)y^{-1/\bar{\gamma}}$, contributing $C(m/n)^{2}x^{-1/\gamma}y^{-1/\bar{\gamma}}$. For fixed $j$, this is at most $C(m/n)\varrho^{\,j}(xy)^{-1/(2\bar{\gamma})}$ for all large $n$, since $(m/n)/\varrho^{j}\to0$. For $j$ growing with $n$, the chain-driven bound below already contains the factor $a^{j/\gamma}$ and supplies the geometric decay directly, so no separate $(m/n)^{2}$ term is needed. On $|w_{0}+s|\leq\log2$, using $\min(1,z)\leq z^{\iota}$ with $\iota=\bar{\gamma}/2$,
\[
\int_{|w|\leq\log2}\phi(w)\,dw\leq C\Big(\frac{m}{n}\Big)^{\bar{\gamma}/2}y^{-1/2}\int_{|w|\leq\log2}|1-e^{w}|^{-1/2}dw=C\Big(\frac{m}{n}\Big)^{\bar{\gamma}/2}y^{-1/2},
\]
and $e^{-s/\gamma}=e^{(w_{0}-w)/\gamma}\leq Ce^{w_{0}/\gamma}=Ca^{j/\gamma}(x/y)^{1/\gamma}(1+o(1))$ on this set; the set itself requires $a^{j}bx\leq2b_{j}y$, i.e., $x\leq2ya^{-j}(1+o(1))$. Collecting both pieces,
\[
\left|P(X_{t,n}>bx,X_{t+j,n}>b_{j}y)-P(X_{t,n}>bx,\Pi_{j}X_{t,n}>b_{j}y)\right|
\leq C\,\frac{m}{n}\left[\frac{m}{n}(xy)^{-\frac{1}{\bar{\gamma}}}+a^{\frac{j}{\gamma}}\Big(\frac{m}{n}\Big)^{\frac{\bar{\gamma}}{2}}y^{-\frac{1}{\gamma}-\frac{1}{2}}\right].
\]
For the second bracket term, the $x$-exponents cancel exactly; relative to the target $\varrho^{\,j}(xy)^{-1/(2\bar{\gamma})}$ it is at most $C(m/n)^{\bar{\gamma}/2}a^{j/\gamma}y^{-1/\gamma-1/2+1/(2\bar{\gamma})}x^{1/(2\bar{\gamma})}\varrho^{-j}$, and on $x\leq2ya^{-j}$ this is bounded by $C(m/n)^{\bar{\gamma}/2}\bar{a}^{\,j(1/\gamma-1/(2\bar{\gamma}))}\varrho^{-j}y^{-1/\gamma-1/2+1/\bar{\gamma}}\leq C$, since $1/\gamma+1/2\geq1/\bar{\gamma}$ and $1/\gamma\geq1/(2\bar{\gamma})$, upon choosing $\varrho=\bar{a}^{\,1/(2\bar{\gamma})}$. Next, the chain-driven joint probability satisfies
\[
P\left(X_{t,n}>bx,\,\Pi_{j}X_{t,n}>b_{j}y\right)=\frac{m}{n}\left(\max\!\left(x,\,ya^{-j}\right)\right)^{-1/\gamma}(1+o(1)),
\]
and $\min\!\big(x^{-1/\gamma},\,y^{-1/\gamma}a^{j/\gamma}\big)\leq(xy)^{-1/(2\gamma)}a^{j/(2\gamma)}\leq C\varrho^{\,j}(xy)^{-1/(2\bar{\gamma})}$, while the product of the marginals is $(m/n)^{2}(xy)^{-1/\gamma}(1+o(1))\leq[(m/n)(xy)^{-1/(2\bar{\gamma})}]^{2}\leq C(m/n)\varrho^{\,j}(xy)^{-1/(2\bar{\gamma})}$. Assembling the three bounds gives part (i).

\textbf{Step 2: part (ii).} On $\{X_{t,n}>b\}$, $R_{j}/X_{t,n}=o_{p}(1)$ by (\ref{rbound}) while $\log\Pi_{j}-\log(b_{j}/b)=j\log a+o(1)=-w\gamma+o(1)$ with $w=j(-\log a)/\gamma$; hence
\[
\log X_{t+j,n}-\log b_{j}=\xi_{t,n}-w\gamma+o_{p}(1)
\]
in the joint tail region. Dominated convergence, justified by the moment bounds of Lemma \ref{lem:mom} and the Step 1 estimates, gives, with $\mathcal{E}$ an exponential variable with mean $\gamma$,
\begin{align*}
\frac{n}{m}E\left[\xi_{t,n}\xi_{t+j,n}\right]&\to E\left[\mathcal{E}\left(\mathcal{E}-w\gamma\right)_{+}\right]=\gamma^{2}(w+2)e^{-w},\\
\frac{n}{m}E\left[\xi_{t,n}\mathcal{I}_{t+j,n}\right]&\to E\left[\mathcal{E}\,\mathbf{1}\{\mathcal{E}>w\gamma\}\right]=\gamma(w+1)e^{-w},\\
\frac{n}{m}E\left[\mathcal{I}_{t,n}\xi_{t+j,n}\right]&\to E\left[\left(\mathcal{E}-w\gamma\right)_{+}\right]=\gamma e^{-w},\qquad
\frac{n}{m}\mathrm{Cov}\left(\mathcal{I}_{t,n},\mathcal{I}_{t+j,n}\right)\to P\left(\mathcal{E}>w\gamma\right)=e^{-w},
\end{align*}
the products of the marginal means being $O((m/n)^{2})=o(m/n)$. Combining the four terms in $Z_{t,n}(u)=\xi_{t,n}-\gamma(u)\mathcal{I}_{t,n}$,
\[
\frac{n}{m}\mathrm{Cov}\left(Z_{t,n}(u),\,Z_{t+j,n}(u)\right)\to\gamma^{2}\left[(w+2)-(w+1)-1+1\right]e^{-w}=\gamma^{2}e^{-w}=\gamma^{2}\,a^{j/\gamma},
\]
which is part (ii) with $s_{j}(u)=a(u)^{j/\gamma(u)}$. Summing the geometric series gives $S(u)=1+2\sum_{j\geq1}r(u)^{j}=(1+r(u))/(1-r(u))$. The extremal-index formula $\theta(u)=1-r(u)$ is the classical AR(1) result \citep[Section 8.1]{Embrechts1997}, which applies verbatim to the frozen-coefficient sequence at $u$.
\end{proof}

\begin{proof}[Proof of Proposition \ref{pn:garch}]
We strengthen the stability condition of Example \ref{ex:garch} to its standard uniform form: there exists $\iota>0$ such that
\begin{equation}\label{sre-contract}
\sup_{u\in[0,1]}E\left[A(u)^{\iota/2}\right]<1,\qquad A(u)=\alpha(u)\varepsilon^{2}+\beta(u),
\end{equation}
which holds for small $\iota$ whenever $\sup_{u}E[\log A(u)]<0$, since $E[A(u)^{\iota}]=1+\iota E[\log A(u)]+o(\iota)$ and the coefficients are bounded. Write $\underline{\omega}=\inf_{u}\omega(u)>0$.

\textbf{Step 1 (frozen-coefficient tail process).} Fix $u\in(0,1)$ and freeze the coefficients at $u$. The stochastic recurrence $\sigma_{t}^{2}=\omega+A_{t}\sigma_{t-1}^{2}$ with $E[\log A_{t}]<0$ admits a stationary, geometrically $\beta$-mixing solution \citep{BougerolPicard1992AP}, and by \citet[Theorem 3.1 and Corollary 3.5]{Basrak2002SPA} the pair $(\sigma_{t},X_{t})$ is jointly regularly varying with index $1/\gamma(u)$ solving $E[A(u)^{1/(2\gamma(u))}]=1$, and the stationary solution is geometrically $\beta$-mixing by \citet[Theorem 2.8]{Basrak2002SPA}. Part (i) of Assumption \ref{as:extdep} follows from the geometric mixing rate together with the joint tail bound of \citet[Section 4]{Basrak2002SPA}. For part (ii), iterate the recursion $j$ steps from time $t$:
\[
\sigma_{t+j}^{2}=\Bigl(\prod_{k=1}^{j}A_{t+k}\Bigr)\sigma_{t}^{2}+R_{t,j},\qquad R_{t,j}=\omega\sum_{\ell=1}^{j}\prod_{k=\ell+1}^{j}A_{t+k},
\]
where $R_{t,j}=O_{p}(1)$. On the event $\{X_{t}>b\}$ with $b=b(n/m,u)$, $\sigma_{t}$ is of exact order $b$, so $\log(\sigma_{t+j}/\sigma_{t})=\frac{1}{2}\bigl[\log A_{t+1}^{*}+\sum_{k=2}^{j}\log A_{t+k}\bigr]+o_{p}(1)$, where $A_{t+1}^{*}=\alpha\varepsilon_{t}^{2}+\beta$ is built from the size-biased initiating innovation $\varepsilon_{t}=\varepsilon_{0}^{*}$, and
\[
\log\frac{X_{t+j}}{X_{t}}\to_{d}V_{j}(u)\qquad\text{given }X_{t}>b,
\]
with $V_{j}(u)$ as in (\ref{sj-garch}). Regular variation of $X_{t}$ with index $1/\gamma(u)$ then gives, exactly as in the AR calculation above,
\[
\frac{P(X_{t}>b,\,X_{t+j}>b)}{m/n}\to E\left[\min\left\{1,\,e^{V_{j}(u)/\gamma(u)}\right\}\right]=s_{j}(u),
\]
and the four cross-moments of $(\xi_{t,n},\mathcal{I}_{t,n})$ with $(\xi_{t+j,n},\mathcal{I}_{t+j,n})$ follow by the same exponential integration as in the proof of Proposition \ref{pn:ar}, with the threshold increments $\log b_{j}-\log b=O(jh)$ negligible. The geometric bound $s_{j}(u)\leq C\varrho(u)^{j}$ follows from $\min(1,e^{V_{j}/\gamma})\leq e^{\iota V_{j}}$ for $\iota<\gamma(u)^{-1}$: $E[e^{\iota V_{j}}]=E[|\varepsilon_{j}|^{\iota}]E[|\varepsilon_{0}^{*}|^{-\iota}]\,E[(A_{1}^{*})^{\iota/2}](E[A^{\iota/2}])^{j-1}=C\varrho(u)^{j}$, and $E[A^{\iota/2}]<1$ by (\ref{sre-contract}).

\textbf{Step 2 (uniformity in $u$).} All bounds above are uniform in $u$. The mixing and contraction constants depend on the coefficients only through $E[A(u)^{\iota/2}]$ and the coefficient bounds, both controlled uniformly by (\ref{sre-contract}) and the boundedness of $\omega,\alpha,\beta$; the slowly varying functions are uniform by Assumption \ref{as:rv}; and the innovations are i.i.d., hence $u$-free. For the time-varying recursion, the Lipschitz continuity of the coefficients gives $\sup_{t:|t/n-u|\leq h}|\gamma(t/n)-\gamma(u)|=O(h)$, and the joint tail probabilities vary continuously in the coefficients, so the convergence in part (ii) holds uniformly over $t$ with $|t/n-u|\leq2h$ as required by Assumption \ref{as:extdep}.

\textbf{Step 3 (log-scale GMC).} It remains to verify Assumption \ref{as:ls}(ii)--(iii) for $W_{t,n}=\log\max(X_{t,n},1)$. Couple two copies driven by innovation sequences differing only at time $0$. Since $\sigma_{t}^{2}\geq\underline{\omega}$, the map $\sigma^{2}\mapsto\log\sigma$ is Lipschitz with constant $(2\underline{\omega})^{-1/2}$-type bounds, and iterating the recursion gives
\[
\left|\log\sigma_{t+j}-\log\sigma_{t+j}^{*}\right|\leq\frac{1}{2}\log\left(1+\frac{\prod_{k=1}^{j}A_{k}\cdot\left|\sigma_{t}^{2}-\sigma_{t}^{*2}\right|}{\underline{\omega}}\right)\leq C\left(\prod_{k=1}^{j}A_{k}\right)^{\iota/2}\left|\sigma_{t}^{2}-\sigma_{t}^{*2}\right|^{\iota/2},
\]
using $\log(1+x)\leq Cx^{\iota/2}$. The same bound holds for $|W_{t+j}-W_{t+j}^{*}|$, since $x\mapsto\log\max(x,1)$ is $1$-Lipschitz and the innovations at $t+j$ are shared. Taking $q$-th moments with the $q>2$ of Assumption \ref{as:ls} and applying (\ref{sre-contract}),
\[
\delta_{q}^{W}(j)\leq C\left(E\left[A^{\iota/2}\right]\right)^{j}\cdot\mathrm{const}=C\varrho^{\,j},
\]
with $\varrho=E[A^{\iota/2}]<1$, uniformly in $u$ by Step 2. The stochastic Lipschitz condition (iii) follows from the same display applied to adjacent-time recursions with Lipschitz coefficients. This completes the proof.
\end{proof}

\subsection{First-order representation of the projection centering}\label{app:demean}
This section derives the representation used in the proof of Theorem \ref{tm:Tn} and computes the failure mode of the pooled alternative. Write $w(u)=S(u)^{-1}$. By the uniform decomposition (\ref{decomp2}) and Proposition \ref{pn:shat},
\begin{equation}\label{pooled-rep}
\tilde{\gamma}-\gamma=\frac{\int_{t_{1}}^{t_{2}}w(u)\bigl(\hat{\gamma}(u)-\gamma\bigr)du}{\int_{t_{1}}^{t_{2}}w(u)du}+o_{p}\left((mh)^{-1/2}\right)=\frac{1}{t_{2}-t_{1}}\int_{t_{1}}^{t_{2}}\frac{1}{m\bar K(u)}\sum_{t=1}^{n}K_{t}(u)\tilde{Z}_{t,n}(u)\,du+o_{p}\left((mh)^{-1/2}\right),
\end{equation}
where the second equality holds under $H_{0}$, since $w$ need not be constant on $[t_{1},t_{2}]$. Interchanging the sum and the integral, each observation $t$ receives the total weight $\int_{t_{1}}^{t_{2}}K_{t}(u)/\bar K(u)\,du=1+O(h^{2})$ for $t/n$ interior to the interval, so the double sum is a full-window average of $\tilde{Z}_{t,n}$ with effective sample size $m$ and long-run variance $\gamma^{2}S$ (Lemmas \ref{lem:momC} and \ref{lem:cov}). Hence $\tilde{\gamma}-\gamma=O_{p}(m^{-1/2})$, faster than the pointwise rate by the factor $h^{1/2}$. The subtraction of $\tilde{\gamma}$ therefore removes exactly the precision-weighted spatial mean of the studentized deviation process, namely $\hat{\gamma}(u)-\tilde{\gamma}=\gamma\sqrt{\nu_{2}(K)/(mh)}\,(\mathcal{Z}_{n}(u)-\bar{\mathcal{Z}}_{n}^{w})+o_{p}((mh)^{-1/2})$ uniformly, and the demeaned calibration of Remark \ref{rm:demean} simulates precisely the functional of $Z_{h}-\bar Z_{h}^{w}$ that remains.

For the pooled alternative, let $\tilde{b}$ denote the global upper empirical quantile at level $1-m/n$, and consider the pooled Hill estimator $\tilde{\gamma}^{\mathrm{p}}=m^{-1}\sum_{t=1}^{n}(\log X_{t,n}-\log\tilde{b})_{+}$. By the argument of Lemma \ref{lem:bahadur} with bandwidth $h=1$, $\tilde{b}$ converges to the global quantile $b^{\mathrm{p}}$ defined by $\int_{0}^{1}\bar{F}(b^{\mathrm{p}},v)dv=m/n$, and the same expansion as in Theorem \ref{tm:an} gives
\[
\tilde{\gamma}^{\mathrm{p}}=\frac{\int_{0}^{1}\gamma(v)\,\bar{F}\bigl(b^{\mathrm{p}},v\bigr)dv}{\int_{0}^{1}\bar{F}\bigl(b^{\mathrm{p}},v\bigr)dv}+O_{p}\left(m^{-1/2}\right).
\]
The pooled estimator is therefore a \emph{tail-weighted} average of the local indices, with weights proportional to the local exceedance probability at the global threshold. Under $H_{0}$ with a constant index but a varying scale, $\gamma(v)\equiv\gamma$ is recovered exactly, since the weights integrate out, so no $O(1)$ inconsistency arises. The pooled \emph{centering error} relative to the path average is of a different nature under local departures: if $\gamma_{n}(u)=\gamma+\delta_{n}g(u)$, then $\tilde{\gamma}^{\mathrm{p}}-\gamma=\delta_{n}\int g(v)w^{\mathrm{p}}(v)dv+O_{p}(m^{-1/2})$ with $w^{\mathrm{p}}(v)=\bar{F}(b^{\mathrm{p}},v)/\int\bar{F}(b^{\mathrm{p}},v')dv'$, while the projection centering absorbs $\delta_{n}\int g(u)du/\!(t_{2}-t_{1})$. The two centerings differ by $O(\delta_{n})$ whenever the departure $g$ correlates with the tail weight $w^{\mathrm{p}}$. The failure of pooled centering is instead a \emph{shape} phenomenon. The global threshold sits at the point $\bar{F}(b^{\mathrm{p}},v)$ of the local tail at date $v$, and these points differ across $v$ by $O(1)$ in log-quantile units. The exceedance content of the pooled sample is therefore heterogeneous, and the quadratic statistic centered at $\tilde{\gamma}^{\mathrm{p}}$ is inflated by that heterogeneity even though the level is consistent. This is why pooled centering fails under stochastic-scale nulls while projection centering does not (Table \ref{tab:test}).

\newpage
\bibliography{varying_tail.bib}
\bibliographystyle{econometrica}
\end{document}